\documentclass[10pt]{wlscirep}
\usepackage{mathptmx}
\usepackage{amsfonts,amssymb,stmaryrd,latexsym,amsmath,braket,booktabs}
\usepackage{graphicx,float} % subfigure has long been deprecated
\usepackage{subcaption} % ---- Add By YZM
\usepackage[export]{adjustbox}
\usepackage{array}
\usepackage{bbm}
\usepackage{bm}
\usepackage{mathrsfs}
\usepackage{slashed}
\usepackage{braket}
\usepackage{verbatim}
\usepackage[utf8]{inputenc}
\usepackage{tabularx}%, tabularray}
\usepackage{siunitx}
\usepackage{xparse}
\usepackage{graphicx}
\usepackage{hyperref}
\usepackage{times}
\usepackage{slashed}
\usepackage{braket}
\usepackage{verbatim}
\usepackage{multirow}
\usepackage[utf8]{inputenc}

\newcommand{\PreserveBackslash}[1]{\let\temp=\\#1\let\\=\temp}
\newcolumntype{C}[1]{>{\PreserveBackslash\centering}p{#1}}
\newcolumntype{R}[1]{>{\PreserveBackslash\raggedleft}p{#1}}
\newcolumntype{L}[1]{>{\PreserveBackslash\raggedright}p{#1}}

\newcommand{\approxpropto}{%
  \ensuremath{%
      \mathrel{\vbox{\offinterlineskip\ialign{%
        \hfil##\hfil\cr
        $\sim$\cr
        \noalign{\kern-0.1ex}
        $\propto$\cr
}}}}}

\newcommand{\blockcomment}[1]{}

\newcommand{\MSU}{Department of Physics and Astronomy, Michigan State University, East Lansing, Michigan 48824}

\newcommand{\FRIB}{Facility for Rare Isotope Beams, Michigan State University, East Lansing, Michigan 48824}

\newcommand{\LANL}{Theoretical Division, Los Alamos National Laboratory, Los Alamos, NM 87545, USA}
\newcommand{\TRIUMF}{TRIUMF, 4004 Wesbrook Mall, Vancouver, BC V6T 2A3, Canada}
\newcommand{\Guelph}{Department of Physics, University of Guelph, Guelph, ON N1G 2W1, Canada}
\newcommand{\ASU}{Department of Physics, Arizona State University, Tempe, Arizona 85287, USA}
\newcommand{\Florida}{Department of Physics and Quantum Theory Project, University of Florida, Gainesville, FL 32611, USA}
\makeatletter
\def\namedlabel#1#2{\begingroup
   \def\@currentlabel{#2}%
   \label{#1}\endgroup
}
\makeatother

\usepackage{bm,color,mathrsfs,latexsym,graphicx,psfrag,mathtools}
\usepackage{color}
\usepackage[dvipsnames]{xcolor}

\usepackage{empheq}%\begin{empheq}[box=\fbox]{align} ... \end{empheq}

\newcommand{\Tr}{\mathop{\mathrm{Tr}}}

\let\oldAA\AA
\renewcommand{\AA}{\text{\normalfont\oldAA}}

\newcommand{\mf}[1]{\mathbf{#1}}
\newcommand{\+}{\dagger}
\newcommand{\mv}[1]{\left< #1 \right>}

\newcommand{\mfk}{\mf{k}}
\newcommand{\mfq}{\mf{q}}
\newcommand{\up}{\uparrow}
\newcommand{\dn}{\downarrow}

\usepackage{subcaption}
\usepackage{cleveref}
\crefname{appendix}{App.\,}{Apps.\,}
\crefname{equation}{Eq.\,}{Eqs.\,}
\crefname{figure}{Fig.\,}{Figs.\,}
\crefname{table}{Tab.\,}{Tabs.\,}
\crefname{section}{Sec.\,}{Secs.\,}
\creflabelformat{appendix}{[#2#1#3]}

\hypersetup{
    pdfborder={0 0 0}
}

\title{Evidence for Multimodal Superfluidity of Neutrons}

\author[1,2]{Yuan-Zhuo~Ma} 
\author[3,4]{Georgios~Palkanoglou}
\author[5]{Joseph~Carlson}
\author[5]{Stefano Gandolfi}
\author[4]{Alexandros~Gezerlis}
\author[1,2]{Gabriel~Given}
\author[1,2]{Ashe~Hicks}
\author[1,2]{Dean~Lee}
\author[6]{Kevin~E.~Schmidt}
\author[7]{Jiabin~Yu} 

\affil[1]{\FRIB}
\affil[2]{\MSU}
\affil[3]{\TRIUMF}
\affil[4]{\Guelph}
\affil[5]{\LANL}
\affil[6]{\ASU}
\affil[7]{\Florida}

\begin{abstract}
We present theoretical and experimental evidence for a new phase of matter in neutron-rich systems that we call multimodal superfluidity. Using \textit{ab initio} lattice calculations, we show that the condensate consists of coexisting s-wave pairs, p-wave pairs in entangled double-pair combinations, and quartets composed of bound states of two s-wave pairs. We identify multimodal superfluidity as a general feature of single-flavor spin-1/2 fermionic systems with attractive s-wave and p-wave interactions, provided the system is stable against collapse into a dense droplet. Beyond neutrons at sub-saturation densities, we demonstrate that this phase appears in generalized attractive extended Hubbard models in one, two, and three dimensions. We elucidate the mechanism for this coexistence using self-consistent few-body Cooper models and compare with Bardeen-Cooper-Schrieffer theory.  We also derive the form of the effective action and show that spin, rotational, and parity symmetries remain unbroken. Finally, we analyze experimental data to show that p-wave pair gaps and quartet gaps are present in atomic nuclei, and we discuss the consequences of this new phase for the structure and dynamics of neutron star crusts.

\end{abstract}

\begin{document}

\DeclareDocumentCommand\asymunc{ m m g }{%
    {\ensuremath{%
        {#1}%
        ^{+#2}%
        \IfNoValueF {#3} {_{-#3}}%
        \IfNoValueT {#3} {_{-#2}}%
        }
    }%
}

\maketitle  

Fermionic superfluidity is a striking example of collective quantum behavior, where interacting particles form correlated pairs that comprise a macroscopic condensate~\cite{Yang:1962zz}. Since the introduction of Bardeen-Cooper-Schrieffer (BCS) theory~\cite{Bardeen1957}, fermionic superfluid phases have generally been classified by the orbital symmetry of the constituent Cooper pairs.  In the absence of explicit symmetry breaking that mixes different channels, the condensate is usually dominated by a single channel such as the s-wave with orbital angular momentum $l=0$, which describes most superfluids and superconductors, or the p-wave state $l=1$, as observed in the well-known example of superfluid ${}^{3}\text{He}$~\cite{osheroff1972new,Leggett1975,Anderson:1973,Balian:1963}. 
Analogously, in heavy nuclei with $N=Z$ one expects either
isospin-singlet ($t=0$) or isospin-triplet ($t=1$) pairing to emerge~\cite{Gezerlis:2011,Palkanoglou:2025}.
In physical systems where interactions in multiple partial wave channels are attractive, the conventional expectation is governed by phase competition, where the channel with the strongest binding completely dominates, or by phase separation, where distinct superfluids occupy different spatial regions~\cite{Sedrakian2019,Sedrakian:2024dgk}. Neutron matter has long been known to exhibit superfluidity and is characterized by attractive interactions in both the singlet s-wave (${}^{1}S_{0}$) and two triplet p-wave channels (${}^{3}P_{0}$ and ${}^{3}P_{2}$) \cite{Migdal:1959noc,Ginzburg:1965cpw,Gezerlis:2014efa}.  Previous approaches have generally treated each of these pairing channels in isolation~\cite{Tamagaki1970,Hoffberg1970,takatsuka1971superfluid,Sauls1989,Baldo:1992kzz,Takatsuka1993,Dean:2003,Schwenk2005,gurarie2007resonantly,Gezerlis2008,Gezerlis:2009iw,Gandolfi:2015jma,Vidana:2021wna,Gandolfi:2022dlx}. While the s-wave attraction is strong at densities below saturation density ($0.16$~fm$^{-3}$), the p-wave attraction is generally assumed to be too weak to produce non-negligible condensation on its own, leading to the prevailing assumption that the ground state is a pure s-wave superfluid.  In contrast with the standard view, however, we report here theoretical and experimental evidence that neutron matter below saturation density exhibits a new quantum phase of matter that we term multimodal superfluidity.

We begin by establishing the existence of multimodal superfluidity through \textit{ab initio} lattice calculations of generalized attractive extended Hubbard models. While our primary focus is on the three-dimensional system, corresponding calculations for one and two dimensions are detailed in Methods. From these results and the underlying symmetries of the quantum system, we deduce the corresponding low-energy effective action. We then perform \textit{ab initio} lattice calculations of realistic neutron matter using next-to-next-to-next-to-leading order (N$^3$LO) chiral effective field theory interactions and the wavefunction matching method described in Ref.~\cite{Elhatisari:2024}.  Based on these neutron matter calculations, we discuss the impact for the structure and dynamics of neutron star crusts. We conclude by identifying experimental signatures of multimodal superfluidity in atomic nuclei and discuss connections to condensed matter physics, ultracold atomic systems, and quantum computing.

We consider a generalized attractive extended (GAE) Hubbard model in three dimensions.  In contrast with previous studies of charge-$4e$ superconductivity~\cite{Berg_2009_Charge4e,Herland_2010_MetallicSuperfluid,Gnezdilov_2022_SolvableModel,Wu_2024_4eSC,Samoilenka_2026_Electron_Quadrupling,Huecker_2026_4eSC} in multi-flavor Hubbard models with on-site lattice interactions~\cite{Lecheminant_2005_Multicomponent_Hubbard,Lecheminant_2007_spin_3over2_attractive_Hubbard,White_2008_multicomponent_Hubbard,Roux_2009_spin_3over2_attractive_hubbard,Katsura_2021_SUN_Attractive_Hubbard,Katsura_2022_SUN_Attractive_Hubbard,Neupert_2024_Charge4eSC_Mutiorbital_Attractive_Hubbard,Gao_2026_primarycharge4esuperconductivitydoping}, here we demonstrate the formation of s-wave pairs, p-wave pairs in entangled double-pair combinations, and quartets for single-flavor spin-1/2 fermions with interactions that extend beyond on-site.  We note that both p-wave double pairs and quartets correspond to charge-$4e$ condensation.  Our GAE Hubbard model Hamiltonian has nearest-neighbor hopping for the kinetic energy term and two-particle interactions with local and nonlocal smearing.  The local smearing parameter $s_\textrm{L}$ controls the strength of the nearest-neighbor attraction of the density-density interaction, while the nonlocal smearing parameter $s_\textrm{NL}$ controls the size of terms where the annihilation and creation operators are at different lattice sites, producing velocity dependent terms that go beyond density-density interactions.  These interactions are independent of spin, and so our Hamiltonian has symmetries associated with particle number conservation, spin rotations, cubic lattice rotations, and parity inversion of spatial positions.  The set of spin rotations forms the $\textrm{SU(2)}$ group of special unitary $2\times2$ matrices with unit determinant.  The details of the Hamiltonian are presented in Methods.

\begin{figure}[H]
    \centering
    \includegraphics[width=0.85\linewidth]{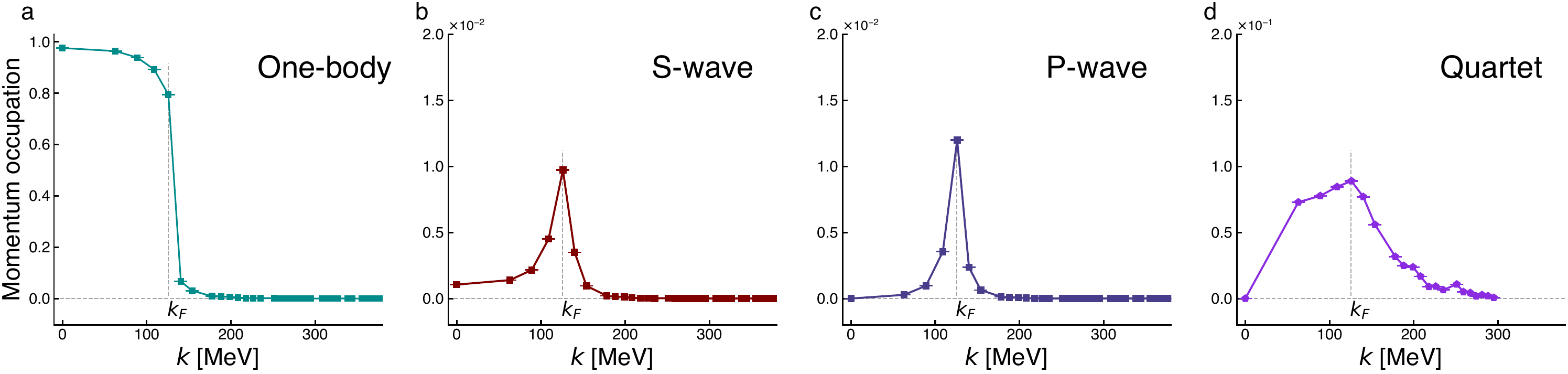}
    \caption{\textbf{Momentum-space distributions for the 3D GAE Hubbard model.} \textit{Ab initio} lattice results for an $L^3=10^3$, $A=66$ system with coupling $c=-1.6\times 10^{-6}$~MeV$^{-2}$, local smearing strength $s_{\text{L}}=0.5$, nonlocal smearing strength $s_{\text{NL}}=0.1$ and Euclidean time $\tau=1.67$ MeV$^{-1}$. The vertical dashed lines indicate the Fermi momentum $k_F$.  \textbf{a:} One-body momentum occupation number as function of momentum $k$. \textbf{b:} S-wave pair probability distribution as a function of momentum. \textbf{c:} P-wave pair probability distribution. \textbf{d:} Quartet probability distribution. \label{fig:SU2_OB_SPQ}}
\end{figure}

We simulate the GAE Hubbard model using \textit{ab initio} auxiliary-field projection Monte Carlo methods, adopting a lattice spacing of $a = 1.97~\mathrm{fm}$ and particle mass of $938.92~\textrm{MeV}$.  We use natural units where $\hslash=c=1$. Full details of the lattice formalism and results in one, two, and three dimensions are provided in Methods. In Fig.~\ref{fig:SU2_OB_SPQ}, we present the one-body momentum distribution and the condensate momentum distributions for s-wave pairs, p-wave pairs, and quartets. These results correspond to a spin-balanced system of $A=66$ fermions on an $L^3=10^3$ periodic cubic lattice at a density of $0.0086$~fm$^{-3}$. To isolate the condensate signals, these momentum distributions are extracted by computing cumulants that subtract background contributions from disconnected, uncorrelated processes.  Here and throughout our paper, the error bars represent one-sigma uncertainties.

In Table~\ref{tab:SU2_Thermodynamic} we show 3D GAE Hubbard model lattice results for the condensate fractions for s-wave pairs, p-wave pairs, and quartets for $A=14, 38, 66, 114$ particles at approximately constant density.  Due to the limits of computational resources, we could not compute the quartet condensate fraction for $114$ particles. Although the results for $A=14$ are strongly affected by finite size effects, the data for $A=38, 66, 114$ are consistent with smooth thermodynamic limits for the condensate fractions of about $1.8(4)\%$ for s-wave pairs, $1.3(3)\%$ for p-wave pairs, and $45(3)\%$ for quartets with densities between $0.0086~\textrm{fm}^{-3}$ and $0.0097~\textrm{fm}^{-3}$.  This can be contrasted with the condensate fraction for fermions in the unitary limit, an idealized limit of spin-1/2 fermions with zero interaction range and infinite scattering length.  In the unitary limit, the interactions are purely in the s-wave channel, and the s-wave pair condensate is $43(2)\%$ \cite{He:2019ipt} and experiments with ultracold $^6$Li atoms have measured the condensate fraction to be 0.46(7) \cite{Zwierlein:2004zz,Zwierlein:2005} and 0.47(7) \cite{Kwon:2020}.  It appears that most of the condensate in multimodal superfluidity has shifted to quartets, while smaller but nonzero condensate fractions remain for s-wave pairs and p-wave pairs.  In Methods we discuss how this balance arises from the competition between the maximization of binding energy and the minimization of Pauli-blocking for composite bosons \cite{Combescot2003,Combescot2008,Combescot2015}.

\begin{table}[htbp]
\centering
\caption{\textbf{Condensate fractions for the 3D GAE Hubbard model.} \textit{Ab initio} lattice results for the condensate fractions for s-wave pairs, p-wave pairs, and quartets for $c = -1.6\times 10^{-6}$ MeV$^{-2}$, $s_{\text{L}}=0.5$ and $s_{\text{NL}}=0.1$.}
\label{tab:SU2_Thermodynamic}
\begin{tabular}{lllllllll}
\toprule
$L$ & $A$ & density (fm$^{-3}$) & $2 \times$S-wave/$A$ &  $2 \times$P-wave/$A$ & $4 \times $Quartets/$A$\\
\midrule
6  & 14  & 0.0084     & 0.0253 (1) & 0.0217 (1) & 0.0034 (3) \\
8  & 38  & 0.0097    & 0.0169 (3)  & 0.0130 (2)  & 0.4347 (83)\\
10 & 66  & 0.0086  & 0.0177 (5)  & 0.0147 (7)  & 0.4759 (207)\\
12 & 114 & 0.0086    & 0.0217 (34) & 0.0105 (13) & not calculated \\
\bottomrule
\end{tabular}
\end{table}

In Ref.~\cite{Lee:2007eu} a theorem was proven stating that for any fermionic system with an $\textrm{SU(2$N$)}$ spin symmetry and purely attractive spin-independent interactions, the ground state energy is minimized by forming spin singlets invariant under the $\textrm{SU(2$N$)}$ symmetry. Since our GAE Hubbard model satisfies these conditions, the condensation of p-wave pairs must not spontaneously break the $\textrm{SU(2)}$ spin symmetry.  To understand how the $\textrm{SU(2)}$ spin symmetry remains intact, it is useful to consider the low-energy effective action describing our multimodal superfluid.  We can produce attractive pairing interactions by introducing auxiliary bosonic fields, $\phi$ and $A_{\mu j}$, where $\phi$ couples to the s-wave spin-singlet parity-even ($^1S_0$) fermion bilinear and $A_{\mu j}$ couples to the p-wave spin-triplet parity-odd ($^3P_J$) fermion bilinears.
Here $\mu = 1,2,3$ denotes the vector index for orbital angular momentum ($l=1$) and $j=1,2,3$ denotes the vector index for spin ($s=1$).  While these fields are introduced as auxiliary fields, after integrating out the gapped fermionic modes they acquire induced dynamics and encode the collective pairing correlations in the superfluid condensate.  

When the p-wave interactions are strong enough to form quartets as bound states of s-wave pairs, another asymptotic state appears in our system and we must define an interpolating field ${Q}_s$ that couples to the quartets.  A key feature of the low-energy effective action in multimodal superfluidity is that ${Q}_s$ has a scalar coupling to two s-wave pairs, $\phi^2$, as well as two p-wave pairs, $\sum_{\mu, j} A_{\mu j} A_{\mu j}$.  As a result, a nonzero expectation value in either the s-wave pair field or the quartet field inevitably induces condensation in the other, and together they drive the formation of the double p-wave composite operator $\sum_{\mu,j}A_{\mu j}A_{\mu j}$. The p-wave pairs form entangled double-pair combinations that are invariant under spin rotations, spatial rotations, and parity inversion, and the $\textrm{SU(2)}$ spin symmetry remains intact.  In Methods, we give details of the effective action as well as lattice results showing that the condensate fractions for the $^3P_J$ pairs are consistent with unbroken $\textrm{SU(2)}$ invariance.

\begin{figure}[H]
    \centering
    \includegraphics[height=6cm]{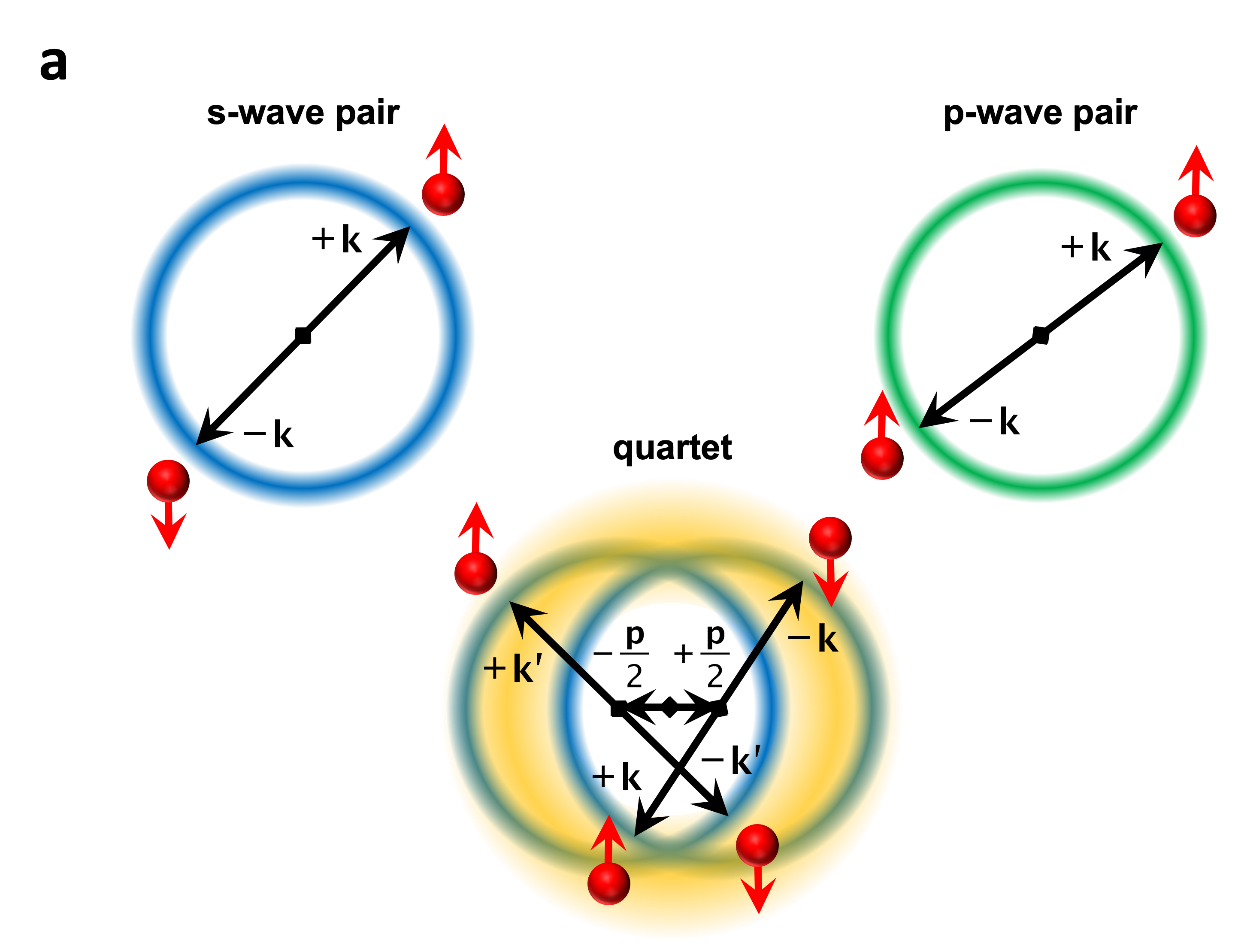}
    \hspace{1cm}
    \includegraphics[height=6cm]{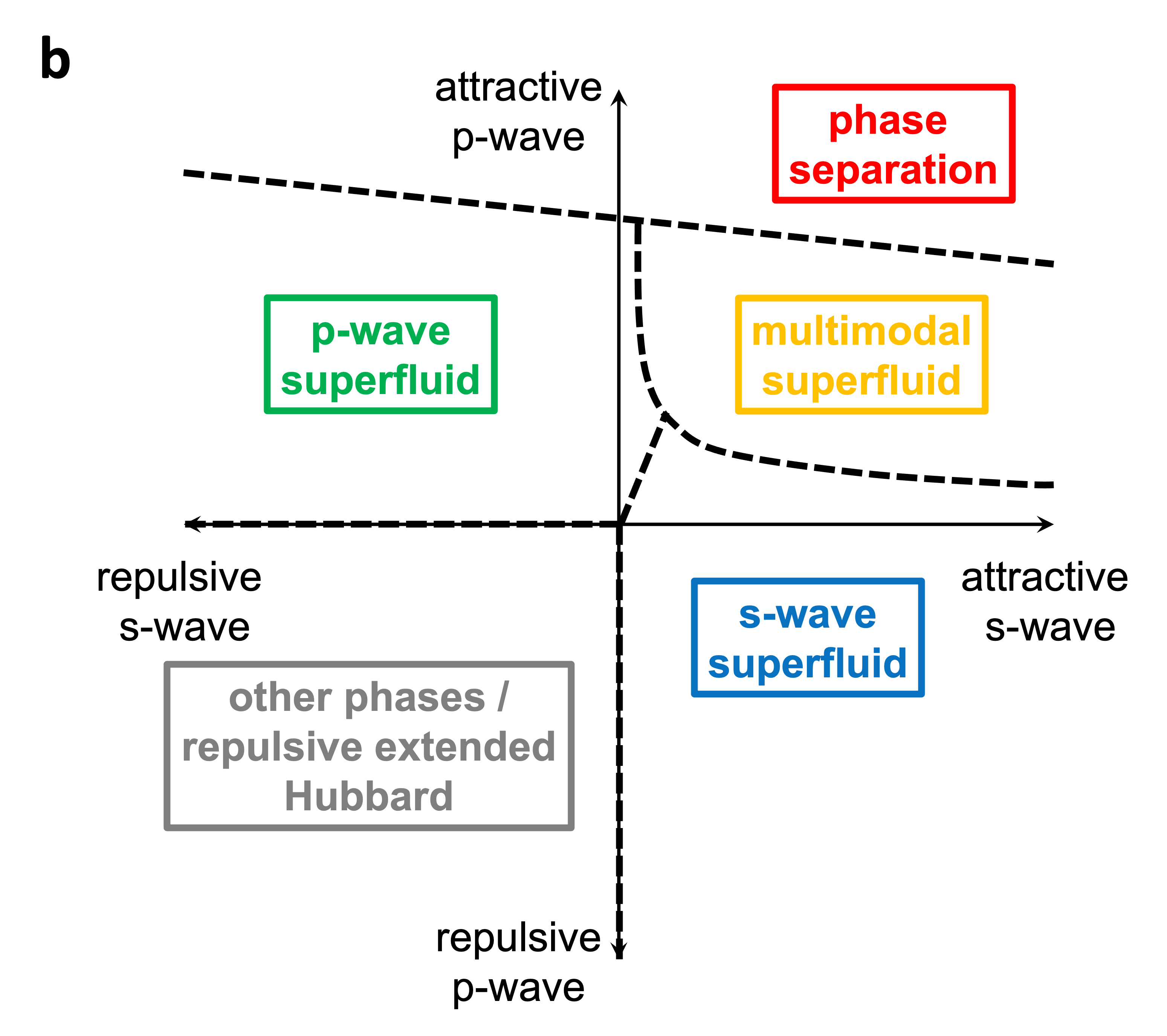}    
    \caption{\textbf{Schematic diagrams for multimodal superfluidity.} \textbf{a:} Illustration of an s-wave pair, p-wave pair, and quartet composed of two s-wave pairs bound together. \textbf{b:} The quantum phase diagram for a spin-1/2 fermionic system at zero temperature for different s-wave and p-wave interactions.  The diagram is only schematic, and the locations of the phase boundaries will depend on the details of the system.\label{fig:multimodal_schematic}}
\end{figure}

In multimodal superfluidity, the ground state wavefunction for the spin-balanced system is a superfluid condensate composed of s-wave pairs, p-wave pairs in entangled double-pair combinations, and quartets formed by the binding of two s-wave pairs.  This is illustrated in Panel a of Fig.~\ref{fig:multimodal_schematic}.  In Methods, we present results showing comparisons with BCS calculations for s-wave pairing for the spin-balanced case as well as p-wave pairing for fully polarized case.  While the BCS calculations are not able to describe multimodal superfluidity for the spin-balanced system, in Methods we describe a semi-analytic approach that illustrates the key microscopic mechanisms such as the binding of two s-wave pairs into a quartet.  In this semi-analytic approach, which we call the self-consistent Cooper model, we perform few-body calculations with particles obeying a BCS quasiparticle dispersion relation with an s-wave pairing gap $\Delta_s$ that is determined self-consistently.  In Methods we show that the self-consistent Cooper model is in good agreement with BCS theory calculations while also reproducing the key features of multimodal superfluidity seen in the \textit{ab initio} many-body lattice results that go beyond the BCS approximation.  In future work, it could be used to explore multimodal superfluidity in other systems and it may be possible to further improve the self-consistent Cooper model using physics-informed machine learning tools for quantum systems \cite{Cook:2024toj}. We note that the binding of s-wave pairs into quartets must occur in a many-body environment and is therefore separate from recent interest in the possibility of a low-energy tetraneutron resonance in vacuum \cite{Shimoura:2002ns,Kisamori:2016jie,Duer:2022ehf}.

In Panel b of Fig.~\ref{fig:multimodal_schematic} we sketch the quantum phase diagram for multimodal superfluidity with the strength and sign of the s-wave and p-wave interactions on the horizontal and vertical axes.  The phase diagram applies to any spin-balanced system of single-flavor spin-1/2 fermions in two or three dimensions. While there is no long range order in one dimension even at zero temperature \cite{Mermin:1966fe,Coleman:1973}, we demonstrate in Methods that many of the same features such s-wave pairs, p-wave pairs, and quartets can also be seen in 1D finite systems.  When the p-wave interaction is too strong and attractive, however, the ground state is no longer a gas and we have phase separation into a high-density self-bound droplet.  We have a s-wave superfluid when only the s-wave attraction is significant, and we have a p-wave superfluid when only the p-wave attraction is significant.  When both the s-wave and p-wave interactions are repulsive, we have some other quantum phase associated with repulsive extended Hubbard models.  Multimodal superfluidity appears when the s-wave and p-wave interactions are both sufficiently attractive and the instability towards phase separation is not yet reached.

\begin{figure}[H]
    \centering
    \includegraphics[width=0.85\linewidth]{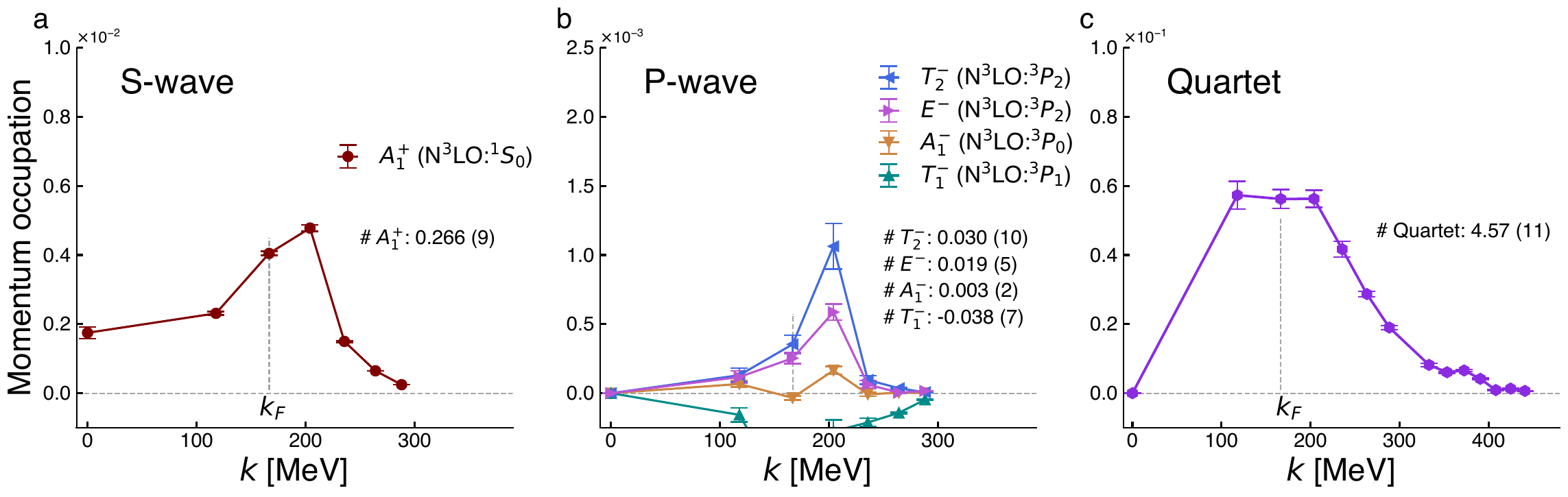}
    \caption{\textbf{Momentum-space distributions for realistic neutron matter} \textit{Ab initio} lattice results for at density $\rho = 0.033$ fm$^{-3}$ obtain using high-fidelity chiral N$^3$LO interactions and Euclidean time $\tau=0.2$ MeV$^{-1}$. The legend labels indicate the pairing channels with their corresponding cubic representations: $^1S_0$ ($A_1^+$), $^3P_0$ ($A_1^-$), $^3P_1$ ($T_1^-$), $^3P_2$ ($T_2^-$,$E^-$). The numerical values indicate the total numbers of pairs or quartets counted in the corresponding momentum distributions. As described in Methods, some additional steps are needed to properly count the actual number of quartets.  The vertical dashed line indicates the Fermi momentum $k_F$. \textbf{a:} S-wave pair distribution as a function of momentum $k$. \textbf{b:} P-wave pair probability distribution as a function of momentum. \textbf{c:} Quartet probability distribution.  \label{fig:N3LO_SPQ}}
\end{figure}

For realistic calculations of neutron matter, we employ high-fidelity chiral effective field theory interactions at N$^3$LO, with low-energy constants fitted to nucleon-nucleon scattering data and lattice spacing $a=1.32$~fm. To overcome the severe Monte Carlo sign problem typically associated with such realistic high-order interactions, we utilize the wavefunction matching method developed to accelerate the convergence of perturbation theory \cite{Elhatisari:2024}.  In Fig.~\ref{fig:N3LO_SPQ}, we plot \textit{ab initio} lattice results for neutron matter at density $\rho = 0.033$ fm$^{-3}$.  We present the condensate momentum distributions for s-wave pairs, p-wave pairs, and quartets. These results correspond to a spin-balanced system of $A=38$ fermions on an $L=8$ periodic cubic lattice.  They show that multimodal superfluidity occurs in realistic neutron matter with the simultaneous condensation of the s-wave $^1S_0$ pairs, p-wave $^3P_0$ and $^3P_2$ pairs, and quartets.  

We note that the $^3P_1$ channel is repulsive and no condensation is expected.  The negative signal we obtain for $^3P_1$ is due to the fact that we are starting with a Hamiltonian with a small amount of $^3P_1$ attraction and are applying a first-order perturbation to calculate the properties of the N$^3$LO chiral Hamiltonian with repulsive $^3P_1$ interactions. This produces a negative overshoot that would be corrected when including higher-order terms in perturbation theory. As detailed in Methods, with $A=38$ fermions on an $L=8$ lattice we obtain condensate fractions of 1.40(5)\% for $^1S_0$ pairs, 0.02(1)\% for $^3P_0$ pairs, 0.26(8)\% for $^3P_2$ pairs, and 48(1)\% for quartets at density $\rho=0.033~\textrm{fm}^{-3}$.  With $A=66$ fermions on an $L=7$ lattice we obtain condensate fractions of 0.68(1)\% for $^1S_0$ pairs, 0.00\% for $^3P_0$ pairs, 0.12(2)\% for $^3P_2$ pairs, and 13(1)\% for quartets at density $\rho=0.085~\textrm{fm}^{-3}$.  In Part a of Table~\ref{tab:combined_energies}, we show \textit{ab initio} lattice results for s-wave pair binding ($2\Delta_S$), p-wave pair binding ($2\Delta_P$), and quartet binding ($4\Delta_Q$) from neutrons using chiral N$^3$LO interactions at densities of $0.033$~fm$^{-3}$ and $0.085$~fm$^{-3}$.  The ground state energies per particle obtained in these simulations match the results reported in Ref.~\cite{Elhatisari:2024}, which used the same interactions. Furthermore, the obtained ground state energies are consistent with other \textit{ab initio} calculations of neutron matter using chiral effective field theory~\cite{Hebeler:2009iv,Gandolfi:2015jma}. Similarly, the $^1S_0$ pairing gaps are in good agreement with existing \textit{ab initio} predictions in the literature~\cite{Gezerlis:2014efa,Gandolfi:2022dlx}. So, while our results showing multimodal superfluidity are starkly different from previous studies, there is consistency on the energy observables that were the main focus of previous \textit{ab initio} calculations.

\begin{table}[t!]
    \centering
    \caption{\textbf{Pairing and Quartet Energies.} \textbf{a:} Neutron matter s-wave pair binding ($2\Delta_S$), p-wave pair binding ($2\Delta_P$), and quartet binding ($4\Delta_Q$) from \textit{ab initio} lattice calculations with chiral N$^3$LO interactions at densities of $0.033$~fm$^{-3}$ and $0.085$~fm$^{-3}$. \textbf{b:} Empirically observed p-wave pair binding in nuclei ($2\Delta_P$) derived from one-neutron separation energies. \textbf{c:} Empirically observed quartet binding in nuclei ($4\Delta_Q$) derived from two-neutron separation energies. Energies are in MeV. One-sigma uncertainties are given in parentheses, and values without uncertainties have errors smaller than the last significant digit.}
    \label{tab:combined_energies}
    \begin{tabular}{lcl}
        \toprule
        \multicolumn{3}{l}{\textbf{a. Neutron matter (chiral N$^3$LO)}} \\
        \midrule
        Density (fm$^{-3}$) &  &$2\Delta_S$ (MeV) \\
        \midrule
        $0.033$ && $5.58(92)$ \\
        $0.085$ && $2.61(76)$ \\
        \addlinespace[0.6em]
        Density (fm$^{-3}$) & channel &$2\Delta_P$ (MeV) \\
        \midrule
        $0.033$&$^3P_0$ & $1.25(86)$ \\
        $0.033$&$^3P_1$ & $0.59(113)$ \\
        $0.033$&$^3P_2[T_2^-]$ & $1.36(86)$ \\
        $0.033$&$^3P_2[E^-]$ & $2.16(90)$ \\
        \addlinespace[0.4em]
        $0.085$&$^3P_0$ & $1.95(95)$ \\
        $0.085$&$^3P_1$ & $-0.24(81)$ \\
        $0.085$&$^3P_2[T_2^-]$ & $2.10(80)$ \\
        $0.085$&$^3P_2[E^-]$ & $2.03(87)$ \\
        \addlinespace[0.6em]
        Density (fm$^{-3}$)&  &$4\Delta_Q$ (MeV) \\
        \midrule
        $0.033$ && $3.90(167)$ \\
        $0.085$ && $2.01(86)$ \\
        \addlinespace[0.6em]
        \multicolumn{3}{l}{\textbf{b. Empirical p-wave pair binding in nuclei}} \\
        \midrule
        System &&  $2\Delta_P$ (MeV) \\
        \midrule
        \multicolumn{2}{l}{$\{^{17}\textrm{O}(\frac{3}{2}^+), ^{18}\textrm{O}(1^+)\}$} & $0.172(12)$ \\
        \multicolumn{2}{l}{$\{^{55}\textrm{Fe}(\frac{5}{2}^-), ^{56}\textrm{Fe}(1^+), ^{57}\textrm{Fe}(\frac{5}{2}^-)\}$} & $0.139(3)$ \\
        \multicolumn{2}{l}{$\{^{57}\textrm{Ni}(\frac{5}{2}^-), ^{58}\textrm{Ni}(1^+), ^{59}\textrm{Ni}(\frac{5}{2}^-)\}$} & $0.245(1)$ \\
        \addlinespace[0.6em]
        \multicolumn{3}{l}{\textbf{c. Empirical quartet binding in nuclei}} \\
        \midrule
        System && $4\Delta_Q$ (MeV) \\
        \midrule
        \multicolumn{2}{l}{$\{^{6}\textrm{He}, ^{8}\textrm{He}\}$} & $1.150$ \\
        \multicolumn{2}{l}{$\{^{18}\textrm{O}, ^{20}\textrm{O}, ^{22}\textrm{O}\}$} & $0.142(28)$ \\
        \multicolumn{2}{l}{$\{^{42}\textrm{Ca}, ^{44}\textrm{Ca}, ^{46}\textrm{Ca}\}$} & $0.236(1)$ \\
        \multicolumn{2}{l}{$\{^{44}\textrm{Ca}, ^{46}\textrm{Ca}, ^{48}\textrm{Ca}\}$} & $0.332(3)$ \\
        \bottomrule
    \end{tabular}
\end{table}

In uniform neutron matter, translational and rotational invariance ensure that momentum and orbital angular momentum remain good quantum numbers, allowing p-wave correlations to develop unimpeded. Finite nuclei, however, present a complex environment that typically includes configuration mixing among orbital sub-shells, and this mixing of orbitals acts as an effective
source of disorder. The survival of s-wave pairing is explained by Anderson’s theorem \cite{Anderson1959}, which states that s-wave pairing is uniquely robust against disorder because it is protected by time
reversal symmetry without any requirement of spatial symmetries.  In contrast, the experimental signatures of p-wave pairing and the predicted quartet condensate are much more subtle, appearing only under specific conditions where the relevant attractive two-body matrix elements are sufficiently strong. To identify these signatures in experimental data, we apply staggered $k$-point difference formulas to the binding energies of nuclei in an isotopic chain.  The details of these calculations are given in Methods.

In Part b of Table~\ref{tab:combined_energies}, we present empirically observed p-wave pair binding energies in atomic nuclei.  In Part c of Table~\ref{tab:combined_energies}, we show empirically observed quartet binding energies.  For each case, we have computed the additional binding using finite difference formulas involving the nuclei listed.  The calculations are performed using AME2020 masses \cite{Wang2021,Huang2021} and the Evaluated Nuclear Structure Data File (ENSDF) from the National Nuclear Data Center (NNDC) \cite{NNDC, Kondev2021}. Details are presented in Methods along with 15 more examples of quartet binding energies in the tin, lead, polonium, and radon isotopic chains.

The multimodal superfluidity of neutron matter appears to have a significant impact on the physics of the neutron star crust. The formation of quartets introduces an additional binding energy that increases the effective quasiparticle gap ($\Delta_{\text{eff}} \simeq \Delta_S + \Delta_Q$). This enhanced gap suppresses the neutron heat capacity, providing a microscopic explanation for the low thermal inertia inferred from rapidly cooling transient sources like KS~1731--260~\cite{Shternin:2007md,Brown2009,Cackett2010}. While such a large effective gap would typically freeze out standard pair-breaking cooling mechanisms~\cite{Page2006}, there are lower-energy excitations which simply break the entanglement of the $^3P_2$ double pairs. This allow coupling to the axial-vector current and efficient neutrino emission that keeps older crusts radiatively active~\cite{Leinson2010}.  Furthermore, the phase-locking among the coexisting condensates requires that macroscopic phase twists generate currents in all sectors simultaneously. By enhancing the free energy cost of phase gradients, this cooperative response increases the superfluid stiffness, offering a physical mechanism to counteract the suppression of the superfluid density caused by entrainment within the crustal lattice~\cite{Link:1992mdl,Chamel:2008ca,Chamel2012,Andersson2012,Chamel2013}. Finally, the coexistence of quartets and pairs topologically mandates the formation of half-quantum vortices bounded by domain walls. This topological frustration creates complex structural phases, such as bound vortex dimers or rigid percolated networks, which introduce novel elastic moduli and pinning forces to trigger and sustain pulsar glitches~\cite{Baym1969,Anderson:1975zze,pines1985superfluidity}. Taken collectively, these effects appear to ameliorate some of the current tensions in our microscopic understanding of the structure and dynamics of neutron star crusts.

In summary, we have presented theoretical and empirical evidence pointing to the existence of multimodal superfluidity, a new phase of matter characterized by the coexistence of s-wave pairs, p-wave pairs in entangled double-pair combinations, and quartets. Through \textit{ab initio} lattice simulations, we found that this composite condensate can emerge in both generalized attractive extended Hubbard models and realistic neutron matter without spontaneously breaking the underlying spin or spatial symmetries.  We have also found experimental evidence for multimodal superfluidity in atomic nuclei with several examples of p-wave pair binding and quartet binding.  There appear to be significant scientific impacts for the properties of neutron star crusts. Multimodal superfluidity could be realized in future experiments using ultracold fermionic dipolar molecules or ultracold fermionic atoms on optical lattices. In the realm of condensed matter, producing coexisting s-wave and p-wave attraction for multimodal superconductivity may require moving beyond conventional electron-phonon interactions. Promising experimental platforms for this phase include strongly correlated systems with magnetic spin fluctuations, special lattice geometries, spin-orbit interactions, topological materials, proximity-coupled heterostructures, and other effects that engage more than one attractive pairing channel and could lead to bound charge-$4e$ quartets. Observable consequences include multiple quasiparticle excitation scales, additional low-energy collective modes corresponding to relative phase oscillations between condensate components, unconventional vortex structures, and Josephson responses beyond the standard charge-2e periodicity. A key theoretical question is the structure of the superfluid current and the relative contributions carried by s-wave pairs, p-wave double pairs, and quartets. Addressing this problem is challenging for classical Monte Carlo methods due to severe sign oscillations, and multimodal superfluidity may therefore provide a natural target for analog or digital quantum simulation.

We note that the low-energy spectrum of a multimodal superfluid contains the usual Goldstone boson associated with the spontaneous breaking of the U(1) particle number symmetry, along with gapped fermionic quasiparticles. However, the multi-component nature of the condensate also gives rise to a richer spectrum of low-energy bosonic excitations. In addition to relative phase oscillations between the condensate components analogous to Leggett modes \cite{Leggett1966}, this spectrum includes collective modes corresponding to the decoupling of the entangled p-wave double pairs. Furthermore, the extra binding of the quartet produces a larger effective quasiparticle gap than that derived from s-wave pairing alone.  While we have not seen any evidence of significant higher-body correlations beyond quartets and p-wave double pairs in the condensate of realistic neutron matter, the condensate fraction of such objects may not be exactly zero.  We have some numerical evidence that it is possible to tune the parameters of generalized attractive extended Hubbard models to significantly increase the binding of sextets and even octets, before hitting the phase boundary where the system collapses into a dense droplet.  The relative populations of pairs, quartets, sextets, octets, etc. are determined by a balance between binding and repulsive Pauli blocking, and this is discussed in Methods.  In this work we have focused entirely on multimodal superfluidity due to attractive s-wave and p-wave interactions. However, it is clear that other combinations of channels could also exhibit multimodal superfluidity.

\smallskip

\noindent {\bf \large Acknowledgments}
We are grateful for discussions with Alex Brown, Nicolas Chamel, Serdar Elhatisari, Hironori Iwasaki, Myungkuk Kim, Youngman Kim, Bing-Nan Lu, Nadya Mason, Ulf-G. Mei{\ss}ner, Chetan Nayak, Sofia Quaglioni, Leo Radzihovsky, Sanjay Reddy, Thomas Schaefer, Young-Ho Song, Jun Ye, Martin Zwierlein, members of the Nuclear Lattice Effective Field Theory Collaboration, and others.  We acknowledge financial support provided as follows:  D.L. and Y.M. [U.S. Department of Energy (DOE) grants DE-S0013365, DE-SC0023175, DE-SC0026198, DE-SC0023658, and U.S. National Science Foundation (NSF) grant PHY-2310620]; G.P. [ TRIUMF
receives federal funding via a contribution agreement
with the National Research Council of Canada; Natural Sciences and Engineering Research Council (NSERC) of Canada and the Canada Foundation for Innovation (CFI).]; J.C. and S.G.[U.S. Department of Energy through the Los Alamos National Laboratory. Los Alamos National Laboratory is operated by Triad National Security, LLC, for the National Nuclear Security Administration of U.S. Department of Energy (Contract No. 89233218CNA000001), and by the Office of Advanced Scientific Computing Research, Scientific Discovery through Advanced Computing (SciDAC) NUCLEI program]; A.G. [Natural Sciences and Engineering Research Council (NSERC) of Canada and the Canada Foundation for Innovation (CFI)]; J.Y.'s work is supported by startup funds at University of Florida.  We also acknowledge computational support as follows: Oak Ridge Leadership Computing Facility computing resources through the INCITE award “Ab-initio nuclear structure and nuclear reactions”, National Energy Research Scientific Computing Center computing resources through Contract No. DE-AC02-05CH11231 using NERSC award NP-ERCAP0036716 and NP-ERCAP0036535, Michigan State University's Institute for Cyber-Enabled Research and High-Performance Computing Center, the Advanced Cyberinfrastructure Coordination Ecosystem: Services \& Support (ACCESS) program through allocation PHY250148, 
Compute Ontario through the Digital Research Alliance of Canada, and
Research Computing at Arizona State University. This research also used resources provided by the Los Alamos National Laboratory Institutional Computing Program, which is supported by the U.S. Department of Energy National Nuclear Security Administration under Contract No. 89233218CNA000001.

\noindent {\bf \large Author Contributions}
All authors participated in regular discussions of the project in
which they helped to formulate theoretical concepts and plan
the required calculations. All authors also contributed to the
analysis of data and the writing and editing of the paper.  Y-Z.M. - Developed lattice algorithms, performed lattice calculations, and wrote first draft of lattice Methods sections;
G.P. - Developed polarized BCS formalism, performed BCS calculations, and wrote first draft of BCS Methods section;
J.C. - Supervised work on BCS calculations, supervised exploratory work using continuum QMC;
S.G. - Supervised and performed exploratory work using continuum QMC;
A.G. - Supervised work on BCS calculations, supervised exploratory work using continuum QMC;
G.G. - Performed exploratory work using continuum QMC and lattice simulations;
A.H. - Performed exploratory work using continuum QMC and lattice simulations;
D.L. - Coordinated project efforts, wrote first draft of main text and several Methods sections on theory and applications;
K.S. - Supervised work on BCS calculations, supervised exploratory work using continuum QMC;
J.Y. - Co-led the effort to construct the effective action and make connections to condensed matter physics.\\
\\
\noindent {\bf \large Data Availability} All of the data produced in association with this work have been stored and are publicly available at \url{https://drive.google.com/drive/folders/1F6_tT97xo9mdhrWujpfYpHq898sv4wCq}.\\
\\
\noindent {\bf \large Code Availability} All of the codes produced in association with this work have been stored and can be obtained upon request from the corresponding author, subject to possible export control constraints.\\
\\
\noindent {\bf \large Competing Interest Statement} 
The authors declare no competing interests.\\
\\
\noindent {\bf \large Inclusion and Ethics} We have complied with community standards for authorship and all relevant recommendations with regard to inclusion and ethics.
\\
\\

\pagebreak

%%%%%%%%%%%%%%
\section*{Methods}
\renewcommand{\thesection}{S\arabic{section}}
\renewcommand{\thefigure}{S\arabic{figure}}
\renewcommand{\theequation}{S\arabic{equation}}
\renewcommand{\thetable}{S\arabic{table}}
\setcounter{figure}{0}
\setcounter{equation}{0}
\setcounter{section}{0}
\setcounter{table}{0}

%\subsection*{Preamble ({in progress})}\label{S-1.0}

The material in the Methods section is organized as follows. We begin by defining the lattice Hamiltonian and the observables used for lattice measurements, detailing the rank-one operator and momentum pinhole methods, alongside rotation and projection techniques for irreducible two-body densities. Next, we present our numerical findings across four systems: the lattice Cooper model, the unitary limit, the generalized attractive extended (GAE) Hubbard model, and realistic neutron matter. To contextualize these findings, we provide Bardeen-Cooper-Schrieffer theory calculations and analyze the system through the framework of composite boson theory, focusing on Pauli repulsion and multimodal coexistence. We then construct an effective action description of multimodal superfluidity and detail its thermodynamics, vortex structure, and astrophysical implications. The section concludes by outlining the expected signatures of this phase in finite nuclei and detailing the difference formulas for pairing and quartetting used to extract them.

% \section{Lattice 1 (placeholder name)}
\section{Lattice Hamiltonian}
Nuclear lattice effective field theory (NLEFT)~\cite{Lee:2008fa, Lahde:2019npb} 
combines chiral effective field theory ($\chi$EFT), lattice field theory, and stochastic Monte Carlo algorithms. 
% Each of these components forms a cornerstone of modern approaches to strongly interacting many-fermion systems across a wide range of disciplines, including nuclear, particle, condensed-matter, and atomic physics.
Building on early developments~\cite{Borasoy:2007, Lee:2008fa}, 
NLEFT has expanded its scope from light nuclei~\cite{Epelbaum:2010prl, Epelbaum:2010epja, Epelbaum:2011, Epelbaum:2012, Epelbaum:2013} 
to medium-mass systems~\cite{Lahde:2014, Elhatisari:2016, Elhatisari:2017}.  
Two-body scattering calculations on the lattice have enabled the construction of high-quality chiral interactions~\cite{Lu:2016, Alarcon:2017, Li:2018, Li:2019}.
More sophisticated $\alpha$–$\alpha$ scattering has been studied within the NLEFT framework~\cite{Elhatisari:2015}.
The complex tensor structures and repulsive components of realistic chiral interactions make direct Monte Carlo simulations with high-quality nuclear potentials computationally challenging due to the severe sign problem. 
To address this issue, a novel ``wavefunction matching'' approach~\cite{Elhatisari:2024} was developed. 
This method exploits the sign-problem-free SU(4)-symmetric interaction~\cite{Lee:2007eu,Lu:2019, Lee:2021} and enables the inclusion of high-fidelity chiral N$^3$LO interactions.
With these high-fidelity chiral interactions, NLEFT has achieved detailed and quantitative predictions of nuclear structure properties, including binding energies~\cite{Elhatisari:2024}, root-mean-square radii~\cite{Zhang:2025}, proton and neutron density distributions~\cite{Meissner:2024}, and low-lying excited states~\cite{Shen:2025}. 
The framework has also provided insights into hypernuclei,~\cite{Hildenbrand:2024} $\alpha$ clustering in finite nuclei,~\cite{Shen:2023,Giacalone:2025}, and infinite neutron matter~\cite{Ren:2024}. 
Finite-temperature properties of infinite neutron and nuclear matter have likewise been investigated~\cite{Lu2020, Ma:2024}.
Recent developments include the application of advanced numerical techniques and supercomputers to heavy nuclei~\cite{Niu:2025, Hildenbrand:2026}, as well as the construction of an improved nuclear interaction~\cite{Wu:2025}.

In the present work, we employ three lattice Hamiltonians: i) an SU(2) Hamiltonian for the generalized attractive extended (GAE) Hubbard model; ii) a simple leading-order chiral Hamiltonian; and iii) a high fidelity chiral Hamiltonian at N$^3$LO.
In the present work, we use the lattice spacing $a = 1.97~\mathrm{fm}$ for the GAE Hubbard model and $a = 1.32~\mathrm{fm}$ for chiral Hamiltonians. 

\subsection{SU(2) Hamiltonian} 
The SU(2)-symmetric Hamiltonian contains of an improved kinetic term and a smeared contact interaction,
\begin{equation} \label{eq:H_SU2}
  H^{\text{SU(2)}}=H_{\text{free}}+\frac{1}{2} c \sum_{\vec{n}}: [\tilde{\rho}^{(1)}(\vec{n})]^2:
\end{equation}
where the $::$ symbols mean normal ordering and
$H_{\text{free}}$ is the free kinetic term with parameter $i_{\text{kin}}$ that controls the improvements of the second-order derivative from the finite difference. When $i_{\text{kin}}=1$, it becomes:
  \begin{equation}
    H_{\rm free} = \frac{1}{2m}\sum_{j=\uparrow,\downarrow}\sum_{\vec{n}} \left[ 2d a^\dagger_j(\vec{n})a_j(\vec{n}) - \sum_{\ell=1}^d a^\dagger_j(\vec{n}+\vec{e}_\ell)a_j(\vec{n}) - \sum_{\ell=1}^d a^\dagger_j(\vec{n}-\vec{e}_\ell)a_j(\vec{n})\right].
\end{equation}
with nucleon mass $m=938.92$ MeV. The dressed density operator $\tilde{\rho}$ includes local and nonlocal smearing,
\begin{equation}
        \tilde{\rho}^{(d)}(\vec{n})=\sum_{j=\uparrow,\downarrow} \tilde{a}_{j}^{\dagger}(\vec{n}) \tilde{a}_{j}(\vec{n})+s_{\mathrm{L}} \sum^d_{\left|\vec{n}-\vec{n}^{\prime}\right|=1} \sum_{j=\uparrow,\downarrow} \tilde{a}_{j}^{\dagger}\left(\vec{n}^{\prime}\right) \tilde{a}_{ j}\left(\vec{n}^{\prime}\right).
\end{equation}
The nonlocally smeared annihilation and creation operators, $\tilde{a}$ and $\tilde{a}^{\dagger}$, with spin $j=0,1$ (up, down) indices are defined as,
\begin{equation}
        \tilde{a}_{j}(\vec{n})=a_{j}(\vec{n})+s_{\text{NL}} \sum_{\left|\vec{n}^{\prime}-\vec{n}\right|=1} a_{j}\left(\vec{n}^{\prime}\right).
\end{equation}
The local and nonlocal strengths are controlled by the smearing parameters $s_{\text{L}}$ and $s_{\text{NL}}$.
In this work, we use different sets of parameters $\{c, s_{\text{L}}, s_{\text{NL}}\}$ for different systems. For instance, we set $c = -5.573\times 10^{-5}$ $\text{MeV}^{-2}$, $s_{\text{L}} = 0.0$,  $s_{\text{NL}} = -0.01$ and $i_{\text{kin}}=3$ for 3D ``Unitary-Limit'' systems.
It should be mentioned that we omit isospin indices for pure neutron systems (or single-flavor Fermi systems). 

\subsection{Wavefunction matching method} 
Wavefunction matching\cite{Elhatisari:2024} is a method which allows for calculations of systems that would otherwise be impossible owing to problems such as Monte Carlo sign cancellations. 
While keeping the observable physics unchanged, wavefunction matching creates a new high-fidelity Hamiltonian $H'$ such that the two-body wavefunctions up to some finite range match that of a simple Hamiltonian $H^{\text{S}}$, which is easily computed. 
This allows for a rapidly converging expansion in powers of the difference $H'-H^{\text{S}}$. 

\subsection{Simple leading-order chiral Hamiltonian} 
In the wavefunction matching framework, the simple leading-order chiral Hamiltonian is constructed for the non-perturbative part of the wavefunction matching procedure \cite{Elhatisari:2024}, 
\begin{equation}
  H^{\text{S}}=H_{\text{free}}+\frac{1}{2} c \sum_{\vec{n}}: \tilde{\rho}^2(\vec{n}): + V^{\Lambda_{\pi}}_{\text{OPE}}.
\end{equation}
In addition to the short-range SU(2) (or SU(4) in nuclear systems) symmetric interaction, we also have a long-range one-pion-exchange (OPE) potential
at leading order $\chi$EFT interaction. 
The one-pion-exchange potential follows a recently developed regularization
method~\cite{Reinert:2017usi},
\begin{align}
V_{\rm OPE}^{\Lambda_{\pi}}  = - &   \frac{g_A^2}{8f^2_{\pi}}\ \, \sum_{{\bf n',n},S',S,I}
:\rho_{S',I\rm }^{(0)}(\vec{n}')f_{S',S}(\vec{n}'-\vec{n}) 
  \rho_{S,I}^{(0)}(\vec{n}):  
\,,
\label{eq:OPEP-full}
\end{align}
\begin{align}
V_{\rm C_{\pi}}^{\Lambda_{\pi}}  = -C_{\pi} \, \frac{g_A^2}{8f^2_{\pi}} 
\sum_{{\bf n',n},S,I}
:\rho^{(0)}_{S,I}(\vec{n}')
f^{\pi}(\vec{n}'-\vec{n})
  \rho^{(0)}_{S,I}(\vec{n}):\,
\label{eq:OPEP-counter}
\end{align}
Here $f^{\pi}$ is a local regulator in momentum space defined as
\begin{align}
f^{\pi}(\vec{n}'-\vec{n})
= &
\frac{1}{L^3}
\sum_{\vec{q}}
e^{-i\vec{q}\cdot(\vec{n}'-\vec{n})-(\vec{q}^2+M^2_{\pi})/\Lambda_{\pi}^2}\,,
\end{align}
$f_{S',S}$ is the locally-regulated pion correlation function,
\begin{align}
f_{S',S}(\vec{n}'-\vec{n}) 
= &\frac{1}{L^3}\sum_{\vec{q}}
\frac{q_{S'}q_{S} \, e^{-i\vec{q}\cdot(\vec{n}'-\vec{n})-(\vec{q}^2+M^2_{\pi})/\Lambda_{\pi}^2}}{\vec{q}^2 + M_{\pi}^2} \,,
\end{align}
and
\begin{align}
C_{\pi} = & -
\frac{\Lambda_{\pi} (\Lambda_{\pi}^2-2M_{\pi}^{2}) + 2\sqrt{\pi} M_{\pi}^3\exp(M_{\pi}^2/\Lambda_{\pi}^{2}){\rm erfc}(M_{\pi}/\Lambda_{\pi})}
{3 \Lambda_{\pi}^3}\,,
\end{align}
with $g_{A}=1.287$ the axial-vector coupling constant (adjusted to account for the Goldberger-Treiman discrepancy)\cite{Fettes:1998ud}, $f_{\pi}=92.2$~MeV the pion decay constant and $M_{\pi}=134.98$~MeV the pion mass. The term given in Eq.~(\ref{eq:OPEP-counter}) is a counterterm introduced to remove the short-distance admixture in the one-pion-exchange potential~\cite{Reinert:2017usi}. In the simple Hamiltonian $H^{\text{S}}$, we set  $\Lambda_{\pi}=180$~MeV and $C_{\pi} = 0$, and the difference $V_{\rm OPE}^{\Lambda_{\pi}=300}-V_{\rm OPE}^{\Lambda_{\pi}=180}$ along with the counterterm $V_{\rm C_{\pi}}^{\Lambda_{\pi}}$ are calculated perturbatively. Here we use the notation
\begin{align}
\rho^{(d)}(\vec{n}) = \sum_{i,j=0,1} 
a^{\dagger}_{i,j}(\vec{n}) \, a^{\,}_{i,j}(\vec{n})
+
s_{\rm L}
 \sum_{|\vec{n}-\vec{n}^{\prime}|^2 = 1}^d 
 \,
 \sum_{i,j=0,1} 
a^{\dagger}_{i,j}(\vec{n}^{\prime}) \, a^{\,}_{i,j}(\vec{n}^{\prime})
.
\label{eqn:appx--001}
\end{align}
and 
\begin{align}
\rho^{(d)}_{S,I}(\vec{n}) = & \sum_{i,j,i^{\prime},j^{\prime}=0,1} 
a^{\dagger}_{i,j}(\vec{n}) \, [\sigma_{S}]_{ii^{\prime}} \, [\sigma_{I}]_{jj^{\prime}} \, a^{\,}_{i^{\prime},j^{\prime}}(\vec{n})
\nonumber \\
& +
s_{\rm L}
 \sum_{|\vec{n}-\vec{n}^{\prime}|^2 = 1}^d 
 \,
  \sum_{i,j,i^{\prime},j^{\prime}=0,1} 
a^{\dagger}_{i,j}(\vec{n}^{\prime}) \, [\sigma_{S}]_{ii^{\prime}} \, [\sigma_{I}]_{jj^{\prime}} \, a^{\,}_{i^{\prime},j^{\prime}}(\vec{n}^{\prime})
\label{eqn:appx--005}
\end{align}
for the density operators, with the isospin indices $i$. 

\subsection{Chiral N$^3$LO Hamiltonian}
{The high fidelity $\chi$EFT Hamiltonian at N$^3$LO level} is used to calculate realistic neutron matter systems.
\begin{equation}
H' = H_{\text{free}} 
   + V_{\rm OPE}^{\Lambda_{\pi}} 
   + V_{\rm C_{\pi}}^{\Lambda_{\pi}} 
   + V_{\rm Coulomb}
   + V_{\rm 3N}^{\rm Q^3}
   + V_{\rm 2N}^{\rm Q^4}
   + W_{\rm 2N}^{\rm Q^4}
   + V_{\rm 2N,WFM}^{\rm Q^4}
   + W_{\rm 2N,WFM}^{\rm Q^4}.
\label{eq:H-N3LO}
\end{equation}
$V_{\rm OPE}^{\Lambda_{\pi}}$ and $V_{\rm C_{\pi}}^{\Lambda_{\pi}}$ are defined in Eqs.~(\ref{eq:OPEP-full}) and (\ref{eq:OPEP-counter}) with ${\Lambda_{\pi}=300}$~MeV. 
$V_{\rm Coulomb}$ denotes the Coulomb interaction, whose expectation value vanishes in pure neutron systems.
$V_{\rm 3N}^{\rm Q^3}$ represents the three-nucleon (3N) interaction.
$V_{\rm 2N}^{\rm Q^4}$ denotes the short-range two-nucleon (2N) interaction at N$^3$LO in $\chi$EFT, while $W_{\rm 2N}^{\rm Q^4}$ is the corresponding Galilean-invariance-restoration (GIR) term at the same chiral order.
$V_{\rm 2N,WFM}^{\rm Q^4}$ refers to the wavefunction-matching interaction defined as $H' - H$, and $W_{\rm 2N,WFM}^{\rm Q^4}$ is its associated GIR correction.
Further details can be found in the Supplemental Material of our previous wavefunction matching work~\cite{Elhatisari:2024}.

% \section{Lattice 2 (placeholder name)}

\section{Lattice Measurements}\label{sec:latt_measure}
\subsection{Off-diagonal long-range order} 
The concept of off-diagonal long-range order (ODLRO) was first proposed in C. N. Yang's work\cite{Yang:1962zz} as an emergent feature in superfluid He II and superconductors. For Fermi systems, we define the one-body density matrix,
% \begin{equation}
% \rho^I_{j'j}\left(\vec{n}^{\prime}, \vec{n}\right)
% =\langle \Psi| a_{j'}^{\dagger}\left(\vec{n}^{\prime}\right) a_j(\vec{n})|\Psi\rangle
% \end{equation}
\begin{equation}
\rho^I_{j'j}\left(\vec{r}^{\prime}, \vec{r}\right)
=\langle \Psi| a_{j'}^{\dagger}\left(\vec{r}^{\prime}\right) a_j(\vec{r})|\Psi\rangle
\end{equation}
and the irreducible two-body density matrix,
\begin{equation}\label{eq:ODLRO}
\begin{aligned}
 \rho_{j_1j_2}^{II}(\vec{r}_1^{\prime}, \vec{r}_2^{\prime}, \vec{r}_2, \vec{r}_1) 
    = \langle\Psi|a_{j_1}^{\dagger}(\vec{r}_1^{\prime}) a_{j_2}^{\dagger}(\vec{r}_2^{\prime}) a_{j_2}(\vec{r}_2) a_{j_1}(\vec{r}_1)| \Psi \rangle 
    &- \langle\Psi|a_{j_1}^{\dagger}(\vec{r}_1^{\prime}) a_{j_1}(\vec{r}_1) | \Psi \rangle
   \langle\Psi|a_{j_2}^{\dagger}(\vec{r}_2^{\prime}) a_{j_2}(\vec{r}_2) | \Psi \rangle \\
   &+ \langle\Psi|a_{j_1}^{\dagger}(\vec{r}_1^{\prime}) a_{j_2}(\vec{r}_2) | \Psi \rangle
   \langle\Psi|a_{j_2}^{\dagger}(\vec{r}_2^{\prime}) a_{j_1}(\vec{r}_1) | \Psi \rangle
\end{aligned}
\end{equation}
Then the off-diagonal long-range order (ODLRO) can be defined as
\begin{equation}
    F_{j_1 j_2} =  \lim_{\vec{R} \to \infty} \rho^{II}_{j_1 j_2} (\vec{R}) = \lim_{\vec{R} \to \infty} \int d{\vec{r}_{12}}  \rho_{j_1j_2}^{II}(\vec{R}, \vec{r}_{12}) = \lim_{\vec{R} \to \infty}   \int d{\vec{r}_{12}} \int  d{\vec{r}_1}  \rho_{j_1j_2}^{II}(\vec{r}_1 + \vec{R}, \vec{r}_1 + \vec{r}_{12} + \vec{R}, \vec{r}_1 + \vec{r}_{12}, \vec{r}_1)
\end{equation}
in which we define $\vec{R}=\vec{r}'_1 - \vec{r}_1$, $\vec{r}_{12}=\vec{r}_2 - \vec{r}_1$, $\vec{r}'_{12}=\vec{r}'_2 - \vec{r}'_1$ and demand $\vec{r}'_{12} = \vec{r}_{12}$.
%It clearly shows that ODLRO annihilates a pair of two particles at one position and creates them at a long distance $\vec{R}$. 
This definition corresponds to annihilating a pair of particles at one location and creating them at a long distant separation $\vec{R}$.
When $\vec{R} \to \infty$, $\rho^{II}_{j_1 j_2}(\vec{r}_{12})$  actually gives the density distribution of the pairing wavefunction with certain spins $j_1j_2$, and the integral of $\vec{r}_{12}$ indicates the number of pairs.

\subsection{Irreducible momentum pairs} 
Following a construction analogous to ODLRO, the irreducible momentum pairs are defined in momentum space. 
Similar to the definition in coordinate space, the one-body density matrix in momentum space is:
\begin{equation}
    \rho^I_{j'j}(\vec{k}^{\prime}, \vec{k})
    =\langle \Psi| a_{j'}^{\dagger}(\vec{k}^{\prime}) a_j(\vec{k})|\Psi\rangle.
\end{equation}
The irreducible paired two-body density in momentum space, 
\begin{equation}\label{eq:ODLRO-k}
\begin{aligned}
 \rho_{j_1j_2}^{II}(\vec{k}_1, \vec{k}_2, \vec{k}_2, \vec{k}_1) 
    = \langle\Psi|a_{j_1}^{\dagger}(\vec{k}_1) a_{j_2}^{\dagger}(\vec{k}_2) a_{j_2}(\vec{k}_2) a_{j_1}(\vec{k}_1)| \Psi \rangle 
    &- \langle\Psi|a_{j_1}^{\dagger}(\vec{k}_1) a_{j_1}(\vec{k}_1) | \Psi \rangle
   \langle\Psi|a_{j_2}^{\dagger}(\vec{k}_2) a_{j_2}(\vec{k}_2) | \Psi \rangle \\
   &+ \langle\Psi|a_{j_1}^{\dagger}(\vec{k}_1) a_{j_2}(\vec{k}_2) | \Psi \rangle
   \langle\Psi|a_{j_2}^{\dagger}(\vec{k}_2) a_{j_1}(\vec{k}_1) | \Psi \rangle
\end{aligned}
\end{equation}
The momentum pairs are defined with zero total momentum $\vec{k}_2 = -\vec{k}_1$.
Thus, the momentum pair measurements (two-body cumulants) are,
\begin{equation}
    \begin{aligned}
         F_{j_1 j_2} =\int d{\vec{k}_1} \rho^{II}_{j_1 j_2} (\vec{k}_1) = \int d{\vec{k}_1}  \rho_{j_1j_2}^{II}(\vec{k}_1, -\vec{k}_1, -\vec{k}_1, \vec{k}_1) 
    \end{aligned}
\end{equation}
{The relation between irreducible two-body density in r-space and irreducible momentum pairs} can be seen from the inverse Fourier transformation of the momentum pair operator (here we omit spin indices for clarity),
\begin{equation}
    \langle \Psi | a^{\dagger}(\vec{k}_1) a^{\dagger}(\vec{k}_2) a(\vec{k}_2) a(\vec{k}_1) | \Psi \rangle 
    = \frac{1}{V^2}\langle \Psi | \int d \vec{r}'_1 d \vec{r}'_2 d \vec{r}_2 d \vec{r}_1  e^{ik_1 \vec{r}'_1 + i\vec{k}_2 \vec{r}'_2 - i\vec{k}_2 \vec{r}_2 - i\vec{k}_1 \vec{r}_1 } a^{\dagger}(\vec{r}'_1) a^{\dagger}(\vec{r}'_2) a(\vec{r}_2) a(\vec{r}_1) | \Psi \rangle 
\end{equation}
Considering $\vec{k}_2=-\vec{k}_1$, $\vec{r}'_1 = \vec{r}_1 + \vec{R}$, $\vec{r}_2 = \vec{r}_1 + \vec{r}_{12}$ and $\vec{r}'_2 = \vec{r}'_1 + \vec{r}'_{12}$,
the equation above can be rewritten as:
\begin{equation}
    \begin{aligned}
    \rho^{II}\left(\vec{k}_1, \vec{k}_2, \vec{k}_2, \vec{k}_1\right) 
    % &=\langle a^{\dagger}(\vec{k}_1) a^{\dagger}(\vec{k}_2) a(\vec{k}_2) a(\vec{k}_1)  \rangle \\
    % &=\frac{1}{V^2} \langle  \int d \vec{r}_1 d \vec{r}_{12} d \vec{r}'_{12} d \vec{R}  e^{i\vec{k}_1 \vec{r}'_1 + i\vec{k}_2 \vec{r}'_2 - i\vec{k}_2 \vec{r}_2 - i\vec{k}_1 \vec{r}_1 } a^{\dagger}(\vec{r}'_1) a^{\dagger}(\vec{r}'_2) a(\vec{r}_2) a(\vec{r}_1) \rangle \\
    &= \frac{1}{V^2}\langle  \int d \vec{r}_1 d \vec{r}_{12} d \vec{r}'_{12} d \vec{R}  e^{i(\vec{k}_1+ \vec{k}_2)\vec{R}+ i\vec{k}_2 (\vec{r}'_{12} - \vec{r}_{12} ) } a^{\dagger}(\vec{r}'_1) a^{\dagger}(\vec{r}'_2) a(\vec{r}_2) a(\vec{r}_1) \rangle    
    \end{aligned}
\end{equation}
Then we will obtain
\begin{equation}
    \begin{aligned}
    \int d \vec{k}_1 d \vec{k}_2 \rho^{II}\left(\vec{k}_1, \vec{k}_2, \vec{k}_2, \vec{k}_1\right) 
     &= \int d \vec{k}_1 d \vec{k}_2 \frac{1}{V^2}\langle  \int d \vec{r}_1 d \vec{r}_{12} d \vec{r}'_{12} d \vec{R}  e^{i(\vec{k}_1+ \vec{k}_2)\vec{R}+ i\vec{k}_2 (\vec{r}'_{12} - \vec{r}_{12} ) } a^{\dagger}(\vec{r}'_1) a^{\dagger}(\vec{r}'_2) a(\vec{r}_2) a(\vec{r}_1) \rangle    \\
     &= \int d \vec{k}_2 \frac{1}{V} \int d \vec{r}_1 d \vec{r}_{12} d \vec{r}'_{12} d \vec{R}  e^{i\vec{k}_2 (\vec{r}'_{12} - \vec{r}_{12} ) } \rho^{II}(\vec{r}_1,\vec{r}_{12}, \vec{r}'_{12}, \vec{R} )   \\
    &=  \int d \vec{r}_1 d \vec{r}_{12} d \vec{R}  \rho^{II}(\vec{r}_1,\vec{r}_{12}, \vec{r}'_{12}, \vec{R} )   \\
    \end{aligned}
\end{equation}
Thus, there exist two differences between momentum pair measurement and ODLRO: i) $\rho^{II}_{\vec{k}}$ averages all $\vec{R}$ over all lattice sites of $\rho^{II}_{\vec{r}}$, whereas ODLRO only takes $\vec{R} \approx L/2$ of $\rho^{II}_{\vec{r}}$; ii) $\rho^{II}_{\vec{k}}$ sums over all relative separations $\vec{r}_{12}$ of $\rho^{II}_{\vec{r}}$, while, while the ODLRO definition, in order to reduce lattice artifacts, restricts the sum to $|\vec{r}_{12}| \leq L/2$.
However these differences decrease as the lattice volume $V$ increases and should vanish in the thermodynamic limit.

\subsection{S-wave and P-wave pairing} 
S-wave and p-wave pairing can be identified both from the measurement of ODLRO and irreducible momentum pairs. 
%We first discussed it in coordinate space.
The pairing wavefunction includes the spin part and the radial part $\Psi(\vec{r}_{12}) = \phi(\vec{r}_{12}) \chi_{12}$.
The two-spin state $\chi_{12}$  can be coupled to total spin $S=0$ and $S=1$.
\begin{table}[h!]
    \centering
\begin{tabular}{cccc}
    \hline
    $S$ & $S_z$ & $\chi_{12}$ & parity of $\chi_{12}$ \\ 
    \hline
    0  & 0   & $\frac{1}{\sqrt{2}}[\vert{\uparrow \downarrow} \rangle - \vert{\downarrow \uparrow} \rangle]$   & $-$   \\
    \hline
    1  & 1   & $\vert{\uparrow \uparrow} \rangle$     & $+$       \\ 
    1  & 0   & $\frac{1}{\sqrt{2}}[\vert{\uparrow \downarrow} \rangle + \vert{\downarrow \uparrow} \rangle]$  & $+$   \\ 
    1  & -1  & $\vert{\downarrow \downarrow} \rangle$            & $+$     \\ 
    \hline
\end{tabular}
\end{table}
Fermi antisymmetry demands $\Psi(\vec{r}_{12}) = - \Psi(\vec{r}_{21})$. 
Correspondingly, the radial part has even and odd parity for spin singlet and triplet:
\begin{equation}
    \Psi(\vec{r}_{12}) = \phi^{\text{Even}}(\vec{r}_{12}) \chi^{s=0}_{12}
\end{equation}
or 
\begin{equation}
    \Psi(\vec{r}_{12}) = \phi^{\text{Odd}}(\vec{r}_{12}) \chi^{s=1}_{12}.
\end{equation}
The lowest radial orbital for spin singlet is s-wave and for spin triplet it is p-wave.
% Then the pairing correlation strength
% \begin{equation}
%     F_{\uparrow\downarrow}(\vec{r}_{12}) = \phi_{\uparrow\downarrow}^*(\vec{r}_{12})\phi_{\uparrow\downarrow}(\vec{r}_{12}); \quad
%     F_{\uparrow\downarrow}^{\text{ex}}(\vec{r}_{12}) = \phi_{\uparrow\downarrow}^*(-\vec{r}_{12})\phi_{\uparrow\downarrow}(\vec{r}_{12});
% \end{equation}
In the $S_z$ representation, $\rho_{ \uparrow \downarrow}$ will have both the $S=0$ and $S=1$ components.
They can be extracted by even/odd parity.
Even parity of the paired two-particle radial wavefunction corresponds to s-wave pairing,
\begin{equation}
    \begin{aligned}
    F^{S}= \lim_{\vec{R}\to \infty} \rho^{II}_S(\vec{R}) 
    =   \lim_{\vec{R}\to \infty} \int d{\vec{r}_{12}} \frac{1}{2} [\rho^{II}_{\uparrow\downarrow}(\vec{R}, \vec{r}_{12}) + \rho^{II-\text{ex}}_{\uparrow\downarrow}(\vec{R}, \vec{r}_{12})] 
    \end{aligned} 
\end{equation}
where $\rho^{II-\text{ex}}(\vec{R}, \vec{r}_{12})$ means we annihilate two particle at $\vec{r}_1$, $\vec{r}_2$ and create them at $\vec{r}'_1 = \vec{r}_2 + \vec{R}$, $\vec{r}'_2 = \vec{r}_1 + \vec{R}$.
Odd parity of the paired two-particle radial wavefunction corresponds to the p-wave pairing,
\begin{equation}
    \begin{aligned}
    F^{P} = \lim_{\vec{R}\to \infty} \rho^{II}_P(\vec{R})
     = \lim_{\vec{R}\to \infty} \int d{\vec{r}_{12}} \frac{1}{2} \big{[} \rho^{II}_{\uparrow \uparrow}(\vec{R},\vec{r}_{12})
    + \rho^{II}_{\downarrow \downarrow}(\vec{R},\vec{r}_{12})
    + [\rho^{II}_{\uparrow\downarrow}(\vec{R},\vec{r}_{12}) - \rho^{II-\text{ex}}_{\uparrow\downarrow}(\vec{R},\vec{r}_{12})] \big{]}
    \end{aligned} 
\end{equation}
% Here the factor $\frac{1}{2}$ in front of $\rho^{II}_{\uparrow \uparrow }$ and $\rho^{II}_{\downarrow \downarrow }$ is due to twice summation when $\vec{r}_{1}$ goes over all lattice sites.
The factor $\frac{1}{2}$ of $\rho^{II}_{\uparrow\downarrow}$ and $\rho^{II}_{\downarrow \downarrow}$ arises from double counting when $\vec{r}_{1}$ runs over all lattice sites.
It should also be noted that $\rho^{II}_{\uparrow \downarrow }$ should always be larger than $\rho^{II-ex}_{\uparrow \downarrow }$. 
Considering $\rho^{II}_{\uparrow \downarrow }=Even+Odd$, and $Even$ or $Odd$ is positive-definite because it is the square of a pairing wavefunction, then $\rho^{II-ex}_{\uparrow \downarrow }=Even-Odd$.
It is similar in the measurement of momentum pairs,  
\begin{equation}
    \begin{aligned} \label{eq:swave_k}
         F^{S} = \int d{\vec{k}_1} \rho^{II}_{S}(\vec{k}_1) = \frac{1}{2} \int d{\vec{k}_1} [\rho^{II}_{\uparrow \downarrow} (\vec{k}_1) + \rho^{II-\text{ex}}_{\uparrow \downarrow} (\vec{k}_1)]
    \end{aligned}
\end{equation}
with $ \rho^{II}_{\uparrow \downarrow} (\vec{k}_1) = \rho_{\uparrow \downarrow}^{II}(\vec{k}_1, -\vec{k}_1, -\vec{k}_1, \vec{k}_1)$, and $ \rho^{II-\text{ex}}_{\uparrow \downarrow} (\vec{k}_1)= \rho_{\uparrow \downarrow}^{II}(-\vec{k}_1, \vec{k}_1, -\vec{k}_1, \vec{k}_1)$.
Then the p-wave pairing strength can be measured by 
\begin{equation}
    \begin{aligned}\label{eq:pwave_k}
         F^{P} = \int d{\vec{k}_1} \rho^{II}_{P}(\vec{k}_1) = \frac{1}{2} \int d{\vec{k}_1} \big{[} \rho^{II}_{\uparrow \uparrow} (\vec{k}_1) + \rho^{II}_{\downarrow \downarrow} (\vec{k}_1) + [\rho^{II}_{\uparrow \downarrow} (\vec{k}_1) - \rho^{II-\text{ex}}_{\uparrow \downarrow} (\vec{k}_1)] \big{]}.
    \end{aligned}
\end{equation}

\subsection{Irreducible momentum quartets} 
The quartets can be measured from the irreducible four-body density (or fourth-order cumulant) operators $\rho^{IV}_{\uparrow \downarrow \uparrow \downarrow}(\vec{k}_1,\vec{k}_2,\vec{k}_3,\vec{k}_4)$, in which we require zero momentum of quartets $\vec{k}_1= -\vec{k}_2 -\vec{k}_3 -\vec{k}_4$ but any sub-combination cannot be zero, like $\vec{k}_1+\vec{k}_2 \neq 0 $, $\vec{k}_1+\vec{k}_2+\vec{k}_3 \neq 0 $ ..., etc. 
Following the previous discussion, we define irreducible one, two, and three density operators as 
\begin{equation}
\begin{aligned}
    \rho^{I}_{1} &= \rho_{1}\\
    \rho_{12}^{II} &= \rho_{12} - \rho_1 \rho_2 \\
    \rho_{123}^{III} 
    &= \rho_{123} - (\rho_1 \rho^{II}_{23}+\rho_2 \rho^{II}_{13} + \rho_3 \rho^{II}_{12}) 
       - \rho_1 \rho_2 \rho_3  \\
    &= \rho_{123} - (\rho_1 \rho_{23} + \rho_2 \rho_{13} + \rho_3 \rho_{12}) + 2\rho_1 \rho_2 \rho_3
\end{aligned}
\end{equation}
in which
\begin{equation}
    \begin{aligned}
        \rho_{1} &= \langle \Psi| a_{j_1}^{\dagger}(\vec{k}_1) a_{j_1}(\vec{k}_1)|\Psi\rangle \\
        \rho_{12} &=  \langle\Psi|a_{j_1}^{\dagger}(\vec{k}_1) a_{j_2}^{\dagger}(\vec{k}_2) a_{j_2}(\vec{k}_2) a_{j_1}(\vec{k}_1)| \Psi \rangle \\
        \rho_{123} &=  \langle\Psi|a_{j_1}^{\dagger}(\vec{k}_1) a_{j_2}^{\dagger}(\vec{k}_2) a_{j_3}^{\dagger}(\vec{k}_3) a_{j_3}(\vec{k}_3) a_{j_2}(\vec{k}_2) a_{j_1}(\vec{k}_1)| \Psi \rangle.
    \end{aligned}
\end{equation}
It should be noticed that we only consider the diagonal one-body density in the quartet calculations, which will be discussed later in the momentum pinhole method.
Then the fourth-order cumulant of four-body density operators can be written as,
\begin{equation} \label{eq:rhoIV_irre}
\begin{aligned} 
    \rho_{1234}^{IV} = \rho_{1234} 
    &- (\rho^{II}_{12} \rho^{II}_{34} + \rho^{II}_{13} \rho^{II}_{24} + \rho^{II}_{14} \rho^{II}_{23}) \\
    &- (\rho_1 \rho^{III}_{234}+\rho_2 \rho^{III}_{134} + \rho_3 \rho^{III}_{124} + \rho_4 \rho^{III}_{123}) \\
    &- (\rho_1 \rho_2 \rho^{II}_{34} + \rho_1 \rho_3 \rho^{II}_{24} + \rho_1 \rho_4 \rho^{II}_{23} + \rho_2 \rho_3 \rho^{II}_{14} + \rho_2 \rho_4 \rho^{II}_{13} + \rho_3 \rho_4 \rho^{II}_{12}) \\
    &- \rho_1 \rho_2 \rho_3 \rho_4
\end{aligned}
\end{equation}
with the ``raw'' four body 
\begin{equation}\label{eq:rhoIV_raw}
    \begin{aligned}
    &\rho_{1234}   
    = \langle \Psi \vert  a_{j_1}^{\dagger}(\vec{k}_1) a_{j_2}^{\dagger}(\vec{k}_2) a_{j_3}^{\dagger}(\vec{k}_3) a_{j_4}^{\dagger}(\vec{k}_4) a_{j_4}(\vec{k}_4) a_{j_3}(\vec{k}_3) a_{j_2}(\vec{k}_2) a_{j_1}(\vec{k}_1) \vert  \Psi \rangle.
    \end{aligned}    
\end{equation} 
Finally, the total irreducible quartet correlation strength can be measured by $F^{{Q}}=\int d{\vec{k}_1} \rho^{IV}(\vec{k}_1)$ (or in discrete form $F^{\text{Q}}=\sum_{\vec{k}_1} \rho^{IV}(\vec{k}_1)$), with
% \begin{equation} \label{eq:rhoIV_k}
%     \begin{aligned}
%         \rho^{IV}(\vec{k}_1) = \int_{\vec{k}_2 + \vec{k}_3+ \vec{k}_4 = - \vec{k}_1} d_{\vec{k}_2, \vec{k}_3, \vec{k}_4} \frac{1}{4} \times [\rho_{\uparrow \downarrow \uparrow \downarrow}^{IV} (\vec{k}_1, \vec{k}_2, \vec{k}_3, \vec{k}_4) + \rho_{\uparrow \downarrow \uparrow \downarrow}^{IV} (\vec{k}_2, \vec{k}_1, \vec{k}_3, \vec{k}_4)]
%     \end{aligned}
% \end{equation}
\begin{equation}\label{eq:rhoIV_k}
\rho^{IV}(\vec k_1)
= \frac{1}{4} \int d\vec{k}_2\, d\vec{k}_3\, d\vec{k}_4\;
\delta^{(3)}(\vec k_1+\vec k_2+\vec k_3+\vec k_4)\,
\Big[
\rho^{IV}_{\uparrow\downarrow\uparrow\downarrow}(\vec k_1,\vec k_2,\vec k_3,\vec k_4)
+\rho^{IV}_{\uparrow\downarrow\uparrow\downarrow}(\vec k_2,\vec k_1,\vec k_3,\vec k_4)
\Big].
\end{equation}
where we count both spin-up and spin-down in $\rho^{IV}(\vec{k}_1)$ and $\frac{1}{4}$ guarantees the norm when we sum all the $\rho^{IV}(\vec{k}_1)$.
It should be mentioned that the two-body pairing strength can be directly translated to the number of pairs, but the quartet strength does not admit a simple one-to-one mapping to the number of quartets. 

\subsection{Quartet number and four-body cumulants}\label{sec:quartet_num}
Before discussing the quartet number, we first introduce the pairing strength and the number of pairs.
The pairing density matrix is defined as:
\begin{equation}
\rho_{\vec{k}, \vec{k}^{\prime}}^{\text {pair }}=\langle b_{\vec{k}}^{\dagger} b_{\vec{k}^{\prime}}\rangle=\langle a_{\vec{k} \uparrow}^{\dagger} a_{-\vec{k} \downarrow}^{\dagger} a_{-\vec{k}^{\prime} \downarrow} a_{\vec{k}^{\prime} \uparrow}\rangle
\end{equation}
with pairing creation and annihilation operator: $b_{\vec{k}}^{\dagger}=a_{\vec{k} \uparrow}^{\dagger} a_{-\vec{k} \downarrow}^{\dagger}$ and $b_{\vec{k}}=a_{\vec{k} \uparrow} a_{-\vec{k} \downarrow}$.
Since each pair can only occupy one momentum channel, the trace of the pairing density matrix yields the total number of pairs:
\begin{equation}
N_{\text{pair}}=\operatorname{Tr} \rho^{\text {pair }}=\sum_{\vec{k}}\langle b_{\vec{k}}^{\dagger} b_{\vec{k}}\rangle=\sum_{\vec{k}}\langle n_{\vec{k} \uparrow} n_{-\vec{k} \downarrow}\rangle.
\end{equation}
Considering a four-momentum channel $\alpha = (\vec{k}_1,\vec{k}_2,\vec{k}_3,\vec{k}_4)$, we define the four-body creation and annihilation operator,
\begin{equation}
    \begin{aligned}
        A^{\dagger}_{\alpha} 
        = a^{\dagger}_{\uparrow}(\vec{k}_1) 
          a^{\dagger}_{\downarrow}(\vec{k}_2)
          a^{\dagger}_{\uparrow}(\vec{k}_3)
          a^{\dagger}_{\downarrow}(\vec{k}_4);
          \quad
        A_{\alpha} 
        = a_{\uparrow}(\vec{k}_1) 
          a_{\downarrow}(\vec{k}_2)
          a_{\uparrow}(\vec{k}_3)
          a_{\downarrow}(\vec{k}_4)
    \end{aligned}
\end{equation}
The quartet creation operator is defined as
$Q^{\dagger}_i = \sum_{\alpha} \Phi^{i}_{\alpha} A^{\dagger}_{\alpha}$, where the quartet wavefunctions satisfy $\sum_{\alpha} |\Phi_{\alpha}|^2 = 1$ and the orthogonal relation $\langle Q^{\dagger}_i Q_j \rangle = \delta_{i,j}$.
The corresponding quartet number operator is $ \hat{N}_Q \equiv Q_{i}^{\dagger} Q_{i}$, with expectation value $\langle \hat{N}_Q \rangle = \sum_{\alpha \beta} \Phi^{i*}_{\alpha} \Phi^i_{\beta} \langle A^{\dagger}_{\alpha} A_{\beta} \rangle $.
The quartet occupations are obtained from the eigenvalue problem of the irreducible four-body density matrix,
\begin{equation}
\sum_\beta \rho_{\alpha \beta}^{IV} \Phi_\beta^{(n)}=\lambda_n \Phi_\alpha^{(n)}
\end{equation}
where the eigenvalue $\lambda_n$ represents the quartet occupation for configuration $\Phi_\beta^{(n)}$. If the largest eigenvalue $\lambda_0\leq N_{Q}$ (how many quartets are occupying $\Phi_\beta^{(0)}$) has the scaling of $\lambda_0 \sim \mathcal{O}(N)$, we can identify the quartet off-diagonal long-range order in the system \cite{Yang:1962zz}. 

Recalling that our four-body density is $\hat{\rho}_{1234} = A^{\dagger}_{\alpha} A_{\alpha}$, then the total irreducible quartet correlation strength $F^{{Q}}$ or the four-body cumulant $N_{\text{cum}}=\sum_{\vec{k}} \rho^{IV}(\vec{k})$ actually counts the diagonal parts from all the quartets in the system.
% Considering we manually generate $N_{gen}$ quartets in the system.
To relate the total irreducible quartet correlation strength to an effective quartet number, we consider ensembles with a controlled number $N_{\text{gen}}$ of generated quartets.
% At the dilute limit, quartets are separate with each other thus $N_{cum}$ will linear dependent on the quartet number $N_Q$.
In the dilute limit, quartets are well separated from one another; consequently, the total four-body cumulant $N_{\text{cum}}$ scales linearly with the quartet number $N_Q$.
% As we increase $N_{gen}$, quartets will coherent response with each other depends on higher order of $N_{gen}$. 
As $N_{gen}$ increases, coherent correlations among quartets emerge, leading to higher-order (nonlinear) contributions in $N_{\text{gen}}$. 
Thus we have 
\begin{equation}
    \begin{aligned} \label{eq:Ncum_Ngen}
        N_{\text{cum}} = \alpha N_{\text{gen}} + \beta N^2_{\text{gen}} + \mathcal{O}(N^2_{\text{gen}})
    \end{aligned}
\end{equation}
The linear term reflects self-correlation or classical inventory, which tells us how many quartets we can measure for a certain $N_{\text{gen}}$. Thus we define $N^{\text{eff}}_{\text{cum}} = \alpha N_{\text{gen}}$ as the number of quartet $N_Q$ in the system. 
The effective quartet number  $N^{\text{eff}}_{\text{cum}}$ is extracted by Monte Carlo sampling of quartet configurations drawn from the momentum distribution $\rho^{IV}(\vec{k})$.
One numerical example is the result in Fig~\ref{fig:SU2_Q_number}.
% \section{Lattice 3 (placeholder name)}
\section{Rank-One Operator Method and Momentum Pinhole Method}
In the framework of the Nuclear Lattice Effective Field theory, the observable operator is measured by calculating the ratio of amplitudes with and without an operator inserted
$\langle {O} \rangle = \det {M}({O})/ \det {M}$, with the amplitudes $\det {M}$ being the Slater determinant of the single-nucleon correlation matrix $M$.
\subsection{Rank-one Operator method} 
The rank-one operator method (RO) was first proposed in our previous work~\cite{Ma:2024}.
In contrast to the commonly used Jacobi method, also discussed in Ref~\cite{Ma:2024}, the RO operator avoids this exponential scaling by using one-body operators that have the form $f^\dagger_{\alpha'}f_{\alpha}$, where $f_{\alpha}$ is the annihilation operator for nucleon orbital $\alpha$ and $f^\dagger_{\alpha'}$ is the creation operator for nucleon orbital $\alpha'$.  Since $f_{\alpha}$ can only annihilate one nucleon and $f^\dagger_{\alpha'}$ can only create one nucleon, it is an operator of rank one.  We conclude that the insertion of the normal-ordered exponential $:\exp(t f^\dagger_{\alpha'}f_{\alpha}):$ yields
\begin{equation}
    \det { M}[:\exp(t f^\dagger_{\alpha'}f_{\alpha}):] = C + t \det { M}[f^\dagger_{\alpha'}f_{\alpha}],
\end{equation}
The absence of higher-order powers of $t$ allows us to compute ${\det M}[f^\dagger_{\alpha'}f_{\alpha}]$ very easily by taking the limit of large $t$ and dividing by $t$,
\begin{equation}
    {\det M}[f^\dagger_{\alpha'}f_{\alpha}] = \lim_{t\rightarrow \infty} \frac{1}{t}{\det M}[:\exp(t f^\dagger_{\alpha'}f_{\alpha}):].
\end{equation}
We can then implement the RO method to measure the densities of one-body and two bodies.
% For momentum measurements we refer the method as ``ROK''.
For the three-body and four-body momentum densities, due to the large number of $k_1, k_2, k_3$ combinations on the lattice, measuring all momentum configurations with the RO method will be challenging. 
This is why the momentum pinhole method was developed.

\subsection{Pinhole method}
{The pinhole method} (PH) was first introduced in Ref.~\cite{Elhatisari:2017} to study clustering in Carbon isotopes. The key idea is using Metropolis algorithm to generate A-body density configurations $\rho_{i_1, j_1, \cdots i_A, j_A}\left(\vec{n}_1, \cdots \vec{n}_A\right)=: \rho_{i_1, j_1}\left(\vec{n}_1\right) \cdots \rho_{i_A, j_A}\left(\vec{n}_A\right):$.
In the $A$-nucleon subspace, we have the completeness identity
\begin{equation}
\sum_{i_1, j_1, \cdots i_A, j_A} \sum_{\vec{n}_1, \cdots \vec{n}_A} \rho_{i_1, j_1, \cdots i_A, j_A}\left(\vec{n}_1, \cdots \vec{n}_A\right)=A !
\end{equation}
The amplitude with one pinhole configuration is 
\begin{equation}
Z_{f, i}\left(i_1, j_1, \cdots i_A, j_A ; \vec{n}_1, \cdots \vec{n}_A ; L_t\right)=\left\langle\Psi_f\left| M^{L_t / 2} \rho_{i_1, j_1, \cdots i_A, j_A}\left(\vec{n}_1, \cdots \vec{n}_A\right) M^{L_t / 2} \right| \Psi_i\right\rangle
\end{equation}
Thus we can evaluate one operator by:
\begin{equation} \label{eq:pinhole_O}
    \begin{aligned} 
        \langle {O} \rangle 
        = \frac{\sum_{i_1, j_1, \cdots i_A, j_A} \sum_{\vec{n}_1, \cdots \vec{n}_A}Z^{{O}}_{f, i}\left(i_1, j_1, \cdots i_A, j_A ; \vec{n}_1, \cdots \vec{n}_A \right) }{\sum_{i_1, j_1, \cdots i_A, j_A} \sum_{\vec{n}_1, \cdots \vec{n}_A}Z_{f, i}\left(i_1, j_1, \cdots i_A, j_A ; \vec{n}_1, \cdots \vec{n}_A \right)}
    \end{aligned}
\end{equation}
where $Z^{{O}}_{f, i}$ is the amplitude with operator insertion.
{The momentum pinhole method} is the extension of the original pinhole method. In this method, we perform the discrete Fourier transform before and after momentum pinhole insertion, with momentum pinhole defined as $\rho_{i_1, j_1, \cdots i_A, j_A}( \vec{k}_1, \cdots \vec{k}_A )=: \rho_{i_1, j_1}(\vec{k}_1) \cdots \rho_{i_A, j_A}(\vec{k}_A):$. Then we can measure one-body momentum distribution, two- and four-body momentum correlations using an equation like Eq. \ref{eq:pinhole_O}.

\begin{figure}[H]
    \centering 
    \includegraphics[width=0.6\linewidth]{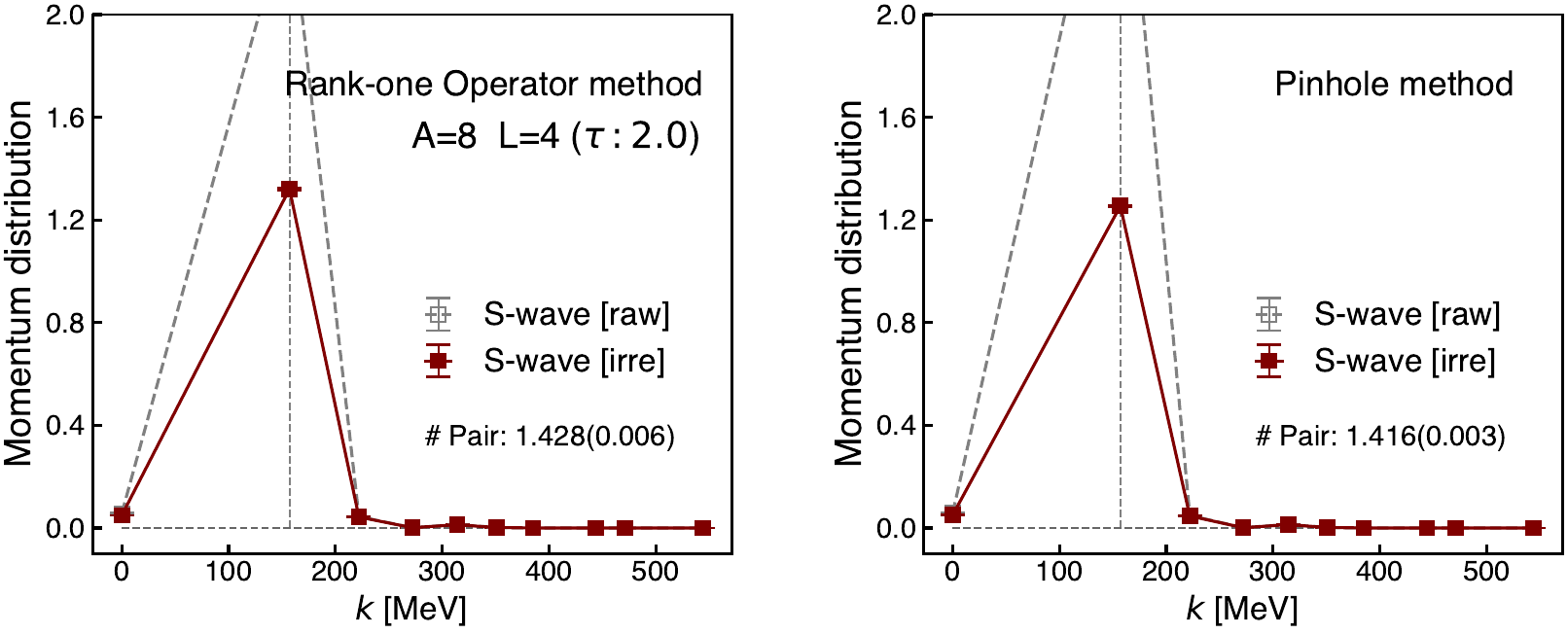}
    \caption{\textbf{S-wave momentum distribution: rank-one vs pinhole methods.} Momentum-space s-wave pair occupation for an $L^3 = 4^3$, $A = 8$ system at Euclidean time $\tau=2.0$ MeV$^{-1}$ using the self-consistent Cooper model dispersion relation. Results are obtained with two measurement methods. Left: rank-one operator method. Right: momentum pinhole method.\label{fig:Cooper_ROK_PHK_S}}
\end{figure}

{The benchmark of two methods\label{sec:ROK_PHK_Bench}} is performed with the self-consistent Cooper model dispersion relation (more details in Sec~\ref{sec:Cooper}) for $A=8$ system, which is shown in Fig~\ref{fig:Cooper_ROK_PHK_S}.   
We observe good consistency between the two measurement methods.
The rank-one operator (RO) method is more convenient for evaluating the exchange term $\rho_{\uparrow\downarrow}^{II-ex}(\vec{k}_1)$, which enters the s-wave and p-wave pairing signals.
In contrast, the pinhole (PH) method is more robust on measuring quartets, because the four-body density can be directly extracted from the generated pinhole configurations, which already contain $A$-body information.
In the following discussion, unless otherwise specified, we employ the RO method for s-wave and p-wave measurements and the PH method for quartet observables.
% \section{Lattice 4 (placeholder name)}

\section{Rotation \& Projection for Irreducible Two-Body Densities} \label{sec:rot_proj}
There are three p-wave pairing channels, $^3P_0$, $^3P_1$, and $^3P_2$.
For SU(2)-symmetric interactions, these three channels are degenerate and contribute identically.
In realistic nuclear systems, however, the different channels exhibit distinct behaviors; a typical example is the repulsive nature of the $^3P_1$ interaction.
To disentangle their individual contributions, one need employ rotation and projection techniques.
We define the rotation operator $R_{\vec{\theta}}$ with an angle $\vec{\theta}=(\theta_x, \theta_y, \theta_z)$. 
The projection operator is defined as,
\begin{equation}
    P^{(\mathrm{r})}=\frac{n_{\mathrm{r}}}{g} \sum_{R_{\vec{\theta}}} \chi^{(\mathrm{r})}(R_{\vec{\theta}}) R_{\vec{\theta}}
\end{equation}
where the sum runs over all $g$ group rotations $R_{\vec{\theta}}$, and $\chi^{(\mathrm{r})}(R)$ is the (real) character of $R_{\vec{\theta}}$ 's matrix representative in the irreducible representation $r$ of dimension $n_r$ \cite{Johnson:1982}.
The double cover of the octahedral group $O$ is $2O$ and it has 48 rotation elements. 
The conjugacy classes and irreducible characters are listed in Table~\ref{tab:character_table}.

\begin{table}[ht]
\centering
\begin{tabular}{c c c c c c c c c}
\toprule
$\chi$ & $I$ ($4\pi$) & $J$ ($2\pi$) & $6C_4$ ($\pi$) & $8C_3$ ($4\pi/3$) & $8C_6$ ($2\pi/3$) & $6C_8'$ ($3\pi/2$) & $6C_8$ ($\pi/2$) & $12C'_4$ ($\pi$) \\
\midrule
A$_1$ & 1 & 1 & 1 & 1 & 1 & 1 & 1 & 1 \\
A$_2$ & 1 & 1 & 1 & 1 & 1 & -1 & -1 & -1 \\
E & 2 & 2 & 2 & -1 & -1 & 0 & 0 & 0 \\
T$_1$ & 3 & 3 & -1 & 0 & 0 & 1 & 1 & -1 \\
T$_2$ & 3 & 3 & -1 & 0 & 0 & -1 & -1 & 1 \\
G$_1$ & 2 & -2 & 0 & -1 & 1 & $-\sqrt{2}$ & $\sqrt{2}$ & 0 \\
G$_2$ & 2 & -2 & 0 & -1 & 1 & $\sqrt{2}$ & $-\sqrt{2}$ & 0 \\
H & 4 & -4 & 0 & 1 & -1 & 0 & 0 & 0 \\
\bottomrule
\end{tabular}
\caption{\textbf{Character table of the double octahedral group $2O$.} Characters of the irreducible representations for the double cover of the octahedral group (48 elements). The conjugacy classes are labeled following the convention of Ref~\cite{Johnson:1982}. Here $C_4$ denotes rotations about coordinate axes and $C_3$ about body diagonals; the angles $\theta$ are the corresponding SU(2) rotation angles used to evaluate characters.\label{tab:character_table}}
\end{table}

When rotating the irreducible two-body density $\rho^{II}$, the rotation operator $R_{\vec{\theta}}$ only acts on creation operators:
\begin{equation}
    R_{\vec{\theta}} \rho^{II}_{j_1j_2,j_1j_2}(\vec{k}'_{12}, \vec{k}_{12})
    =  \rho^{II}_{j_1(\vec{\theta})j_2(\vec{\theta}),j_1j_2}(\vec{k}'_{12}(\vec{\theta}), \vec{k}_{12})
\end{equation}
where $\vec{k}'_{12}(\vec{\theta}) = R^{-1}_{\vec{\theta}}\vec{k}'_{12}$, $\vec{k}'_{12} = \vec{k}'_2 - \vec{k}'_1 $ is the relative momentum of $a_{s'_1}^{\dagger}(\vec{k}'_{1})$ and $a_{s'_2}^{\dagger}(\vec{k}'_{2})$, and $j_1(\vec{\theta})=R^{-1}_{\vec{\theta}}j_1$.
The octahedral group on a lattice has different representations compared with the SU(2) group.
The reduction from continuous SU(2) to a discretized $2O$ is listed in table \ref{tab:SU2_O2}.
\begin{table}[h!]
    \centering
    \caption{\textbf{Reduction of SU(2) angular momentum to $2O$.}
Decomposition of SU(2) irreducible representations with total angular momentum $j \le 3$ 
into irreducible representations of the double octahedral group $2O$. \label{tab:SU2_O2}}
\begin{tabular}{c|l}
\hline \multicolumn{2}{l}{$j$} \\
\hline 0 & $\mathrm{A}_1$ \\
\hline 1/2 & $\mathrm{G}_1$ \\
\hline 1 & $\mathrm{T}_1$ \\
\hline 3/2 & H \\
\hline 2 & $\mathrm{E}+\mathrm{T}_2$ \\
\hline 5/2 & $\mathrm{G}_2+\mathrm{H}$ \\
\hline 3 & $\mathrm{A}_2+\mathrm{T}_1+\mathrm{T}_2$ \\
\hline
\end{tabular}
\end{table}

% \section{Lattice 5 (placeholder name)}

\section{Lattice Results A: Lattice Cooper Model} \label{sec:Cooper}
The Cooper model was first introduced by Leon Cooper in 1956~\cite{Cooper:1956}, demonstrating that an arbitrarily weak attractive interaction can induce a two-body bound state with energy below the Fermi surface. This bound state, now known as a Cooper pair, provides the microscopic foundation of the Bardeen–Cooper–Schrieffer (BCS) theory of superconductivity~\cite{Bardeen1957}.

In the original Cooper Model, two fermions with opposite spins are placed above a ``Fermi sea'' and interact attractively within a narrow energy window near the Fermi surface. This can be achieved by multiplying the interaction by a theta function $V_{\text{eff}}(\epsilon) = -V \cdot[\theta(\epsilon - \epsilon_F) - \theta(\epsilon - (\epsilon_F + \hslash \omega_D))]$, where $\epsilon_F$ is the Fermi energy and $\hslash \omega_D$ denotes the Debye frequency, which sets the ultraviolet cutoff of the interaction.
In the Lattice framework, the momentum $\vec{k} = (\vec{k}_x, \vec{k}_y, \vec{k}_z)$ is discretized with the first Brillouin zone $-\frac{\pi}{a}<k_{i}\leq\frac{\pi}{a}$, where $a$ is the lattice spacing. The Debye frequency is naturally given by the lattice cutoff. For the Fermi surface $k_F$, we let $|\vec{k}| \geq k_F$, which means the two particles are located on the Fermi surface. In the thermodynamic limit, the difference with $|\vec{k}| > k_F$ is negligible. The two-body Schr\"odinger equation in momentum space then reads
\begin{equation} \label{eq:TwoBody_Schrodinger}
    ( 2E_{\vec{k}} - \sum_{\vec{k} \geq k_{F}} V_{\vec{k}',\vec{k}})  \phi(\vec{k})  =   E\phi(\vec{k})
\end{equation}
where $E_{\vec{k}}$ is the single-particle kinetic energy.
For a simplified situation of $V_{\vec{k}',\vec{k}}$, we let $s_{\text{L}}=0$ and $s_{\text{NL}}=0$ in Eq.~\ref{eq:H_SU2}, leading to the constant of $V_{\vec{k}',\vec{k}} = \frac{c}{L^3}$. The resulting Cooper equation takes the form:
\begin{equation} \label{eq:CooperSolution}
    \frac{1}{\lambda} =  \sum_{\vec{k} \geq k_{F}} \frac{1}{2E_k - E}
\end{equation}
with $\lambda = - \frac{c}{L^3}$.

\subsection{Self-consistent Cooper model}
{The self-consistent Cooper model} is considered in this work to calculate few-body two-spin fermion systems to display the multi-modal superfluid in a straightforward setting. We introduce the pairing gap $\Delta$ into the original Cooper dispersion relation as: $\tilde{E}_{\vec{k}} = \sqrt{(E_{\vec{k}} -E_{F} )^2 + \Delta^2} + E_{F}$
which takes the pairing gap definition from BCS theory $\Delta = - \sum_{|\vec{k}| \geq k_F}  {V}_{\vec{k},\vec{k}} \langle a_{-\vec{k}} a_{\vec{k}} \rangle$.
Analogous to Eq~\ref{eq:CooperSolution}, we have 
\begin{equation} 
    \frac{1}{\lambda} =  \sum_{\vec{k}} \frac{1}{2\tilde{E}_k - \tilde{E}}
    =  \sum_{\vec{k}} \frac{1}{2 \sqrt{(E_{\vec{k}} -E_{F} )^2 + \Delta^2} + 2E_F- \tilde{E}}
    = \sum_{\vec{k}} \frac{1}{2 \tilde{e}_{\vec{k}} +  \tilde{e}_{\text{pair}}}
\end{equation}
To recover the BCS many-body ground state, we impose $\tilde{e}_{\text{pair}}=0$, reflecting the fact that adding a pair of particles does not change the ground-state energy. It should be mentioned that $\tilde{e}_{\text{pair}}$ is not the energy to break a pair or the excitation energy, but instead equals twice the chemical potential $\tilde{e}_{\text{pair}} = \tilde{E}(N+2) - \tilde{E}(N)$. Under this condition, the self-consistent Cooper model reproduces the standard BCS gap equation,
\begin{equation} \label{eq:Cooper_Extended}
    \frac{1}{\lambda} 
    =  \sum_{\vec{k}} \frac{1}{2 \sqrt{(E_{\vec{k}} -E_{F} )^2 + \Delta^2}}.
\end{equation}
In the BCS theory, the first excitation energy or the minimum energy to break a pair is $2 \Delta$, which can also be treated as the pair binding energy relative to the continuum. Thus, in this self-consistent Cooper model, it can also be calculated by $\Delta = (E_{\text{free}} - \tilde{E})/2$, where $E_{\text{free}}$ is the eigenvalue with the interaction turned off. The comparison of these two ways of calculating $\Delta$ is listed in Table~\ref{tab:Cooper_Delta}, for systems with the Fermi surface located around 100 MeV.
When taking the self-consistent dispersion relation for $\tilde{E}_{\vec{k}}$ as well as $\Delta = - \sum_{|\vec{k}| \geq k_F}  {V}_{\vec{k},\vec{k}} \langle a_{-\vec{k}} a_{\vec{k}} \rangle$ and $\tilde{e}_{\text{pair}}=0$ into the original two-body Schr\"odinger equation Eq~\ref{eq:TwoBody_Schrodinger}, we can obtain $\phi(\vec{k}) = {\Delta}/{2 \tilde{e}_{\vec{k}}}$. Comparing the solution with the BCS theory $u_k v_k = {\Delta}/{2 E_{\vec{k}}}$, we can see that the wavefunction of the self-consistent Cooper model can be identified as the pairing amplitude of the BCS theory.
% \begin{table}[htbp]
% \centering
% \caption{\textbf{Comparison of analytical and lattice s-wave pairing gaps.}
% S-wave pairing gaps $\Delta_S$ for lattice sizes $L=4$–$7$ with lattice spacing 
% $a = 1.97\,\mathrm{fm}$, coupling $c = -3.0\times10^{-5}\,\mathrm{MeV}^{-2}$, 
% $s_{\mathrm{L}} = 0.0$, $s_{\mathrm{NL}} = 0.0$, and $i_{\mathrm{kin}} = 1$.
% $\Delta_{\mathrm{analy}}$ denotes the analytical result from Eq.~\ref{eq:Cooper_Extended}, 
% while $\Delta = (E_{\mathrm{free}} - \tilde{E})/2$ is extracted from the eigenvalue 
% of the two-body Schr\"odinger equation in the self-consistent Cooper model.
% }
% \label{tab:Cooper_Delta}
% \begin{tabular}{lcccc}
% \toprule
% % \midrule
% Lattice size & $L=4$ & $L=5$ & $L=6$ & $L=7$ \\
% \midrule
% $k_F$(MeV) & 157.0796 & 125.6637 & 104.7198 & 89.75979 \\
% \midrule
% $\Delta_{\text{analy}}$(MeV) & 4.479666 & 2.562412 & 1.527938 & 0.966821 \\
% % $E_{\text{free}} - {E}$ & 8.490618 & 4.880726 & 2.928111 & 1.863105 \\
% $\Delta$ (MeV) & 4.479666 & 2.562412 & 1.527938 & 0.966822 \\
% \bottomrule
% \end{tabular}
% \end{table}

\begin{table}[htbp]
\centering
\caption{\textbf{Comparison of analytical and lattice s-wave pairing gaps.}
S-wave pairing gaps $\Delta_S$ for lattice sizes $L=6$-$10$ with lattice spacing 
$a = 1.97\,\mathrm{fm}$, coupling $c = -3.0\times10^{-5}\,\mathrm{MeV}^{-2}$, 
$s_{\mathrm{L}} = 0.0$, $s_{\mathrm{NL}} = 0.0$, and $i_{\mathrm{kin}} = 1$.
$\Delta_{\mathrm{analy}}$ denotes the analytical result from Eq.~\ref{eq:Cooper_Extended}, 
while $\Delta = (E_{\mathrm{free}} - \tilde{E})/2$ is extracted from the eigenvalue 
of the two-body Schr\"odinger equation in the self-consistent Cooper model.
}
\label{tab:Cooper_Delta}
\begin{tabular}{lccccc}
\toprule
% \midrule
Lattice size & $L=6$ & $L=7$ & $L=8$ & $L=9$ & $L=10$ \\
\midrule
$k_F$(MeV) & 104.72 & 126.94 & 111.07 & 120.92 & 108.83 \\
\midrule
$\Delta_{\text{analy}}$(MeV) & 1.5279 & 2.7090 & 1.7583 & 2.1308 & 1.3820 \\
$\Delta$ (MeV)               & 1.5279 & 2.7090 & 1.7583 & 2.1308 & 1.3820 \\
\bottomrule
\end{tabular}
\end{table}

While the fully spin-polarized system spin-1/2  system is not the main focus of this work, it is a subject of significant interest \cite{Schwenk2004,Stein:2015bpa,Tews:2019qhd}.
For the p-wave pairing gap in the spin-polarized system, we estimate its magnitude by solving the polarized two-body Cooper problem using the axial ansatz, corresponding to the Anderson–Brinkman–Morel (ABM) state \cite{Anderson:1973}. 
Considering the multipole expansion of pairing gap with $l=1$, $\Delta(\vec{k}) = \sum_m \Delta_m Y_{1m}(\hat{k})$, the axial state $m=\pm 1$ has more binding than the polar state  $m=0$. 
For the axial state, we adopt the extended dispersion relation as
$\tilde{E}^P_{\vec{k}} = \sqrt{(E_{\vec{k}} -E_{F} )^2 + (\Delta^{ax}_P)^2 \sin^2 \theta} + E_{F}$, where $\sin^2 \theta = \frac{k^2_x+k^2_y}{k^2_x+k^2_y + k_z^2}$. 
The p-wave eigenvalue $\tilde{E}_P$ is obtained by solving the two-body Schr\"odinger equation with odd-parity wavefunctions. 
The self-consistency condition is then given by $\Delta^{ax}_P = E_{\text{free}} - \tilde{E}_P$.
To generate a nonvanishing p-wave interaction on the lattice, the local smearing parameter $s_{\text{L}}$ must be turned on. 
Using the same SU(2)-symmetric interaction as in the many-body calculations, we compute the axial p-wave pairing gap for the polarized Cooper problem. 
In Table~\ref{tab:Cooper_Delta_Axial_Ansatz}, we first present results for three systems with $L=6,8,10$, chosen to have the same Fermi momenta as the polarized many-body calculations shown in Table~\ref{tab:SU2_Polarized_E}.
For comparison with polarized BCS calculations, we additionally consider three systems with Fermi momenta above 220 MeV.
Since the polarized Cooper problem exhibits more pronounced finite-volume effects than the spin-symmetric case, these calculations are performed in larger lattice volumes, as summarized in Table~\ref{tab:Cooper_Delta_Axial_Ansatz}.
By comparing Table~\ref{tab:Cooper_Delta_Axial_Ansatz}, Table~\ref{tab:SU2_Polarized_E}, and Fig.~\ref{fig:kf_gap_MeV}, we observe that the polarized self-consistent Cooper model yields results consistent with both the BCS axial solutions and the polarized lattice many-body simulations.
Moreover, it can be seen that polarized p-wave gap is numerically very small and therefore even very small finite volume errors are quite significant, in contrast with the spin-balanced system.

\begin{table}[htbp]
\centering
\caption{\textbf{Axial (ABM) p-wave pairing gaps from the polarized Cooper problem.}
The axial (Anderson–Brinkman–Morel) ansatz results for the p-wave pairing gap  $\Delta^{\mathrm{ax}}_{P}$ for lattice sizes $L=6, 8, 10$ and $15, 16, 17$.  
Calculations are performed with lattice spacing $a = 1.97$ fm and coupling parameters 
$c = -1.6 \times 10^{-6}\,\mathrm{MeV}^{-2}$, 
$s_{\text{L}} = 0.5$, and $s_{\text{NL}} = 0.1$.
$\Delta^{\mathrm{ax}}_P = (E_{\mathrm{free}} - \tilde{E})/2$ is extracted from the eigenvalue 
of the polarized two-body Schr\"odinger equation in the self-consistent Cooper model.
}
\label{tab:Cooper_Delta_Axial_Ansatz}
\begin{tabular}{lccccccc}
\toprule
% \midrule
Lattice size & $L=6$ & $L=8$ & $L=10$ & & $L=15$ & $L=16$ & $L=17$  \\
\midrule
$k_F$(MeV) & 104.72 & 111.07 & 125.66 & & 225.57 & 222.14 & 221.76  \\
\midrule
$\Delta^{\mathrm{ax}}_P$ (MeV) & 0.2077 & 0.2094 & 0.0550 & & 0.0358 & 0.0412 & 0.0471 \\
\bottomrule
\end{tabular}
\end{table}

\subsection{Self-consistent dispersion relation in few-body systems}
% {Self-consistent dispersion relation for few-body systems} is considered to show the multimodal pairing in a clear way. 
For the few-body calculations, we first calculate the s-wave pairing order parameter $\Delta$ from the self-consistent Cooper Model, as shown in the third row of Table~\ref{tab:SU2_Delta}, in which we choose the second lattice momentum as the Fermi level. 
Then we insert $\Delta$ into the self-consistent dispersion relation $\tilde{E}_{\vec{k}}$. 
Finally, we solve the many-body Schr\"odinger equation and do the measurements within the NLEFT framework.   
For these calculations, we use the lattice setup $a=1.97$ fm, $c=-1.6\times 10^{-6}$ MeV$^{-2}$, $s_{\text{L}}=0.5$ and $s_{\text{NL}}=0.1$, which is the same as that used for the 3D SU(2) calculations in Sec.~\ref{Sec:3D_SU2}.
In Table~\ref{tab:SU2_Delta} we also list the lattice s-wave and p-wave pairing gap by comparing two-body $E_{\uparrow\downarrow}^{\text{latt}}$ and $E_{\uparrow\uparrow}^{\text{latt}}$ with $E_{\text{free}}$. 
The quartet gaps are obtained by performing four-body lattice calculations.
A comparison of the $L=6$ results in Tables~\ref{tab:SU2_Delta} and \ref{tab:SU2_L6_Energy} shows good agreement between the self-consistent Cooper model and the corresponding many-body calculations. For the s-wave channel, the extracted gaps are $1.61(3)$ vs $1.33(4)$ MeV; for the p-wave channel, $0.32(6)$ vs $0.36(4)$ MeV; and for the quartet contribution, $0.32(1)$ vs $0.37(4)$ MeV.

\begin{table}[htbp]
\centering
\caption{\textbf{Comparison of Cooper model and lattice energy gaps.}
S-wave pairing gaps $\Delta$ obtained from the self-consistent Cooper model using the diagonalization method, compared with those extracted from Monte Carlo lattice calculations of the same two-body Cooper problem. The lattice gaps are defined as $(E_{\mathrm{free}} - E^{\mathrm{latt}}_{\uparrow\downarrow})/2$ for s-wave pairing and $(E_{\mathrm{free}} - E^{\mathrm{latt}}_{\uparrow\uparrow})/2$ for p-wave pairing. The quartet (four-body) gap $(2E_2 - E^{\mathrm{latt}}_4)/4$ is also listed, where $E_2$ denote the spin-symmetrized two-body energy from the self-consistent Cooper model and $E^{\mathrm{latt}}_4$ is the four-body energy from Monte Carlo lattice calculations with self-consistent dispersion relation. Results are shown for lattice sizes $L = 4$–$8$ with coupling parameters $c = -1.6\times10^{-6}\,\mathrm{MeV}^{-2}$, $s_{\text{L}} = 0.5$, and $s_{\text{NL}} = 0.1$.
}
\label{tab:SU2_Delta}
\begin{tabular}{lccccc}
\toprule
% \midrule
Lattice size & L=4 & L=5 & L=6 & L=7 & L=8\\
\midrule
$k_F$(MeV) & 157.08 & 125.66 & 104.72 & 89.76 & 78.54\\
\midrule
$\Delta$ (MeV) & 2.1372 & 1.8763 & 1.6147 & 1.3760 & 1.1713 \\
\midrule
$(E_{\text{free}} - E_{\uparrow\downarrow}^{\text{latt}})/2$ &  2.10 (1)  & 1.87 (4)  & 1.61 (3)  & 1.36 (1) & 1.19 (4) \\
$(E_{\text{free}} - E_{\uparrow\uparrow}^{\text{latt}})/2$& 0.56 (2)  & 0.39 (6)  & 0.32 (6)  & 0.27 (18) & 0.02 (25) \\
\midrule
$(2 E_{2} - E_{4}^{\text{latt}})/4$  & 1.23 (1) & 0.63 (1) & 0.32 (1) & 0.17 (2) & 0.10 (1)\\
\bottomrule
\end{tabular}
\end{table}

\begin{figure}[H]
    \centering 
    \includegraphics[width=0.9\linewidth]{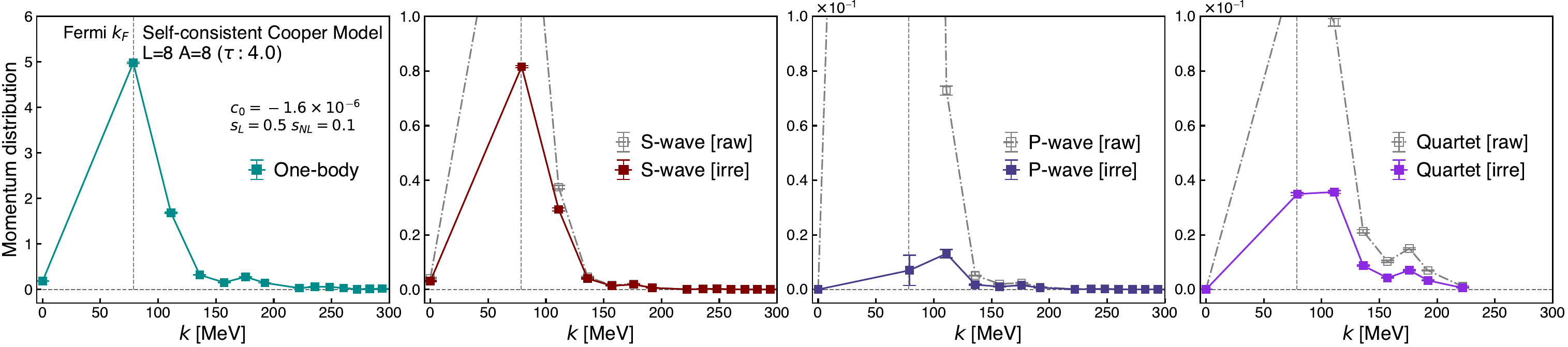}
    \caption{\textbf{Momentum-space pairing and quartet distributions for $A=8$.} \textit{Ab initio} lattice results for an $L^3 = 8^3$, $A = 8$ system using the self-consistent Cooper model dispersion relation. From left to right, the panels show the momentum distribution of s-wave pairing, p-wave pairing and quartets.
    \label{fig:A8_SPQ}}
\end{figure}

Then, with the self-consistent dispersion relation, we perform the $A=8$ spin-balanced neutron system in a $L^3=8^3$ lattice box and implement the pinhole method to measure the one-body momentum distribution, s-wave and p-wave pairing and quartet correlations.
The results are shown in Fig~\ref{fig:A8_SPQ}. 
Different from the original Cooper model (totally blocked for $|\vec{k}|<|\vec{k}_F|$), the new dispersion relation $\tilde{E}_{\vec{k}} = \sqrt{(E_{\vec{k}} -E_{F} )^2 + \Delta^2} + E_{F}$ provides a higher kinetic energy to make $|\vec{k}|<|\vec{k}_F|$ disfavored, which can be seen in the one-body momentum distribution subfigure. 

\begin{figure}[H]
    \centering    
    \includegraphics[width=0.9\linewidth]{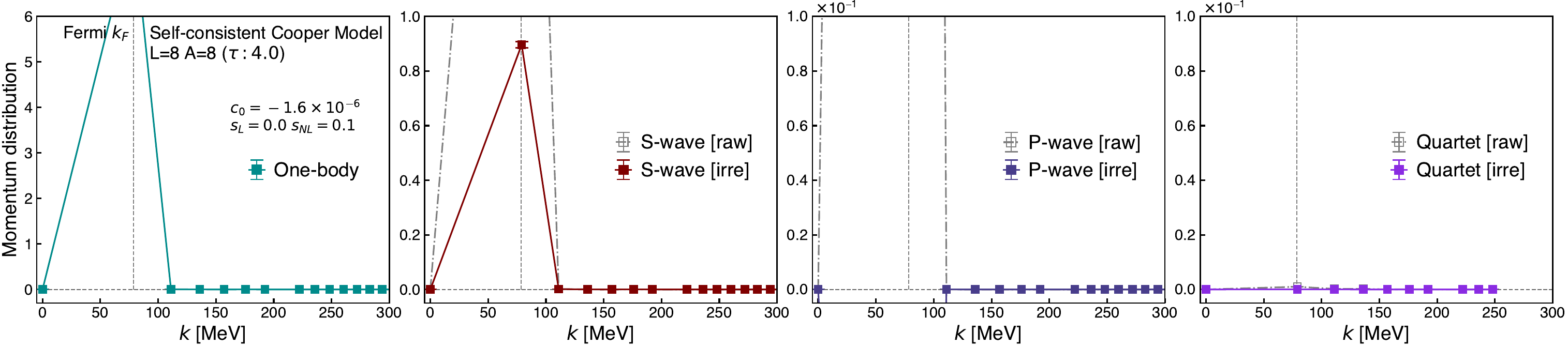}
    \caption{\textbf{Momentum-space pairing and quartet distributions with zero local smearing.} Same as Fig.~\ref{fig:A8_SPQ}, but with the local smearing parameter set to $s_{\text{L}} = 0$. The absence of finite-range interaction suppresses the p-wave and quartet signals, while the s-wave pairing is enhanced. \label{fig:A8_SPQ_sL0}} 
\end{figure}

In the s-wave, p-wave and quartet subfigure of Fig~\ref{fig:A8_SPQ}, ``raw'' stands for the two-body or four body correlations, while ``irre'' represents the irreducible two-body or four-body densities (two-body or four-body cumulants) defined in section~\ref{sec:latt_measure}.
It can be seen that, besides a strong s-wave pair signal at ${k}_F$, there also exist some momentum spread above and below Fermi level. Even though the scale is smaller, we can still measure p-wave and quartet signals in this eight-body system. Due to Pauli blocking p-wave pairing has exact zero value at $k=0$, same as quartet signals.
One essential element for p-wave pairing and quartetting is the interaction should be finite range. In the Hamiltonian of Eq.~\ref{eq:H_SU2}, it was controlled by the local smearing parameter $s_{\text{L}}$, where $s_{\text{L}}=0.0$ stands for a zero-range interaction.
To verify this argument, we perform the same calculation but with local smearing $s_{\text{L}}=0.0$ and show the results in Fig~\ref{fig:A8_SPQ_sL0}. With this interaction, we can see that s-wave pair increase a lot and, at the same time, p-wave and quartet signal drop to zero.  
It should be noticed that the negative irreducible p-wave signal is actually a consequence of strong s-wave pairing. Considering spin-${\uparrow\uparrow}$ component of p-wave pairing (the other two are same with SU(2) symmetry) $\rho^{II}_{\vec{k}\uparrow, -\vec{k}\uparrow} = \rho_{\vec{k}\uparrow, -\vec{k}\uparrow} - \rho_{\vec{k}\uparrow}\rho_{-\vec{k}\uparrow}$: if the ground state is only the mixture four s-wave pairs $\rho^{II}_{\vec{k}_i\uparrow, -\vec{k}_i\downarrow}=1$ (with $i,j=1,2,3,4$ and $\vec{k}_i \neq \vec{k}_j$), then we will have $\rho_{\vec{k}_i\uparrow} = \rho_{-\vec{k}_i\uparrow} = 1$ but $\rho_{\vec{k}_i\uparrow, -\vec{k}_i\uparrow} = 0$, leading to a large negative value of p-wave signal $\rho^{II}_{\vec{k}\uparrow, -\vec{k}\uparrow}$.
This negative amplitude can also be explained as ``Canonical Suppression''. Consider a system without any interaction, the probability of finding a particle at $\vec{k}$ is $\rho_{\vec{k}} = \rho_{-\vec{k}}= \frac{A}{M}$ and the joint probability of finding a pair ($\vec{k}$, $-\vec{k}$) is $\rho_{\vec{k}, -\vec{k}} = \frac{A}{M}\frac{A-1}{M-1}$, with particle number $A$ and number of states $M$. 
Thus, the two-body cumulant $\rho^{II}_{\vec{k}, -\vec{k}} = \frac{A}{M}\frac{A-1}{M-1} - (\frac{A}{M})^2 \approx -\frac{A}{M^2} + \frac{A(A-1)}{M^3} = -\frac{A}{M^2} + \mathcal{O}(\frac{1}{M^3}) $, in which we consider $\frac{1}{M-1} \approx \frac{1}{M} (1+\frac{1}{M})$.
Since number of states $M \propto L^3$, if we keep the density fixed, the negative value of the cumulant will decrease as $1/L^3$ and vanish at the thermodynamic limit. 

\begin{figure}[H]
    \centering
    \includegraphics[width=0.9\linewidth]{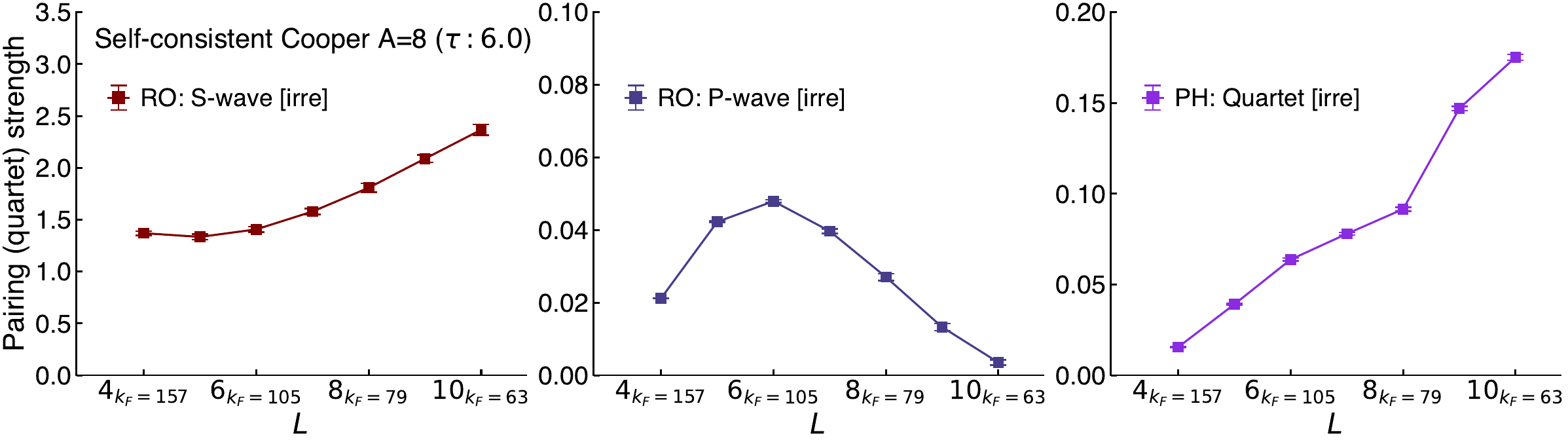}
    \caption{\textbf{Lattice size dependence of pairing and quartet strengths.} Irreducible pairing (quartet) strengths for the $A=8$ system with the self-consistent dispersion relation as a function of the box size $L$, obtained from \textit{ab initio} lattice calculations. From left to right, the panels show the s-wave pairing, p-wave pairing and quartet cumulants.
    \label{fig:Cooper_L}}
\end{figure}

With the discussion above, we perform the $A=8$ self-consistent Cooper Model for different box sizes (specifically, $L^3=4^3,5^3,6^3,7^3,8^3,9^3,10^3$). 
The results are shown in Fig~\ref{fig:Cooper_L}. 
It should be mentioned that, as discussed in Sec~\ref{sec:ROK_PHK_Bench}, the momentum ``pinhole'' method (PHK)  does not calculate the off-diagonal momentum densities and the momentum ``rank-one operator'' method (ROK) is more useful for measuring s-wave and p-wave signals.
Thus, in the remainder of the sections we use ROK to measure the s-wave and p-wave signals and use PHK for measuring quartets. 
To exclude the effects of ``Canonical Suppression'', we drop the negative contribution of p-wave signals. 
The energy levels in a small lattice box are discretized with a large gap and will become more dense in larger boxes.
% The increase of momentum modes will lead to an increase of s-wave and p-wave pairs as long as quartets. 
Increasing the number of momentum modes enhances the formation of both s-wave and p-wave pairs, as well as quartets.
Especially, quartets are hard to form in smaller boxes because squeezed wavefunctions are more favored for double s-wave pairs.
The decreasing of p-wave signal starting from $L^3=6^3$ could come from two aspects: 1) the smaller p-wave interaction as the p-wave gap shown in Tab~\ref{tab:SU2_Delta}; 2) it cannot compete with quartets, i.e., p-wave contribution can be ``eaten'' by quartets.  

% \section{Lattice 6 (placeholder name)}
\section{Lattice Results B: Unitary Limit}
To reach the unitary limit, we turn off the local smearing $s_{\text{L}}$, leaving only the s-wave interactions. 
Following the same low energy coupling constants as in Ref\cite{He:2019ipt}, we calculate the $^1S_0$ phase shift on the lattice.
The lattice phase shifts at low momenta are in excellent agreement with the unitary limit, which corresponds to $90^{\circ}$.
We also calculate the Bertsch parameter, which is the ratio between the ground state energy and the free Fermi gas, $\xi = E_0/E_{\text{FG}}$.
The estimated number $0.3776\pm 0.0004$ is obtained from the last three Euclidean time data.
It is not only consistent with the values $0.372(2)$ from Ref\cite{He:2019ipt}, but also with the results $\xi=0.374(5), 0.372(3), 0.375(5)$ obtained in Ref\cite{Carlson:2011}, using three different lattice actions, as well as the experimental value of 0.377(6) obtained using ultracold $^6$Li atoms \cite{Mukherjee:2017ygo}.
\begin{figure}[H]
  \centering
  \begin{subfigure}{0.35\textwidth}
    \centering
    \includegraphics[width=\linewidth]{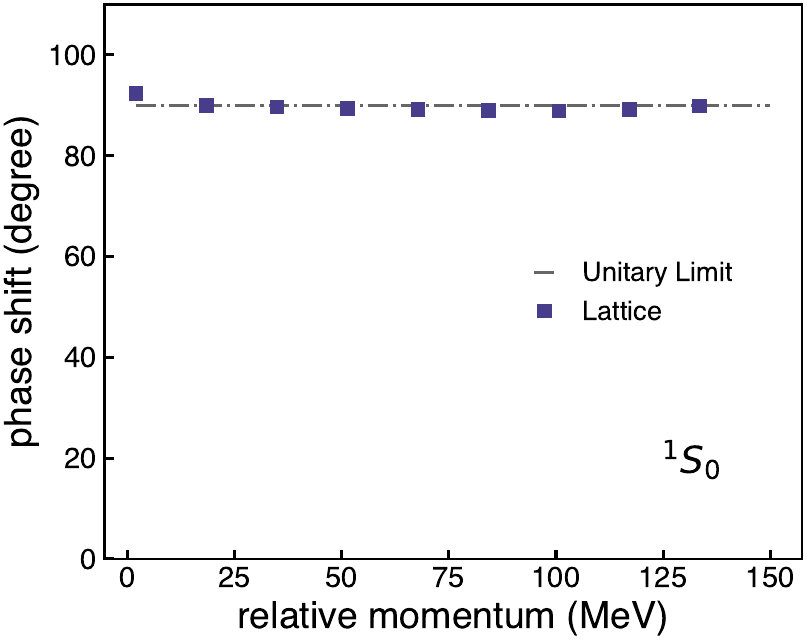}
    %\label{fig:sub1}
  \end{subfigure}\hspace{0.05\textwidth}
  \begin{subfigure}{0.36\textwidth}
    \centering
    \includegraphics[width=\linewidth]{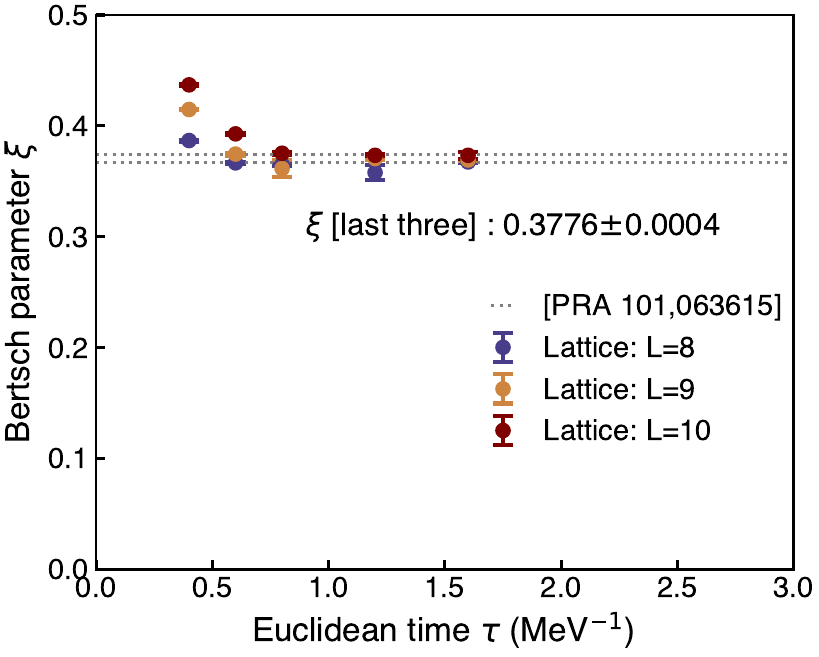}
    %\label{fig:sub2}
  \end{subfigure}
  \caption{\textbf{Unitary-limit benchmark results I.} Phase shift (left) and the Bertsch parameter $\xi$  (right) at the unitary limit from \textit{ab initio} lattice calculations. \label{fig:Unitary_P_B}}
\end{figure}

\begin{figure}[H]
  \centering
  \begin{subfigure}{0.35\textwidth}
    \centering
    \includegraphics[width=\linewidth]{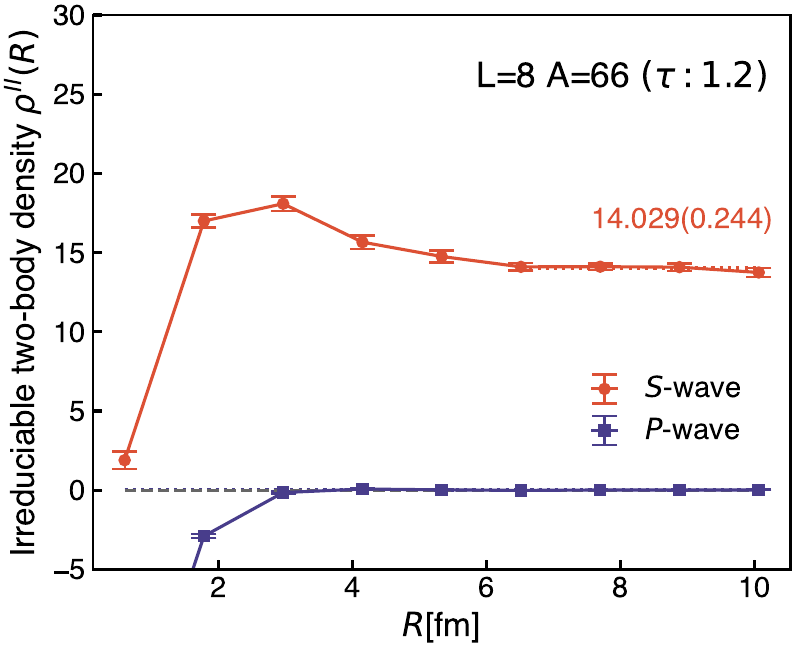}
  \end{subfigure}\hspace{0.05\textwidth}
  \begin{subfigure}{0.36\textwidth}
    \centering
    \includegraphics[width=\linewidth]{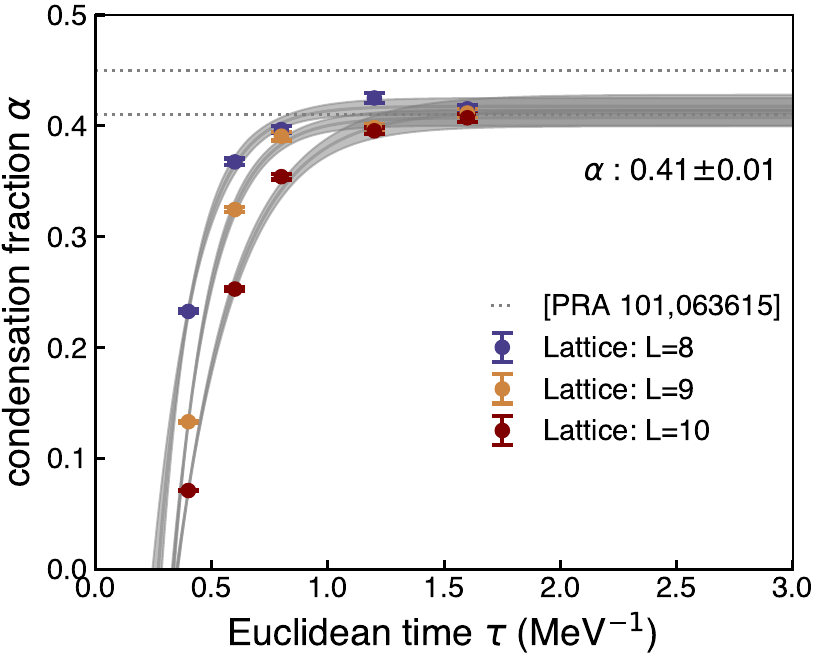}
  \end{subfigure}
  \caption{\textbf{Unitary-limit benchmark results II.} Off-diagonal long-range order as a function of separation $R$ at Euclidean time $\tau=1.2$ MeV$^{-1}$ (left), and the condensate fraction $\alpha$ (right), both obtained from \textit{ab initio} lattice calculations.\label{fig:Unitary_ODLRO}}
\end{figure}

Then, we measure the Off-Diagonal Long-Range Order (ODLRO) at the unitary limit in Fig. \ref{fig:Unitary_ODLRO}. 
Recall that the ODLRO value can be extracted from the two-body irreducible density $ F =  \lim_{\vec{R} \to \infty} \rho^{II} (\vec{R})$, and the pairing condensate fraction can be calculated by $\alpha = F/(A/2)$ where $A$ is the number of particles. 
In the left panel of Fig. \ref{fig:Unitary_ODLRO}, we can see that there is no p-wave irreducible two-body density at the unitary limit.
At large distances, the s-wave irreducible two-body density reaches a plateau of $14.029(0.244)$, which is the value of ODLRO.
In the right panel of Fig. \ref{fig:Unitary_ODLRO}, we show a more sophisticated analysis with the data from different lattice box sizes and Euclidean time.
The estimated pairing condensate fraction of $0.41\pm0.01$ is consistent with previous analyses \cite{He:2019ipt} at the unitary limit.  It is also consistent with experiments using ultracold $^6$Li atoms, which have measured the condensate fraction to be 0.46(7) \cite{Zwierlein:2004zz,Zwierlein:2005} and 0.47(7) \cite{Kwon:2020}.

% \section{Lattice 7 (placeholder name)}
\section{Lattice Results C: Generalized Attractive Extended (GAE) Hubbard Model}
\subsection{1D GAE Hubbard Model}
In 1+1 dimensional systems, Coleman’s theorem \cite{Coleman:1973} forbids spontaneous breaking of continuous global symmetries.
Considering the pairing operator $O_{\text{pair}}=\psi\psi$, under a U(1) transformation $e^{i\alpha}$ it transforms as $\psi\psi \to e^{2i\alpha}\psi\psi$, carrying a global phase with charge 2.
Hence a non-zero value of $\langle O_{\text{pair}}\rangle$ signals the spontaneous symmetry breaking.
If a true long-range order were present, the correlator would approach a nonzero constant, $\lim _{|x| \rightarrow \infty}\langle O_{\text{pair}}^{\dagger}(x) O_{\text{pair}}(0)\rangle=C$. 
Assuming cluster decomposition (in the absence of long-range interactions), this implies $\lim _{|x| \rightarrow \infty}\langle O_{\text{pair}}^{\dagger}(x) O_{\text{pair}}(0)\rangle=|\langle O_{\text{pair}}\rangle|^2$, hence $\langle O_{\text{pair}}\rangle \neq 0$ , which would signal spontaneous symmetry breaking.
Coleman’s theorem therefore excludes such a possibility and one must have $\lim _{|x| \rightarrow \infty}\langle O_{\text{pair}}^{\dagger}(x) O_{\text{pair}}(0)\rangle= 0 $.
Consequently, one-dimensional quantum systems can not have a true long-range order.
Instead, the low-energy physics of one-dimensional quantum systems is universally described by Luttinger liquid theory \cite{Haldane:1981}, characterized by algebraic correlations $\sim\ln{x}$.

\begin{figure}[H]
  \centering
  \begin{subfigure}{0.35\textwidth}
    \centering
    \includegraphics[width=\linewidth]{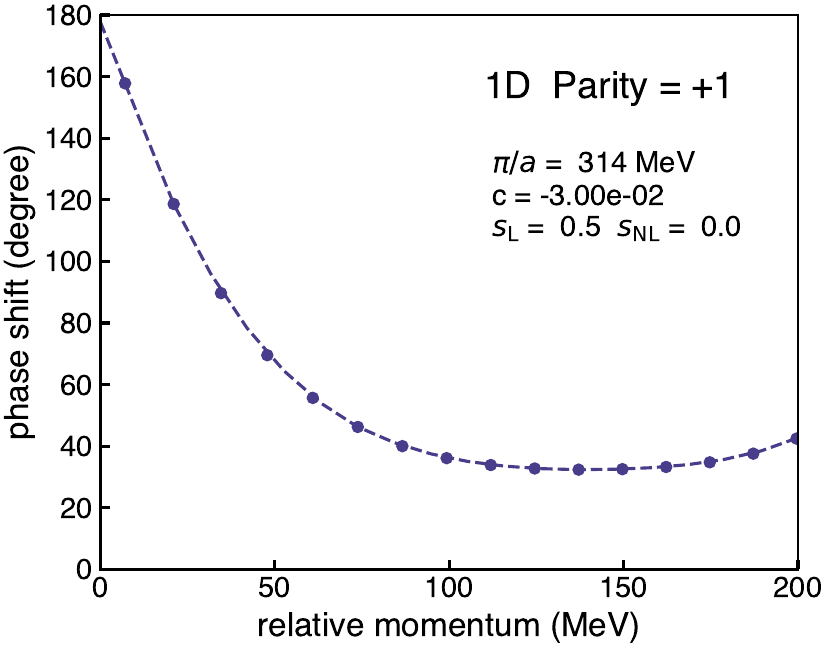}
    %\label{fig:sub1}
  \end{subfigure}\hspace{0.05\textwidth}
  \begin{subfigure}{0.35\textwidth}
    \centering
    \includegraphics[width=\linewidth]{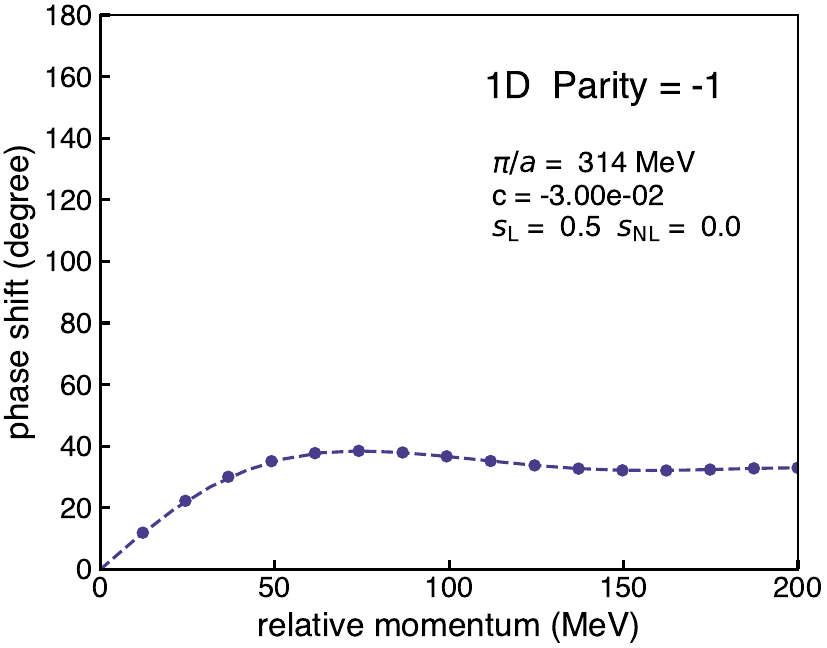}
    %\label{fig:sub2}
  \end{subfigure}
  \caption{\textbf{1D SU(2) scattering phase shifts.} Two-body phase shifts for the SU(2)-symmetric interaction in one dimension. Left: even-parity channel. Right: odd-parity channel. \label{fig:SU2_1D_PhaseShift}}
\end{figure}

We first consider an SU(2) interaction (
% $c=-3.0\times 10^{-6}$ MeV$^{-2}$
$c=-3.0\times 10^{-2}$
, $s_{\text{L}}=0.5$, $s_{\text{NL}}=0.0$ and $i_{\text{kin}}=1$) in a one-dimensional system, which corresponds to the 1D GAE Hubbard model with nearest-neighbor interactions. 
Figure~\ref{fig:SU2_1D_PhaseShift} shows the phase shifts of two-body scattering in the 1D system. With the nearest-neighbor smearing parameter $s_{\text{L}}$ turned on, we can have odd-parity or p-wave attractive interactions. 

\begin{figure}[H]
    \centering
    \includegraphics[width=0.75\linewidth]{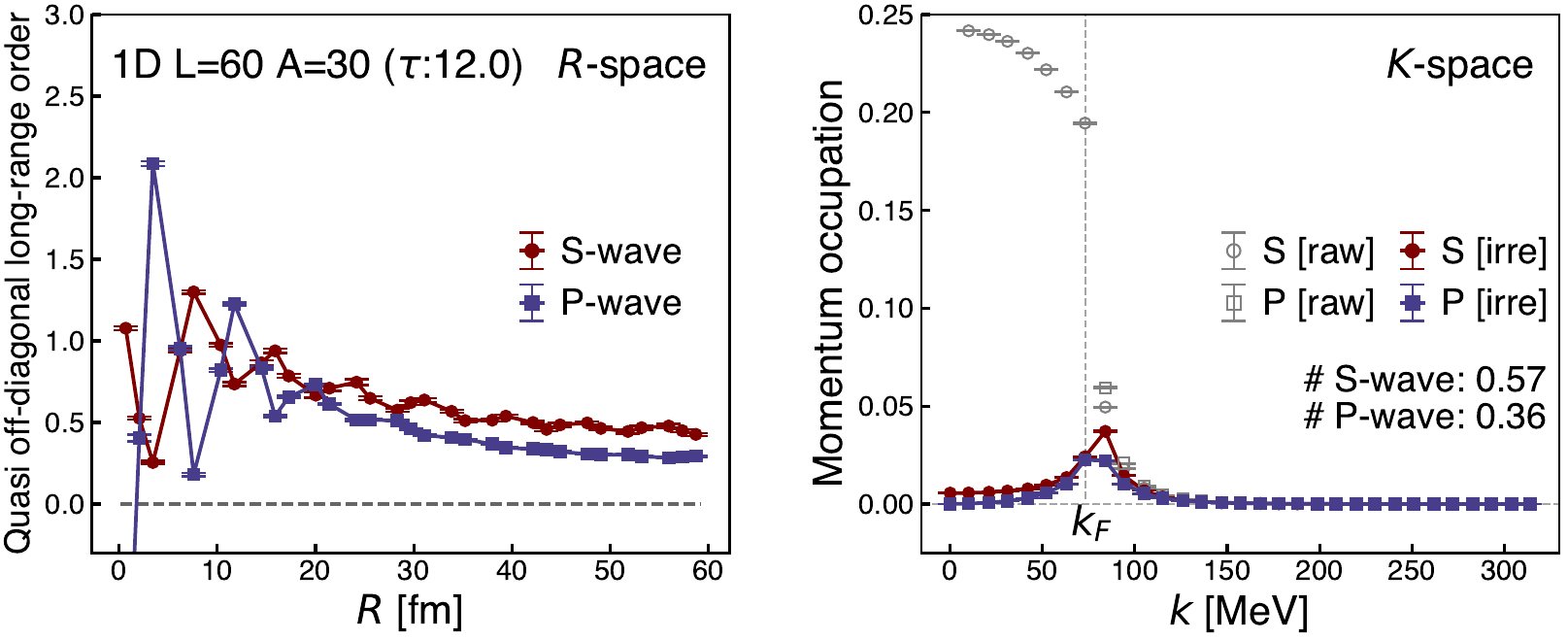}
    \caption{\textbf{Quasi-ODLRO and pair occupation in 1D.}
    Quasi off-diagonal long-range order and momentum-space pair occupation for a one-dimensional system with $L=60$, $A=30$, obtained from \textit{ab initio} lattice calculations. Left: quasi off-diagonal long-range order (qODLRO) in coordinate space. Right: momentum-space pair occupation after subtraction of the one-body background. ``Raw'' and ``irre'' denote the two-body density before and after irreducible (cumulant) subtraction, respectively.
    \label{fig:SU2_1D_SP_L60_A30}}
\end{figure}

With this interaction setup, we measured the ``quasi'' off-diagonal long-range order (qODLRO) in the coordinate space and pairs in the momentum space, shown in Fig~\ref{fig:SU2_1D_SP_L60_A30} . 
We notice that the very slow decay in the left panel shows the behavior of qODLRO.
In the right panel, after subtracting the one-body ``background'', a clear peak shows up just above the Fermi surface, indicating the s-wave and p-wave pairs are most strongly concentrated at the momenta around the Fermi surface.
% We take the last 5 fm average for qODLRO and obtain s-wave for $0.45$ and p-wave for $0.29$ which is in reasonable agreement with the pair measurements in momentum space $0.57$ for s-wave and $0.36$ for p-wave.
If we average qODLRO over a $5$ fm interval ending at the largest value of largest $R$ calculated, we obtain $0.45$ for the s-wave and $0.29$ for the p-wave. These are in reasonable agreement with the pair measurements in momentum space, which were $0.57$ for s-wave and $0.36$ for p-wave.
Since pairing in momentum space is more convenient to measure, in the following 2D and 3D calculations, we mostly concentrate on the pairing and quartet measurement in momentum space.

\begin{figure}[H]
    \centering
    \includegraphics[width=0.43\linewidth]{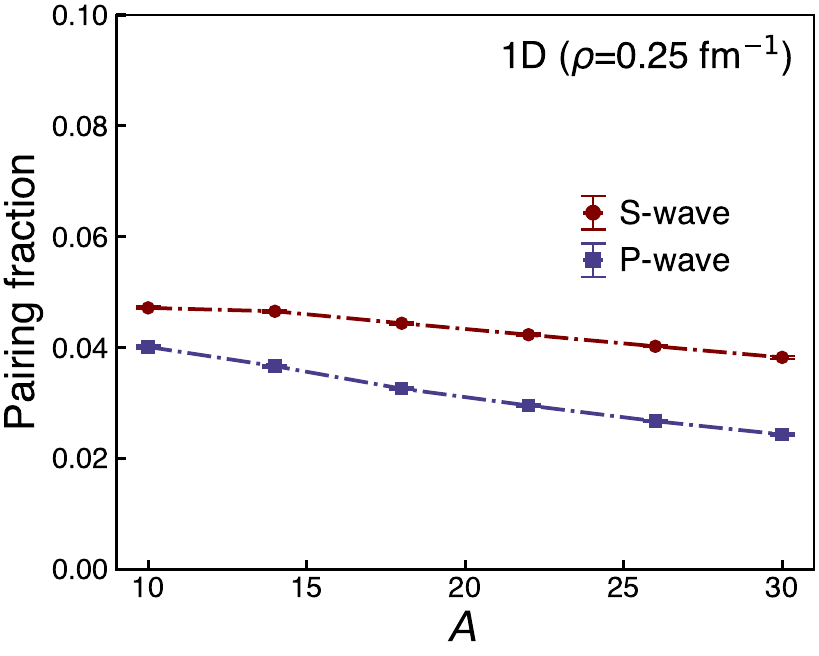}
    \caption{\textbf{1D s-wave and p-wave pairing fractions.} S-wave and p-wave pairing fraction in one-dimensional systems at density $\rho=0.25$ fm$^{-1}$ from \textit{ab initio} lattice calculations. \label{fig:SU2_1D_SP_TDC}}
\end{figure}

\begin{figure}[H]
  \centering
  \begin{subfigure}{0.375\textwidth}
    \centering
    \includegraphics[width=\linewidth]{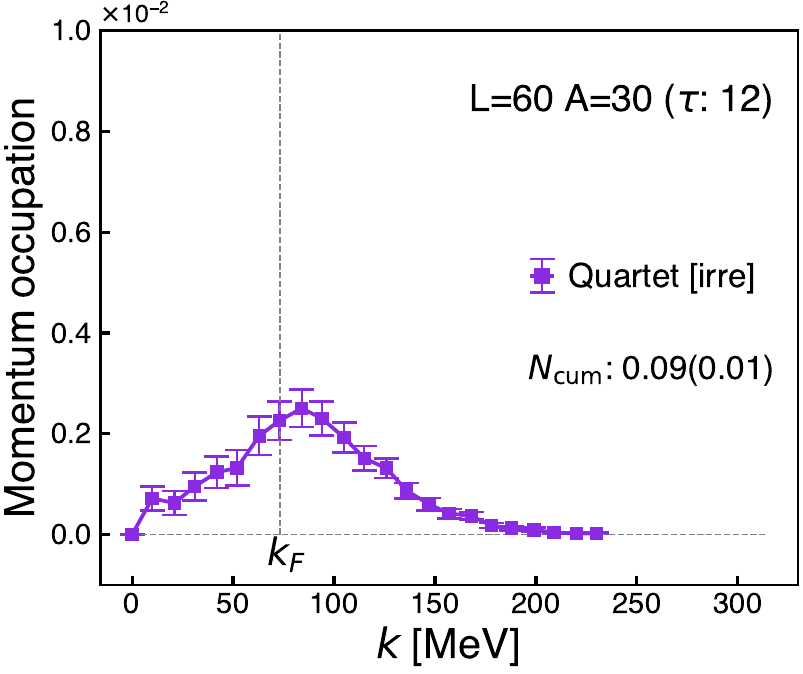}
    %\label{fig:sub1}
  \end{subfigure}\hspace{0.05\textwidth}
  \begin{subfigure}{0.375\textwidth}
    \centering
    \includegraphics[width=\linewidth]{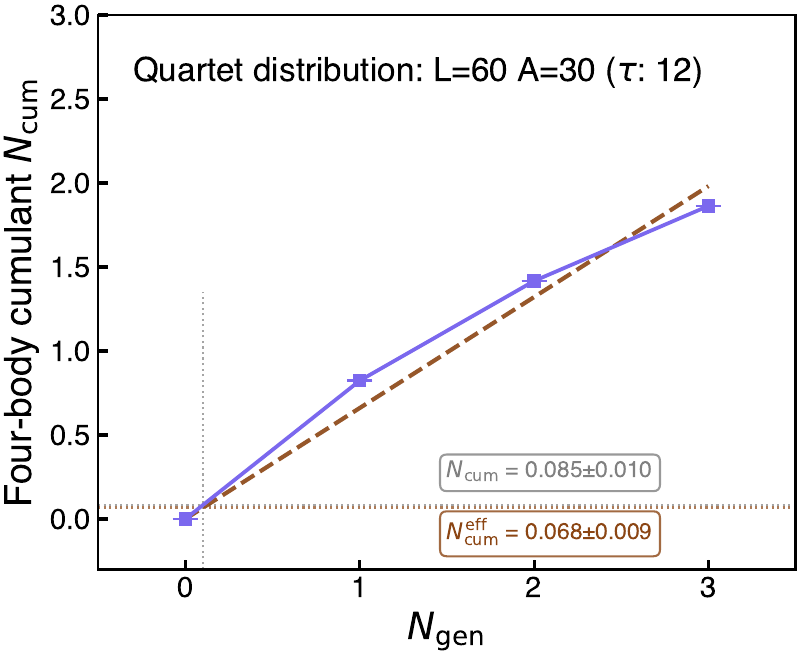}
    %\label{fig:sub2}
  \end{subfigure}
  \caption{\textbf{Quasi-quartets in 1D.} Quasi quartet measurement for a one-dimensional $L=60$, $A=30$ system from \textit{ab initio} lattice calculations. ${N_{\text{cum}}}$ stands for the sum of all four-body cumulants.  \label{fig:SU2_1D_QT_L60}}
\end{figure}

As we discussed above, true long-range order is absent in one-dimensional systems. In the thermodynamic limit, both the s-wave and p-wave signals decay to zero. 
This behavior is illustrated in Fig~\ref{fig:SU2_1D_SP_TDC}, where we keep the density fixed and increase the particle number. The data are taken with Euclidean time $\tau = 24$ MeV$^{-1}$, which already shows a good convergence.  
In addition, we observe a small but finite quasi-quartet signal in the $L=60, A=30$ system, as shown in Fig~\ref{fig:SU2_1D_QT_L60}.

\subsection{2D GAE Hubbard Model}
\begin{figure}[H]
  \centering
  \begin{subfigure}{0.32\textwidth}
    \centering
    \includegraphics[width=\linewidth]{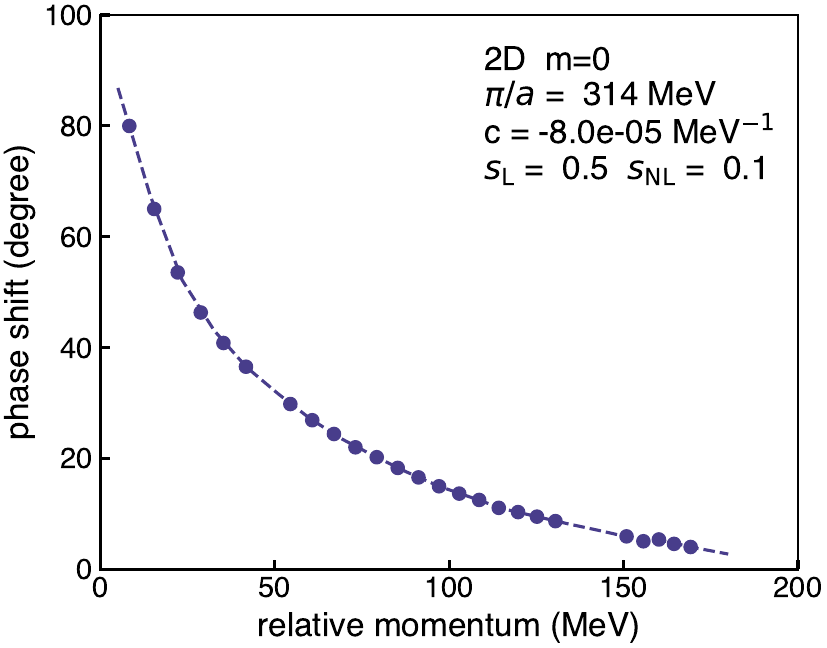}
    %\label{fig:sub1}
  \end{subfigure}\hspace{0.05\textwidth}
  \begin{subfigure}{0.32\textwidth}
    \centering
    \includegraphics[width=\linewidth]{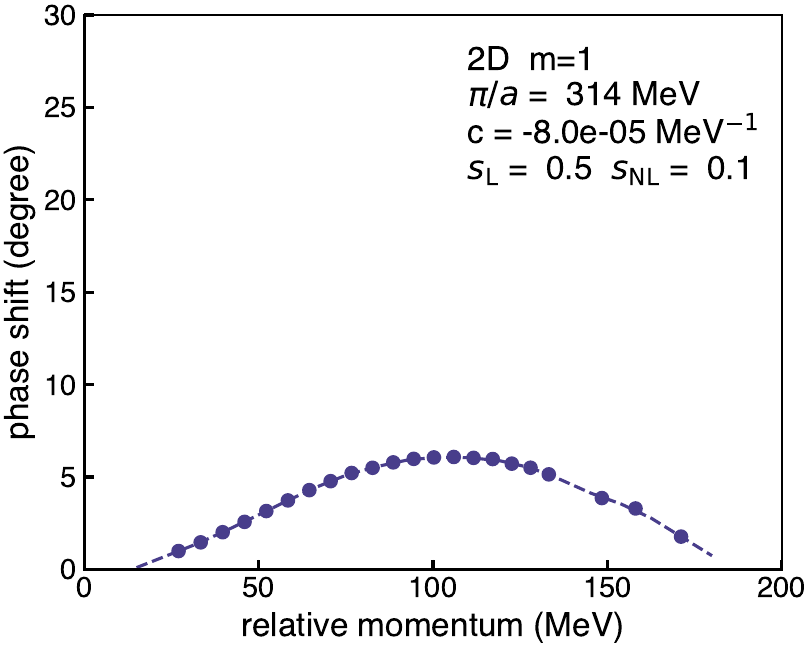}
    %\label{fig:sub2}
  \end{subfigure}
  \caption{\textbf{2D SU(2) scattering phase shifts.} Two-body scattering phase shifts for the SU(2)-symmetric interaction in two dimensions. Left: angular momentum channel $m=0$. Right: angular momentum channel $m=1$. \label{fig:SU2_2D_PhaseShift}}
\end{figure}

\begin{figure}[H]
    \centering
    \includegraphics[width=0.85\linewidth]{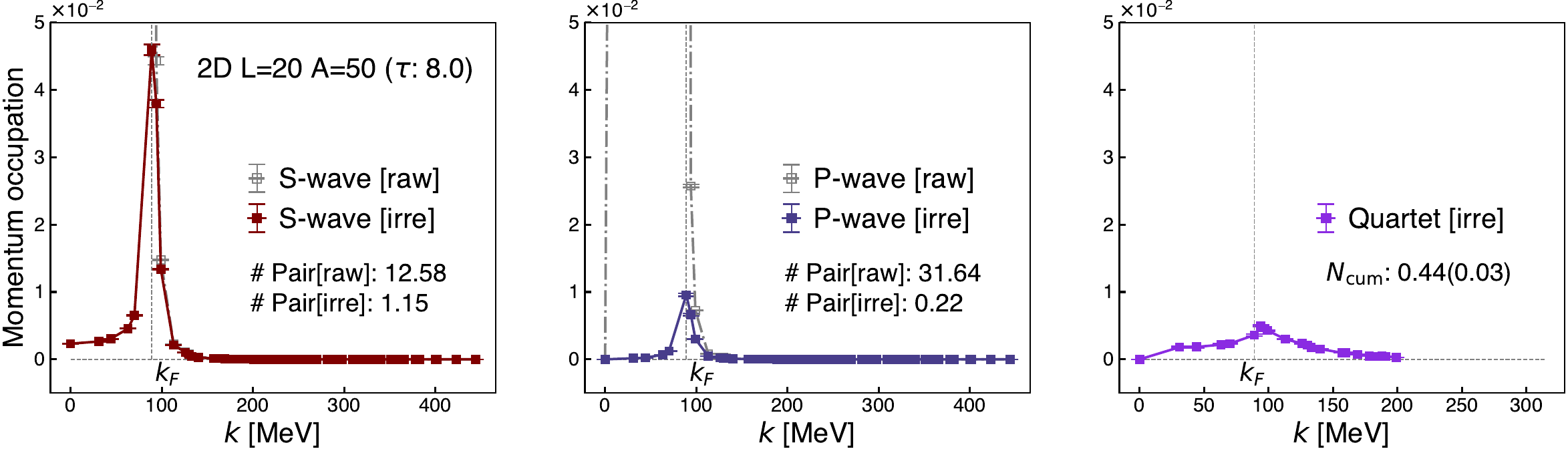}
    \caption{\textbf{2D momentum-space pairing and quartet distributions.} Momentum-space pair and quartet distributions occupations for a two-dimensional system with  $L^2=20^2$, $A=50$ at Euclidean time $\tau=8$ MeV$^{-1}$, obtained from \textit{ab initio} lattice calculations. Left: s-wave pair occupation. Middle: p-wave pair occupation. Right: irreducible quartet number distribution. ``Raw'' and ``irre'' denote the two-body density before and after irreducible (cumulant) subtraction, respectively. $N_{\text{cum}}$ denotes the sum of all four-body cumulants.
    \label{fig:SU2_2D_SP_L20_A50}}
\end{figure}

We employ a similar interaction with $c=-8.0\times 10^{-5}$ MeV$^{-1}$, $s_{\text{L}}=0.5$, $s_{\text{NL}}=0.1$ and $i_{\text{kin}}=1$ for the calculation of the two-dimensional system. In addition to the local smearing, we also turn on the nonlocal smearing $s_{\text{NL}}$, which acts as a regulator to suppress high-momentum attractions, as illustrated in Fig~\ref{fig:SU2_2D_PhaseShift}. 
Using this interaction, we perform lattice calculations for a two-dimensional system with $L^2=20^2$ and $A=50$. 
In Fig~\ref{fig:SU2_2D_SP_L20_A50} we present the measured pairing signals in the s-wave, p-wave, and quartet channels. In addition to a pronounced s-wave pairing signal near the Fermi surface, non-negligible p-wave and quartet signals are also observed. More detailed investigations and systematic discussions of two-dimensional systems will be left for future work.

\subsection{3D GAE Hubbard Model} \label{Sec:3D_SU2}
\begin{figure}[H]
  \centering
  \begin{subfigure}[t]{0.245\textwidth}
    \centering
    \includegraphics[width=\linewidth]{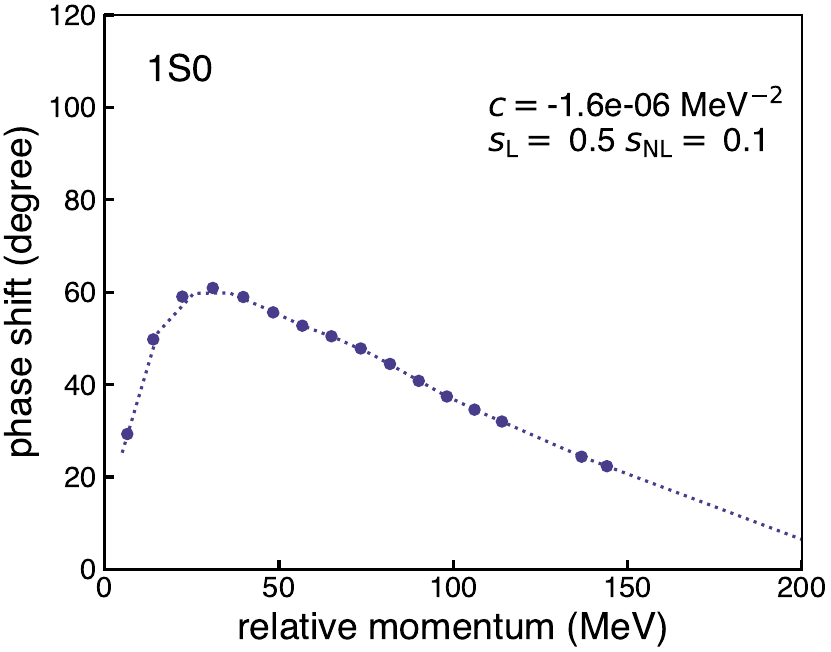}
    %\label{fig:sub1}
  \end{subfigure}\hfill
  \begin{subfigure}[t]{0.245\textwidth}
    \centering
    \includegraphics[width=\linewidth]{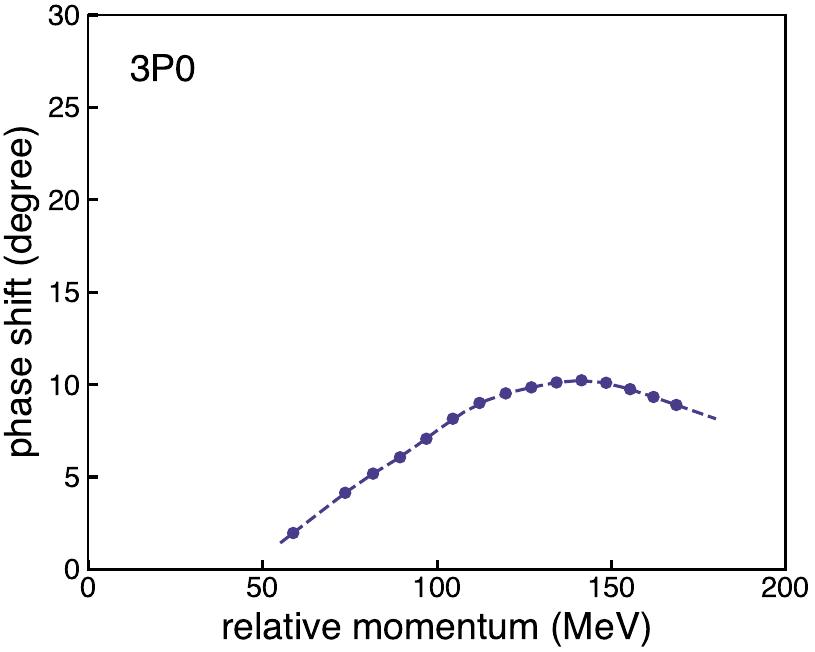}
    %\label{fig:sub2}
  \end{subfigure}\hfill
  \begin{subfigure}[t]{0.245\textwidth}
    \centering
    \includegraphics[width=\linewidth]{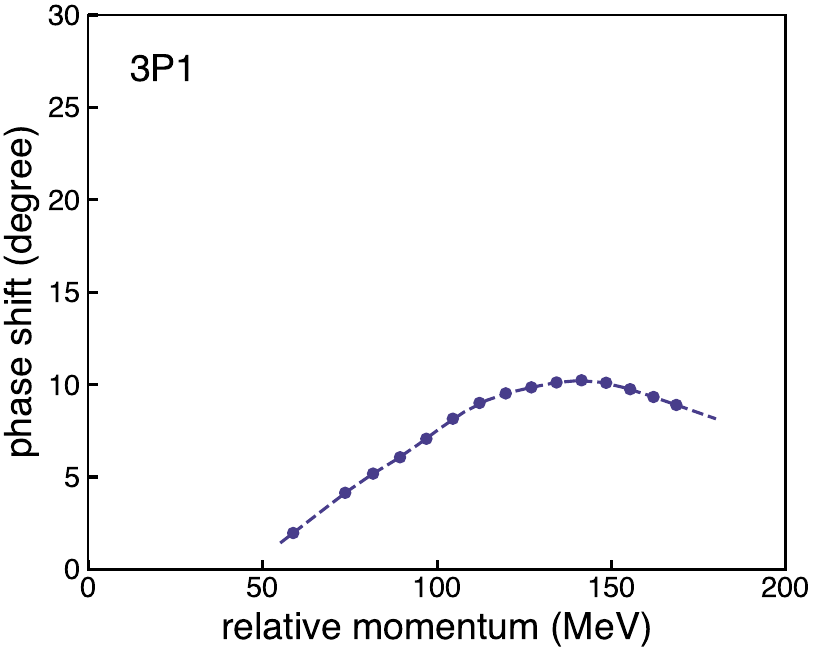}
    \label{fig:sub3}
  \end{subfigure}\hfill
  \begin{subfigure}[t]{0.245\textwidth}
    \centering
    \includegraphics[width=\linewidth]{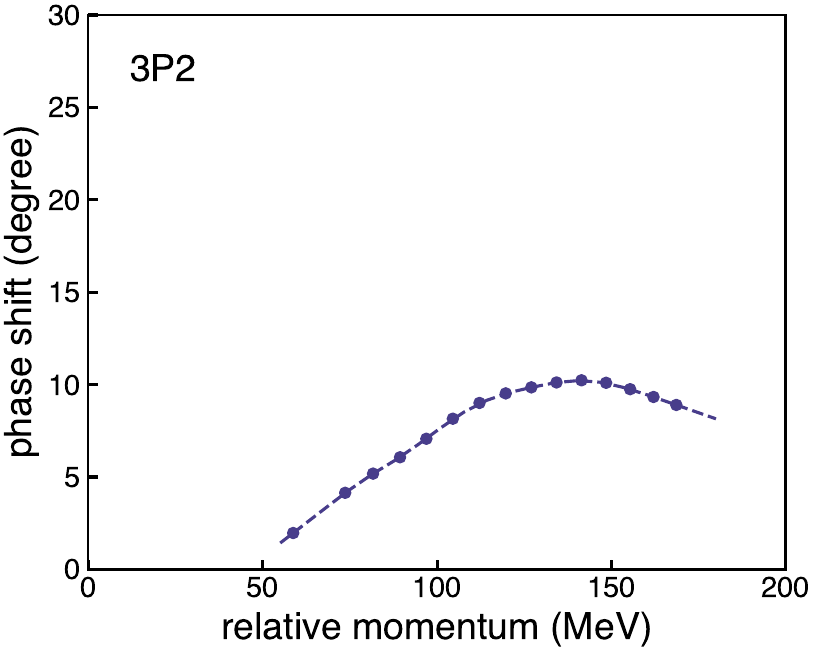}
  \end{subfigure}
  \caption{\textbf{3D SU(2) scattering phase shifts.} Two-body scattering phase shifts for the SU(2)-symmetric interaction in three dimensions. From left to right: $^{1}S_0$, $^{3}P_{0}$, $^{3}P_{1}$, and $^{3}P_{2}$ channels. \label{fig:SU2_PhaseShift}}
\end{figure}

For the three-dimensional Fermi gas, we set $c = -1.6\times 10^{-6}$ $\text{MeV}^{-2}$, $s_{\text{L}} = 0.5$ and $s_{\text{NL}} = 0.1$. The lattice spacing for the 3D GAE Hubbard model is $a=1.97$ fm. We also set $i_{\text{kin}}=1$ which makes the kinetic term only have the onsite and nearest hopping contributions. 
With this interaction setup, the phase shifts are calculated in Fig \ref{fig:SU2_PhaseShift}. 
As expected, the SU(2)-symmetric interaction leads to identical phase shifts in the $^{3}P_{0}$, $^{3}P_{1}$ and $^{3}P_{2}$ channels.
\paragraph{} In the remainder of this section, we will first discuss the multimodal superfluidity of a $A=66$ spin-balanced system in a $L^3=10^3$ lattice box with $a_t=1/300$ MeV$^{-1}$ and $L_t=300$. We then discuss the Euclidean time extrapolation. Finally, we examine the approach to the thermodynamic limit by performing calculations for different lattice sizes $L$ and particle numbers $A$ while keeping the density approximately fixed.

\begin{figure}[H]
    \centering
    \includegraphics[width=0.85\linewidth]{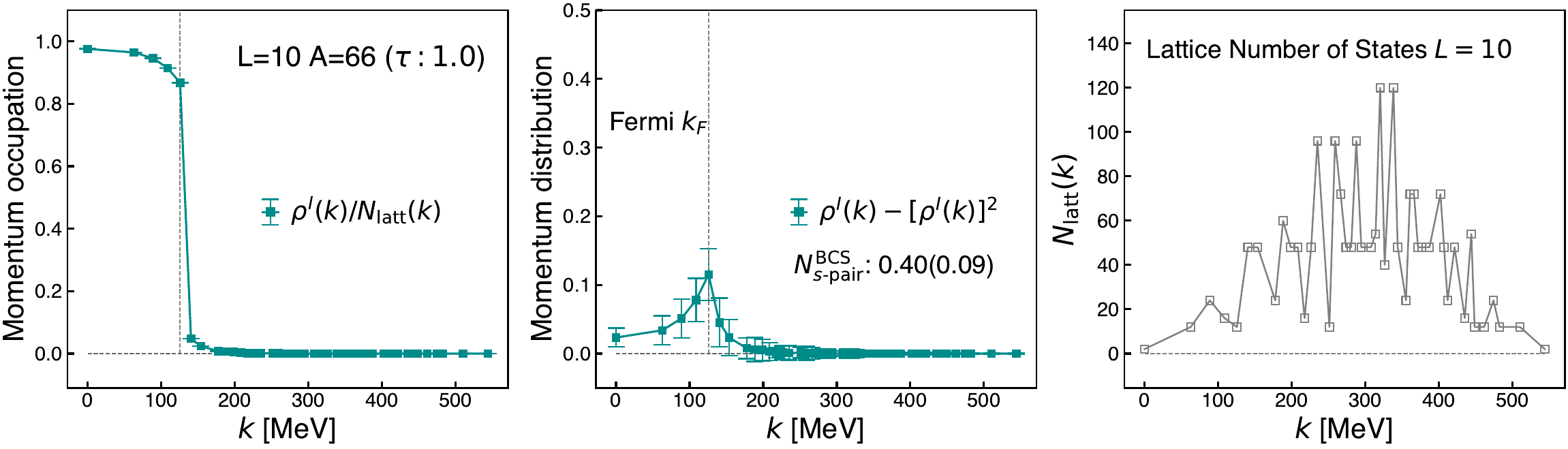}
    \caption{\textbf{Momentum-space one-body observables for $L^3=10^3$, $A=66$.} One-body momentum-space observables for an $L^3 = 10^3$, $A = 66$ system at Euclidean time $\tau = 1.0\,\mathrm{MeV}^{-1}$, obtained from \textit{ab initio} lattice calculations.  Left: normalized one-body momentum occupation $\rho^I(k)/N_{\text{latt}}(k)$. Middle: fluctuation measure $[\rho^{I}(\vec{k}) - [\rho^{I}(\vec{k})]^2]$. Right lattice state number $N_{\text{latt}}(k)$ for $L^3=10^3$.
    \label{fig:SU2_L10A66_1B}}
\end{figure}

The one-body momentum distribution is measured with the ``rank-one operator method''. In Fig~\ref{fig:SU2_L10A66_1B} we show the one-body momentum occupation of the $L^3=10^3$, $A=66$ system in the left panel. The occupation rate is calculated by taking the ratio $\rho^{I}(k)/N_{\text{latt}}(k)$, with $\rho^{I}$ being the one-body momentum density and $N_{\text{latt}}(k)$ the number of states in the lattice shown in the right panel of Fig~\ref{fig:SU2_L10A66_1B}.
Instead of a sharp step function, it shows a smooth depletion around the Fermi momentum, in qualitative agreement with the BCS picture. According to BCS theory, the number of pairs can be calculated by $N_{s\text{-pair}} = \sum_k (u_k v_k)^2 = \sum_k (v^2_k - v^4_k)$, with $v_k$, $u_k$ being the coefficients of the Bogoliubov transformation and $2v^2_k = \rho^{I}_{\uparrow}(\vec{k}) + \rho^{I}_{\downarrow} (-\vec{k})$. Then we have $N^{\text{BCS}}_{s\text{-pair}} = \sum_{\vec{k}}\big{(}\rho^{I}(\vec{k}) - [\rho^{I}(\vec{k})]^2\big{)}$, which provides an estimation of s-wave pairing number from the BCS ansatz. 
The momentum distribution of $\rho^{I}(\vec{k}) - [\rho^{I}(\vec{k})]^2$ is shown in the middle panel of Fig~\ref{fig:SU2_L10A66_1B} with $N^{\text{BCS}}_{s\text{-pair}} =0.40$ the s-wave BCS estimate from the one-body density.  

\begin{figure}[H]
    \centering
    \includegraphics[width=0.65\linewidth]{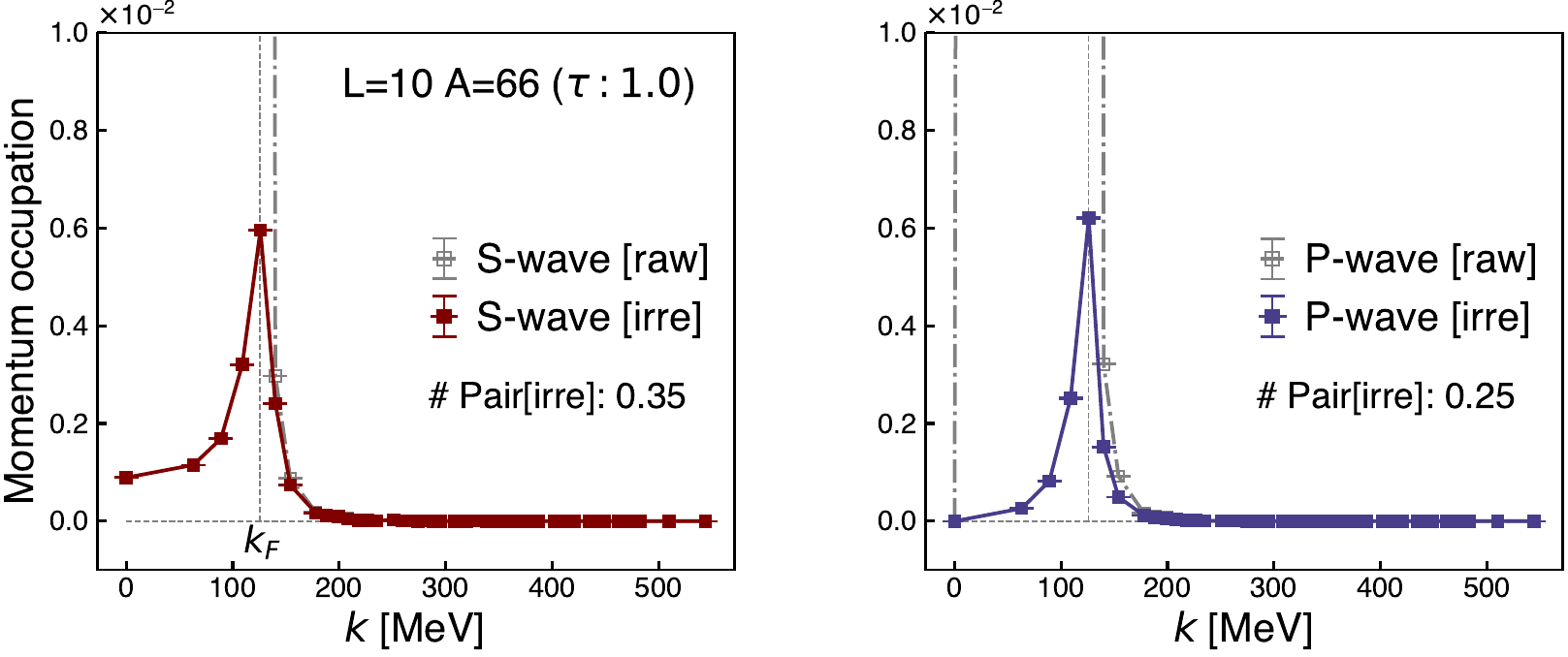}
    \caption{\textbf{Momentum-space pairing for $L=10$, $A=66$.} S-wave and p-wave momentum-space pair occupations for an $L^3 = 10^3$, $A = 66$ system at Euclidean time $\tau = 1.0\,\mathrm{MeV}^{-1}$, obtained from \textit{ab initio} lattice calculations. Left: s-wave pair occupation. Right: p-wave pair occupation. ``Raw'' and ``irre'' denote the two-body density before and after irreducible (cumulant) subtraction, respectively.
    \label{fig:SU2_L10A66_SP}}
\end{figure}

In Fig~\ref{fig:SU2_L10A66_SP}, we plot the two-body density occupations in momentum space, where ``raw'' and ``irre'' denote the original and irreducible two-body densities (two-body cumulants), respectively. Analogous to a one-body occupation, the two-body occupation is calculated by taking the ratio $\rho^{II}(k)/N_{\text{latt}}(k)$.
After subtracting the ``background'' from one-body density products, a pronounced peak emerges at the Fermi surface for both s-wave and p-wave channels. By summing contributions from all momentum modes, we extract the total pairing numbers, yielding 0.353 for s-wave pairing and 0.249 for p-wave pairing. It is noteworthy that the BCS ansatz constructed from one-body densities (middle panel of Fig~\ref{fig:SU2_L10A66_1B}) can provide a reasonable estimation for the s-wave pairing (left panel of Fig~\ref{fig:SU2_L10A66_SP}). However, there is no mechanism in the standard BCS theory to estimate the {\it{ab initio}} measurement of the p-wave pairing. It should also be noticed that due to Pauli blocking, there is no p-wave pair present at ${k}=0$.

\begin{figure}[H]
    \centering
    \includegraphics[width=0.42\linewidth]{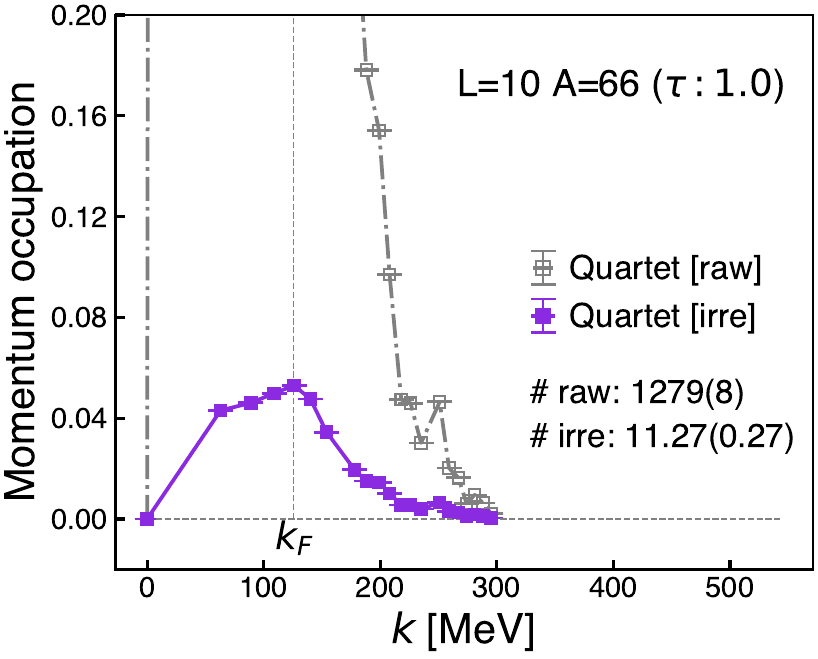}
    \caption{\textbf{Momentum-space quartet distribution for $L^3=10^3$, $A=66$.} Momentum-space quartet occupation for the $L^3=10^3$, $A=66$ system at Euclidean time $\tau=1.0$ MeV$^{-1}$, obtained from \textit{ab initio} lattice calculations. ``raw'' and ``irre'' denote the four-body density before and after irreducible (cumulant) subtraction, respectively. ``\# raw'' stands for the sum of all four-body density and ``\# irr'' stands for the sum of all four-body cumulants $N_{\text{cum}}$. 
    \label{fig:SU2_L10A66_QT}}
\end{figure}

\paragraph{}Four-body quartet superfluidity is the novel correlation mechanism discovered in quantum many-body systems. 
To characterize this effect, we measure the four-body cumulants $\rho_{\uparrow\downarrow\uparrow\downarrow}^{IV}$ as described in Eq~\ref{eq:rhoIV_irre} and Eq~\ref{eq:rhoIV_raw}. Due to the enormous number of four momentum combinations, the direct measurement of each combination with the ``rank-one operator'' method is impractical. Instead, we first generate momentum pinhole configurations during the Euclidean-time propagation and subsequently analyze the quartet signal from these sampled configurations.  
In Fig~\ref{fig:SU2_L10A66_QT}, we present the momentum-space quartet distribution with $\rho^{IV}({k})$ defined in Eq~\ref{eq:rhoIV_k}.
The dominant peak is located near Fermi surface, similar to the s-wave and p-wave cases, but exhibits a substantially broader momentum distribution. The definition of a quartet requires any sub-system to have non-zero momenta. Then the non-zero quartet signal starts at $k=62.8$ MeV which is the first non-zero state of $L^3=10^3$. 
In the present analysis, we only allow $|\vec{k}|<300$ MeV, which already results in a total of 7707504 distinct $\{\vec{k}_1,\vec{k}_2,\vec{k}_3,\vec{k}_4\}$ momentum combinations.
% After summing up all these contributions we obtain an remarkable irreducible quartet number 6.77(0.17), which means about $10.69/(66/4)\approx 65 \%$ particles in the system get involved in forming quartets.

\begin{figure}[H]
    \centering
    \includegraphics[width=0.65\linewidth]{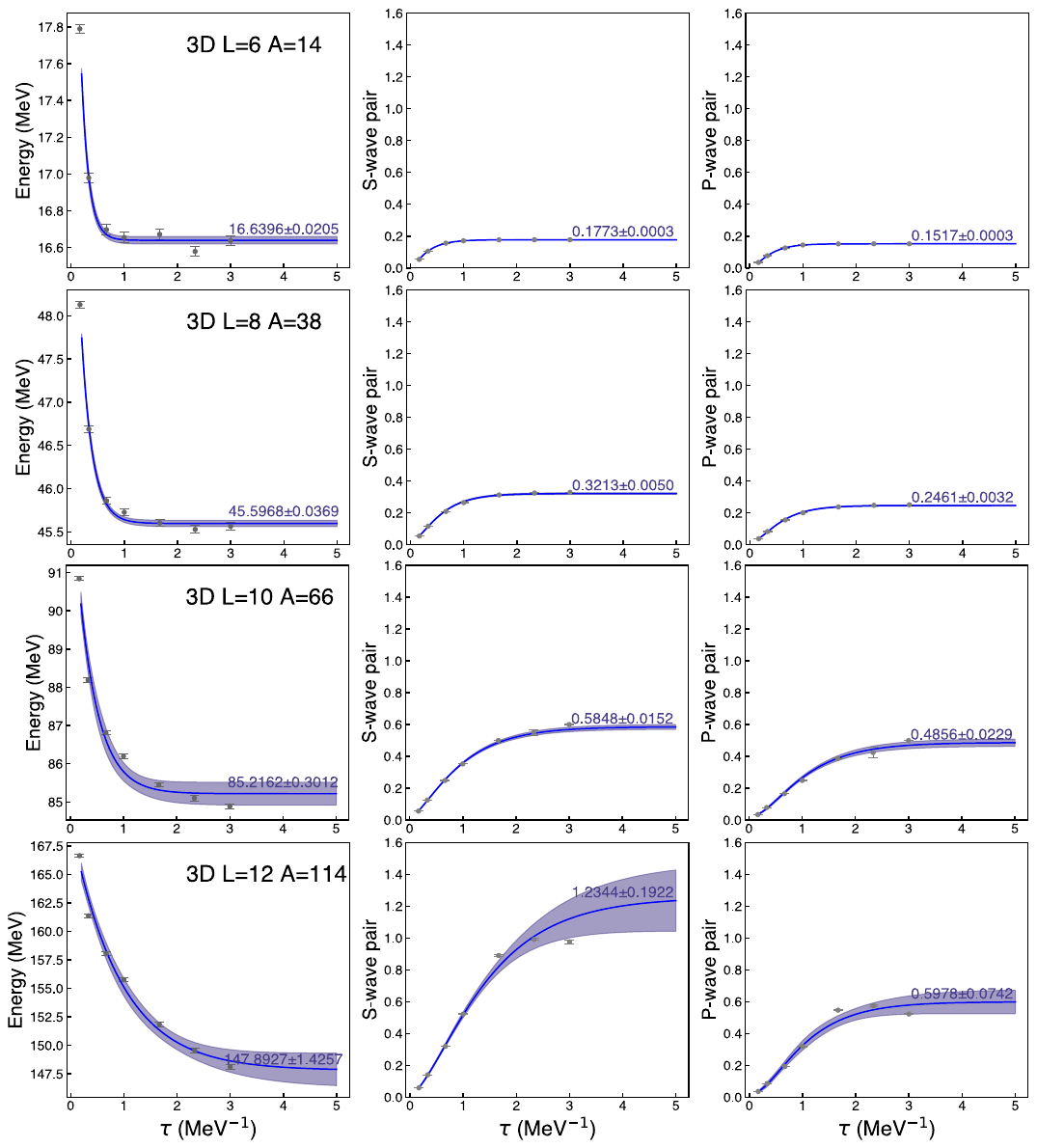}
    \caption{\textbf{Euclidean-time extrapolation of energy and pairing observables.} Euclidean-time extrapolation of the energy and two-body pairing observables for three-dimensional systems with $L^3=6^3,8^3,10^3,12^3$, obtained from \textit{ab initio} lattice calculations. From left to right, the columns show the total energy, the s-wave pairing observable, and the p-wave pairing observable, respectively.
    \label{fig:SU2_Euclidean_ESP}}
\end{figure}

The above calculations are performed at a fixed Euclidean time $\tau=a_t \times L_t=1.0$ MeV$^{-1}$. To extract the true ground state properties, we performed Euclidean time extrapolation. Two exponential functions $E(\tau) = c + c_1 \exp(-D_0 \tau)$ and $O(\tau) = r_0 + r_1 \exp(-D_0 \tau) + r_2 \exp(-D_0 \tau / 2)$ are used in the fitting process. We consider four different systems in Fig~\ref{fig:SU2_Euclidean_ESP}, $(L^3,A)=(6^3,14)$, $(8^3,38)$, $(10^3,66)$ and $(12^3,114)$, which are chosen due to their closed-shell configurations and similar densities.
For the largest system with $L^3=12^3$, $A=114$, the convergence in Euclidean time is noticeably slower, and the computational cost is significantly higher. Given the current computational resources, we only perform calculations with $\tau$ up to $3.0$ MeV$^{-1}$.

\begin{figure}[H]
    \centering
    \includegraphics[width=0.75\linewidth]{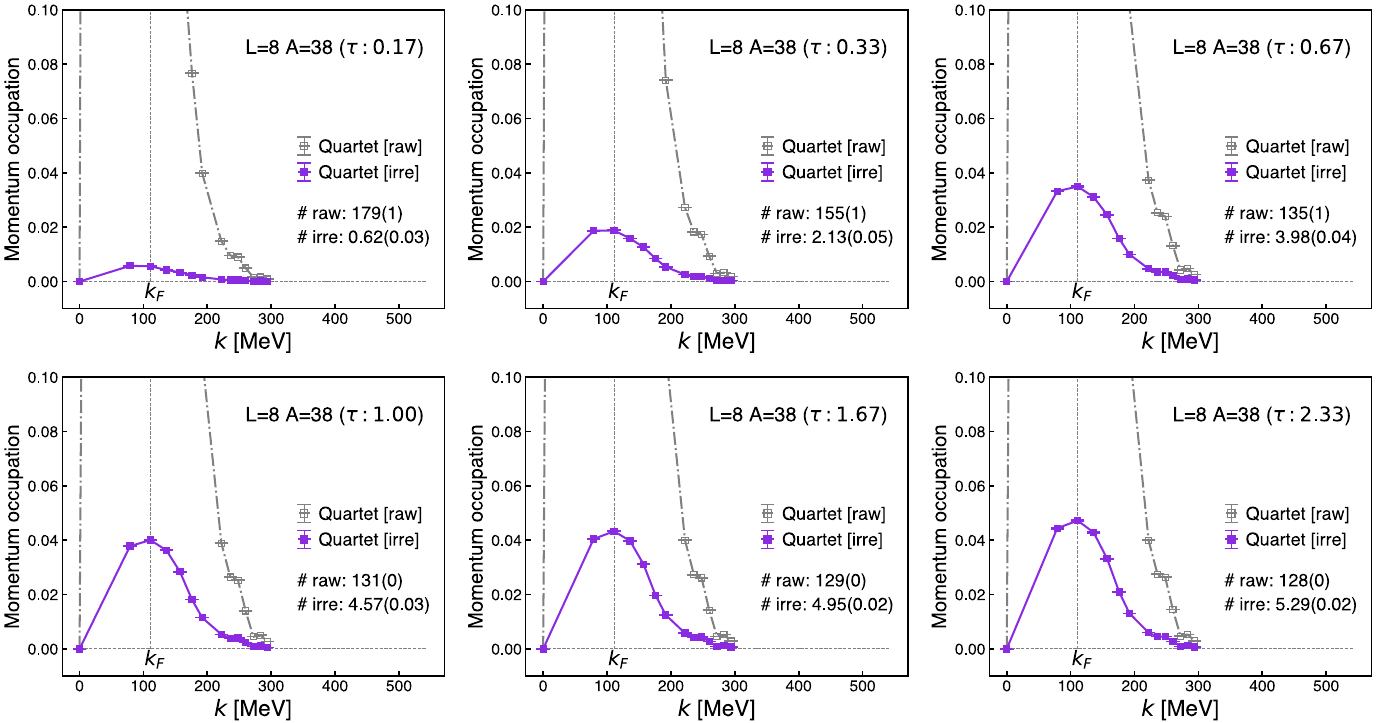}
    \caption{ \textbf{Euclidean-time dependence of the quartet distribution.} Momentum-space quartet occupation for the $L=8, A=38$ system at different Euclidean time $\tau$ in units of MeV$^{-1}$, obtained from \textit{ab initio} lattice calculations. Each panel corresponds to a different Euclidean times $\tau$. ``raw'' and ``irre'' denote the four-body density before and after irreducible (cumulant) subtraction, respectively. ``\# raw'' stands for the sum of all four-body density and ``\# irr'' stands for the sum of all four-body cumulants $N_{\text{cum}}$.
    \label{fig:SU2_Euclidean_QT_sub}}
\end{figure}

For the quartet calculation, we keep $|\vec{k}|<300$ MeV. From Fig~\ref{fig:SU2_Euclidean_QT_sub}, we can see that with several different Euclidean time $\tau$, all the irreducible quartets drop to zero at $|\vec{k}|\approx$300 MeV. Interestingly, as $\tau$ is increased, irreducible quartets increase prominently while the ``raw'' quartets drop gradually. A similar behavior can be also observed in s-wave and p-wave signals. 
At $\tau=0$, the initial many-body wavefunction is just a single Slater determinant. As $\tau$ is increased, many-body correlation will be built in, leading to the increase of irreducible s-wave and p-wave and quartets and the decrease of uncorrelated components. 

\begin{figure}[H]
    \centering
    \includegraphics[width=0.78\linewidth]{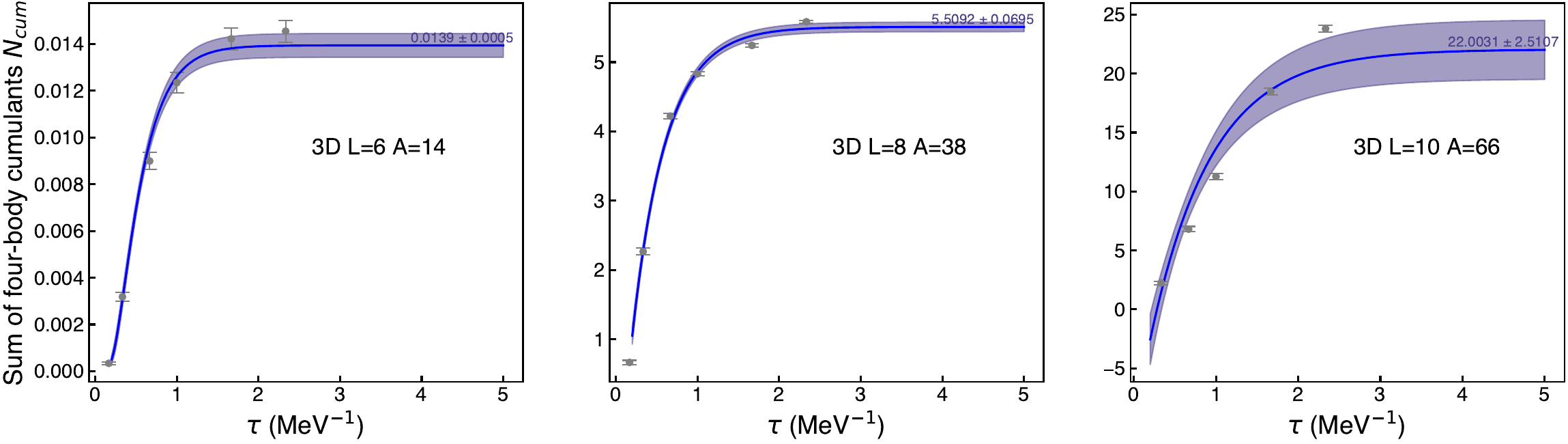}
    \caption{\textbf{Euclidean-time extrapolation of quartet cumulants.} Euclidean-time extrapolation of the sum of four-body cumulants $N_{\text{cum}}$ for three-dimensional systems with $L^3=6^3,8^3,10^3$, obtained from \textit{ab initio} lattice calculations.
    \label{fig:SU2_Euclidean_QT}}
\end{figure}

\begin{figure}[H]
    \centering
    \includegraphics[width=0.75\linewidth]{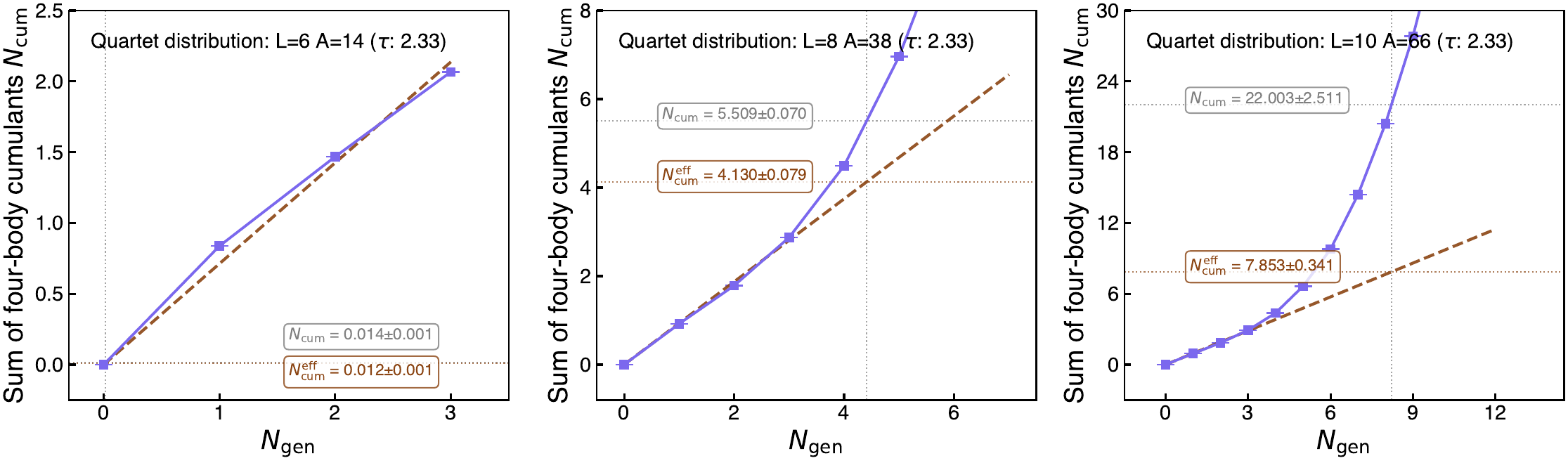}
    \caption{\textbf{Linear correlation between four-body cumulants and effective quartet numbers.} Numerical correlation between the sum of four-body cumulants $N_{\text{cum}}$ and the generated quartet number $N_{\text{gen}}$, obtained from random Monte Carlo sampling. The dashed lines indicate the linear contribution in Eq~\ref{eq:Ncum_Ngen} and $N^{\text{eff}}_{\text{cum}}$ is the estimated quartet number. 
    \label{fig:SU2_Q_number}}
\end{figure}

Due to the substantial computational cost associated with generating momentum pinhole configurations and measuring four-body quartets in large systems, we restrict the quartet measurements to systems with $L^3\leq10^3$. 
In Fig~\ref{fig:SU2_Euclidean_QT}, we present the Euclidean time extrapolation of the sum of four-body cumulants (quartet correlation strength) for three systems: $(L^3,A)=(6^3,14)$, $(8^3,38)$ and $(10^3,66)$.
It should be mentioned that the sum of four-body cumulants can be directly linked to the number of quartets.
As discussed in Sec.~\ref{sec:quartet_num}, we use the calculated quartet momentum density distribution $\rho^{IV}({k})$ to randomly sample momentum pinhole hole configurations for a certain number of generated quartet $N_{\text{gen}}$, and then measure the sum of four-body cumulants $N_{\text{cum}}$. Relevant results are shown in Fig~\ref{fig:SU2_Q_number}.
In the dilute limit, $N_{\text{cum}}$ has a linear relation with $N_{\text{gen}}$. 
We therefore perform a linear fit for $N_{gen}<=3$ and extract the effective cumulants $N^{\text{eff}}_{\text{cum}}$, which corresponds to the linear part of the total cumulants. 
This effective cumulants $N^{\text{eff}}_{\text{cum}}$ can be interpreted as the measured quartet number in the system.

The thermodynamic-limit behavior is examined by comparing systems with similar densities but different lattice sizes.
From Table~\ref{tab:SU2_Thermodynamic_all}, we observe that the averaged s-wave and p-wave pairing signals, together with the quartet contribution, reach a plateau starting from the $L^3 = 8^3$, $A = 38$ system.
The corresponding pairing fractions are shown in Table~\ref{tab:SU2_Thermodynamic_all}, where approximately $1.7\%$ of particles form s-wave pairs and $1.3\%$ form p-wave pairs. At the same time, more than $40\%$ of particles contribute to quartet correlations.
Several related energy gaps are summarized in Table~\ref{tab:SU2_L6_Energy}, Table~\ref{tab:SU2_L8_Energy} and Table~\ref{tab:SU2_L10_Energy}.
Interestingly, these energy gaps are also in qualitative agreement with the self-consistent Cooper-model results shown in Table Table~\ref{tab:SU2_Delta}.

\begin{table}[htbp]
\centering
\caption{\textbf{Thermodynamic-limit behavior of s-wave pairing, p-wave pairing and quartets.} S-wave pairs, P-wave pairs, and quartets in the three-dimensional Hubbard model with $c = -1.6\times10^{-6}\,\mathrm{MeV}^{-2}$, $s_{\mathrm{L}} = 0.5$, and $s_{\mathrm{NL}} = 0.1$. The last three columns list the fractions corresponding to S-wave pairs, P-wave pairs, and quartets, respectively.}
\label{tab:SU2_Thermodynamic_all}
\begin{tabular}{lllllllll}
\toprule
$L$ & $A$ & density (fm$^{-3}$) &  s-wave pairs &  p-wave pairs  &  Quartets & $2 \times$S-wave/$A$ &  $2 \times$P-wave/$A$ & $4 \times $Quartets/$A$\\
\midrule
6  & 14  & 0.00844  & 0.177 (1)  & 0.152 (1)  & 0.012 (1)   & 0.0253 (1) & 0.0217 (1) & 0.0034 (3) \\
8  & 38  & 0.00966  & 0.321 (5)  & 0.246 (3)  & 4.130 (79)  & 0.0169 (3)  & 0.0130 (2)  & 0.4347 (83)\\
10 & 66  & 0.00859  & 0.585 (15)  & 0.486 (23)  & 7.853 (341) & 0.0177 (5)  & 0.0147 (7)  & 0.4759 (207)\\
12 & 114 & 0.00859  & 1.234 (192)  & 0.598 (74)  & -           & 0.0217 (34) & 0.0105 (13) &        - \\
\bottomrule
\end{tabular}
\end{table}

% \begin{figure}[H]
%     \centering
%     \includegraphics[width=0.42\linewidth]{Figures/SU2_Fraction_SP.pdf}
%     \caption{\textbf{System-size dependence of pairing fractions.} Irreducible s-wave and p-wave pairing fractions as functions of particle number $A$, obtained from \textit{ab initio} lattice calculations. The shaded bands indicate the averages over different lattice sizes.
%     \label{fig:SU2_TDC_SP}}
% \end{figure}

\begin{table}[htbp]
\centering
\caption{\textbf{SU(2) lattice energies for $L^3 = 6^3$.} Lattice SU(2) energies calculated with $c = -1.6\times10^{-6}\,\mathrm{MeV}^{-2}$, $s_{\text{L}} = 0.5$, and $s_{\text{NL}} = 0.1$. For $A = 14$, the density is $\rho = 8.4356\times10^{-3}\,\mathrm{fm}^{-3}$, corresponding to a Fermi momentum $k_F = 104.72\,\mathrm{MeV}$. }
\label{tab:SU2_L6_Energy}
\begin{tabular}{llll}
\toprule
$A$ 
& $E_{\text{SU2}}$ (MeV) 
& 2$\Delta$ (MeV) 
& 4$\Delta_{\text{Q}}$ (MeV) \\
\hline
$14$                       &  16.64 (2) &      & \\
$15$                       &  21.26 (2) &     & \\
$16_{\uparrow \downarrow}$ &  23.23 (7) & 2.65 (8) & \\
$16_{\uparrow \uparrow}$   &  25.17 (6) & 0.71 (7) & \\
$18$                       &  28.35 (3) &  & 1.47 (15)\\
\bottomrule
\end{tabular}
\end{table}

\begin{table}[htbp]
\centering
\caption{\textbf{SU(2) lattice energies for $L^3 = 8^3$.} Lattice SU(2) energies calculated with $c = -1.6\times10^{-6}\,\mathrm{MeV}^{-2}$, $s_{\text{L}} = 0.5$, and $s_{\text{NL}} = 0.1$. 
For $A = 38$, the density is $\rho = 9.6595\times10^{-3}\,\mathrm{fm}^{-3}$, corresponding to a Fermi momentum $k_F = 111.07\,\mathrm{MeV}$.}
\label{tab:SU2_L8_Energy}
\begin{tabular}{l l l l l}
\toprule
$A$ 
& $E_{\text{SU2}}$ (MeV) 

& 2$\Delta$ (MeV) 
& 4$\Delta_{\text{Q}}$ (MeV) \\
\hline
$38$                       &  45.57 (4) &      & \\
$39$                       &  49.23 (3) &     & \\
$40_{\uparrow \downarrow}$ &  49.57 (2) & 1.33 (7) & \\
$40_{\uparrow \uparrow}$   &  50.59 (3) & 0.31 (7)& \\
$42$                       &  53.16 (13) &  & 0.41 (14)\\
\bottomrule
\end{tabular}
\end{table}

\begin{table}[htbp]
\centering
\caption{\textbf{SU(2) lattice energies for $L^3 = 10^3$.} Lattice SU(2) energies calculated with $c = -1.6\times10^{-6}\,\mathrm{MeV}^{-2}$, $s_{\text{L}} = 0.5$, and $s_{\text{NL}} = 0.1$. 
For $A = 66$, the density is $\rho = 8.5898\times10^{-3}\,\mathrm{fm}^{-3}$, corresponding to a Fermi momentum $k_F = 125.66\,\mathrm{MeV}$.}
\label{tab:SU2_L10_Energy}
\begin{tabular}{llll}
\toprule
$A$ 
& $E_{\text{SU2}}$ (MeV) 
& 2$\Delta$ (MeV) 
& 4$\Delta_{\text{Q}}$ (MeV) \\
\hline
$66$                       &  84.88 (11) &      & \\
$67$                       &  88.27 (13) &     & \\
$68_{\uparrow \downarrow}$ &  89.07 (19) & 2.59 (34) & \\
$68_{\uparrow \uparrow}$   &  90.65 (21) & 1.01 (35) & \\
$70$                       &  92.10 (28) &  & 1.17 (48)\\
\bottomrule
\end{tabular}
\end{table}

\subsection{Polarized 3D GAE Hubbard Model} \label{Sec:3D_SU2_Polarized}
\begin{figure}[H]
    \centering
    \includegraphics[width=0.85\linewidth]{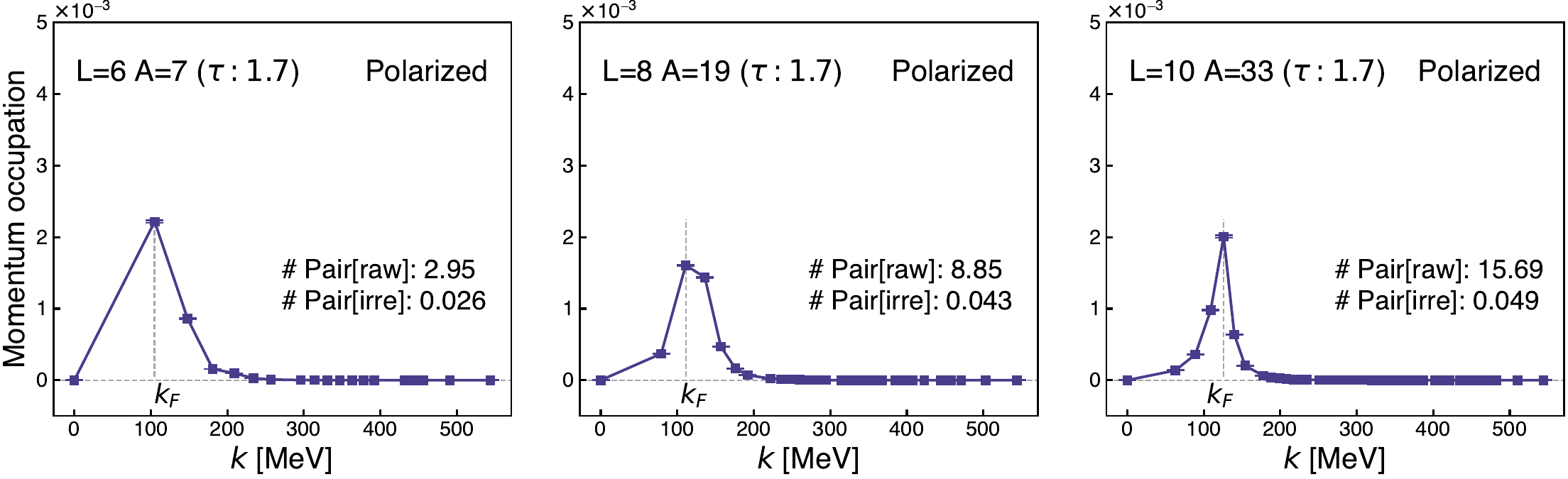}
    \caption{\textbf{P-wave pairing in polarized 3D systems.} Momentum-space p-wave pairing occupations for three three-dimensional polarized systems with $(L^3, A) = (6^3, 7)$, $(8^3, 19)$, and $(10^3, 33)$ at Euclidean time $\tau = 1.7\,\mathrm{MeV}^{-1}$, obtained from \textit{ab initio} lattice calculations. ``raw'' and ``irre'' denote the two-body density before and after irreducible (cumulant) subtraction, respectively. 
    \label{fig:SU2_Polarized_Pwave}}
\end{figure}

For comparison, we also calculate the polarized three-dimensional Hubbard model on a lattice. 
To keep the same Fermi surface as in spin-symmetric systems, $(L^3,A)= (6^3,7)$, $(8^3,19)$ and $(10^3,33)$ are chosen. The p-wave pairing momentum occupation in polarized system are shown in Fig~\ref{fig:SU2_Polarized_Pwave}. In Table~\ref{tab:SU2_Polarized_E}, we list the p-wave pairing gap through a three-point formula.
For these systems, s-wave and quartet superfluidity are not allowed but can p-wave pairs can form. By comparing the p-wave signals in Table~\ref{tab:SU2_Thermodynamic_all}, the polarized systems with half the particle number are much smaller than those of spin-symmetric systems. 
The higher signals in spin-symmetric system can be understood in two aspects: i) the higher density in spin-symmetric system is beneficial for p-wave pairs; ii) the quartet opens a new channel forming two p-wave pairs from two s-wave pairs.  

\begin{table}[htbp]
\centering
\caption{\textbf{SU(2) lattice energies for polarized systems.} Lattice SU(2) energies calculated with $c = -1.6\times10^{-6}\,\mathrm{MeV}^{-2}$, $s_{\text{L}} = 0.5$, and $s_{\text{NL}} = 0.1$. Results are shown for $(L^3, A) = (6^3, 7)$, $(8^3, 19)$, and $(10^3, 33)$, corresponding to densities $\rho = 4.2178\times10^{-3}$, $4.8297\times10^{-3}$, and $4.2949\times10^{-3}\,\mathrm{fm}^{-3}$ with Fermi momenta $k_F = 104.72$, $111.07$, and $125.66\,\mathrm{MeV}$, respectively. For $L^3 = 6^3$, the pairing gap is not calculated due to the limited particle number.}
\label{tab:SU2_Polarized_E}
\begin{tabular}{lll}
\toprule
$(L,A)$ 
& $E_{\text{SU(2)}}$ (MeV) 
& 2$\Delta_P$ (MeV)  \\
\hline
$(6,7)$   & 26.21 (1) &      \\
$(6,8)$   & 54.57 (1) &      \\
$(6,9)$   & 43.13 (1) & - \\
\addlinespace[0.6em]
$(8,19)$   & 75.42 (2) &      \\
$(8,20)$   & 82.55(2) &      \\
$(8,21)$   & 89.45 (2) & 0.23 (4) \\
\addlinespace[0.6em]
$(10,33)$   & 126.62 (2) &      \\
$(10,34)$   & 133.97 (2) &      \\
$(10,35)$   & 141.19 (2) & 0.13 (5) \\
\bottomrule
\end{tabular}
\end{table}

\section{Lattice Results D: Realistic Neutron Matter}
\begin{figure}[H]
    \centering
    \includegraphics[width=0.78\linewidth]{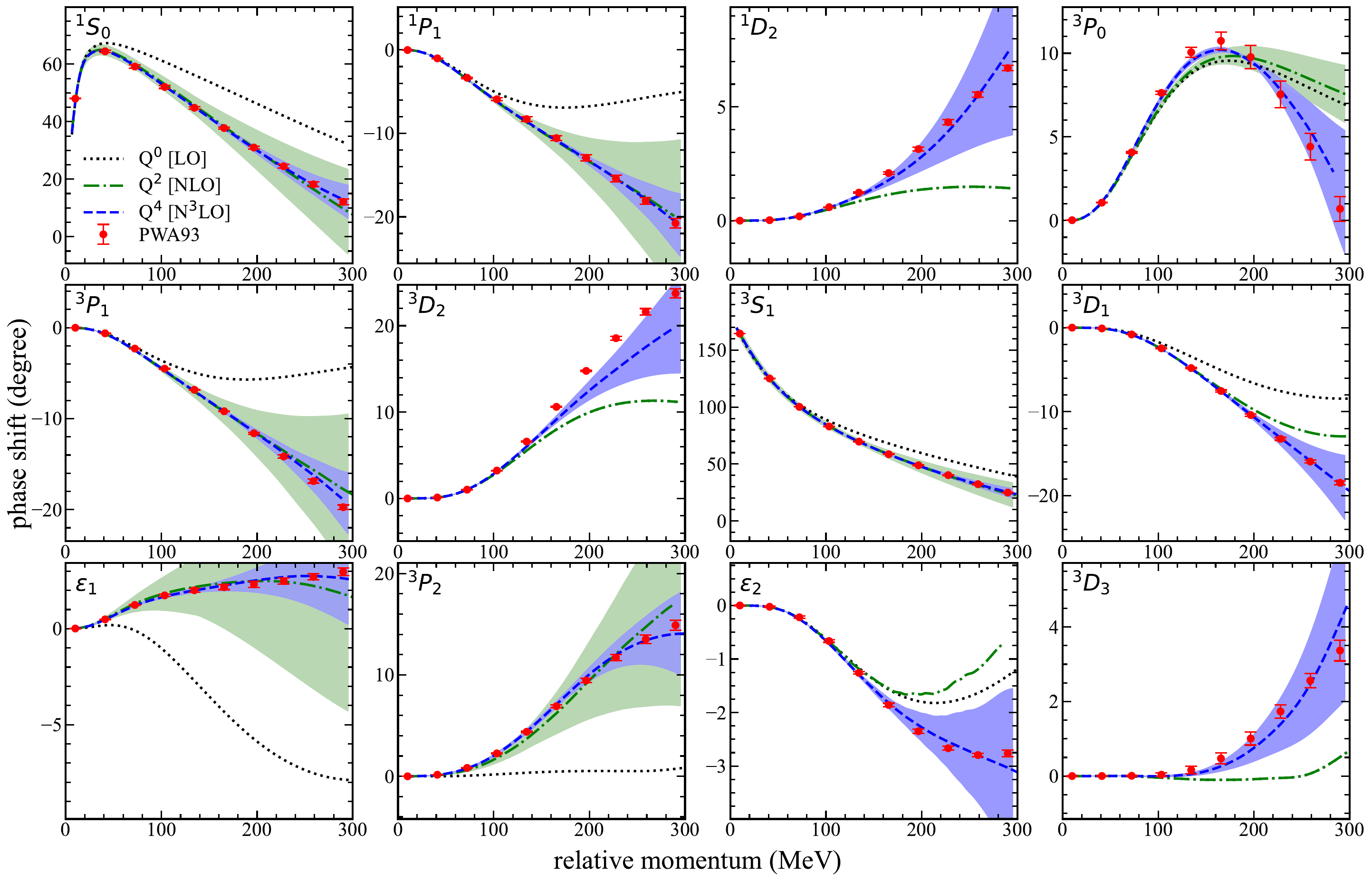}
    \caption{\textbf{N$^3$LO neutron–proton scattering phase shifts.} Neutron-proton scattering phase shifts and mixing angles with error bands, from Ref~\cite{Elhatisari:2024}. \label{fig:N3LO_PhaseShifts}}
\end{figure}

The realistic neutron matter calculations are performed using the wavefunction matching Hamiltonian derived from chiral effective field theory ($\chi$EFT). In particular, we employ a high-fidelity $\chi$EFT interaction constructed up to next-to-next-to-next-to-leading order (N$^3$LO). The two-body low-energy constants (LECs) are determined by fits to two-nucleon scattering data, as shown in Fig.~\ref{fig:N3LO_PhaseShifts}.
Due to the severe ``sign problem'', the direct implementation of this high fidelity Hamiltonian $H$ is prohibited. The wavefunction matching technique provides an elegant solution by introducing a set of unitary transformations to make the transformed Hamiltonian $H'$ as close to some easily computable Hamiltonian $H^{S}$. Then the small gap of $H' - H^{S}$ can be handled with perturbation theory.
The three-body LECs are tuned according to several finite nuclei binding energies, more details can be found in Ref.~\cite{Elhatisari:2024}. 

The anti-symmetry of the two-nucleon system demands that $(-1)^{T+S+L} = -1$, where $T,S,L$ represent the two-nucleon relative isospin, spin and angular momentum. In this work, we consider pure neutron systems $T=1$ with the lowest angular momentum for even and odd parity, thus $L=0,1$. Thus, the relevant channels are $^1S_0$, $^3P_0$, $^3P_1$ and $^3P_2$. In  Fig~\ref{fig:N3LO_PhaseShifts}, strong $^1S_0$ attractive interaction is the origin of s-wave pairing in nuclear systems. The attractive $^3P_0$ and $^3P_2$ also indicate the existing of p-wave pairing. In principle, any repulsive interaction will lead to zero value of irreducible two-body densities $\rho^{II}$. Because we only use 1st order perturbative wavefunction corrections to include high order chiral interactions in $\rho^{II}$, the strong repulsive $^3P_1$ interaction will have negative effects if we measure p-wave pairing as a whole. Thus, according to the discussion of Sec~\ref{sec:rot_proj}, we perform rotation \& projection onto $\rho^{II}$ to obtain the different component of $^3P_0$, $^3P_1$ and $^3P_2$.

\subsection{Benchmark of rotation and projection} 
\begin{figure}[H]
    \centering
    \includegraphics[width=0.78\linewidth]{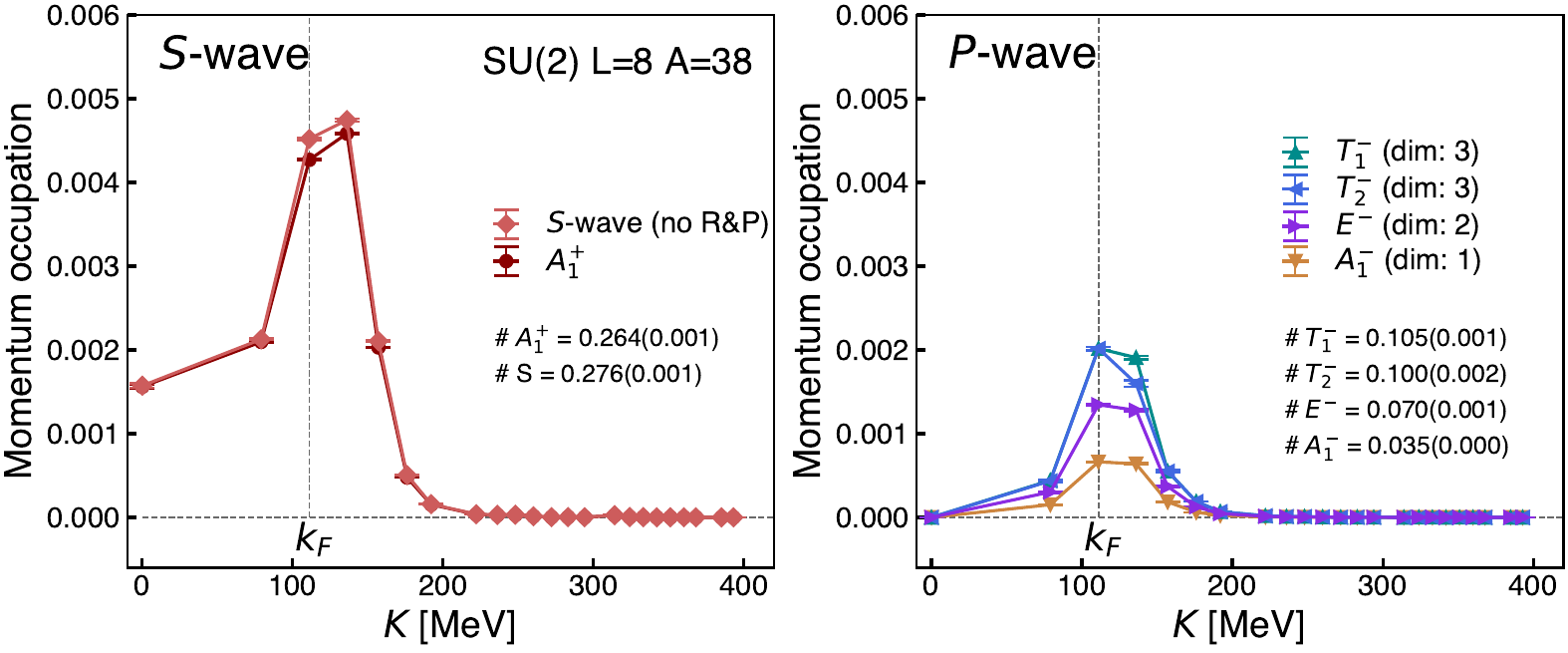}
    \caption{\textbf{Rotation and projection test in the SU(2) system.} Momentum-space s-wave (left) and p-wave (right) pair occupations for the SU(2)-symmetric system with $L=8$ and $A=38$ at $\tau=1.0$ MeV$^{-1}$, obtained from \textit{ab initio} lattice calculations with rotation and projection. Results without rotation and projection are labeled as ``no R\&P''. For the p-wave channel, different irreducible representations of the octahedral group are shown, where ``dim'' denotes the dimension of each representation. The quantities labeled by ``\#'' indicate the summed pair numbers over all momentum modes for each channel.
    \label{fig:SU2_SP_ROT_L8A38}}
\end{figure}

Before performing the calculation with high fidelity chiral interactions, we first benchmark the rotation and projection procedure using the SU(2) Hamiltonian.
In this case, the three p-wave channels ($^3P_0$, $^3P_1$, and $^3P_2$) are identical, as shown in Fig.~\ref{fig:SU2_PhaseShift}.
Therefore, this symmetry must be preserved after applying the rotation and projection.
In Fig.~\ref{fig:SU2_SP_ROT_L8A38}, we present the projected s-wave and p-wave pairing signals for the system with $L^3 = 8^3$ and $A = 38$.
For the s-wave channel, the projected $A_1^+$ result is nearly identical to the original s-wave pairing signal, confirming the consistency of the projection procedure. 
Since the SU(2)-symmetric interaction treats all p-wave channels equivalently, the superfluid contributions in each p-wave direction should be identical. 
In analogy with the SU(2) group, where an angular momentum $J$ multiplet has dimension $2J+1$, the irreducible representations of the octahedral group also carry specific dimensions: $A_1$ is one-dimensional ($x^2+y^2+z^2$), $E$ is two-dimensional ($\{2z^2-x^2-y^2, x^2-y^2\}$), $T_1$ is three-dimensional ($\{x,y,z\}$) and $T_2$ is three-dimensional ($\{xy,yz,zx\}$). 
These representations are consistent with the decomposition of SU(2) angular momentum multiplets. For example, the $^3P_2$ channel decomposes as $E \oplus T_2$, whose total dimension is $2 + 3 = 5$, matching the expected $2J+1=5$ for $J=2$.
If the SU(2) symmetry is preserved, the p-wave signal should should follow the dimensional ratio $A^-_1:E^-:T^-_1:T^-_2=1:2:3:3$. 
Indeed, in the right panel of Fig~\ref{fig:SU2_SP_ROT_L8A38}, we find that the total contributions of $A^-_1, E^-, T_2^-, T^-_1$ are $0.035:0.070:0.100(0.002):0.105$, which are in excellent agreement with a $1:2:3:3$ ratio, thereby confirming that the rotation and projection procedure correctly preserves the underlying SU(2) symmetry. 

With the high fidelity chiral interaction at N$^3$LO level, we carried out two sets of calculation: 1) $L=8, A=38$, the density is $20\%$ of the saturation density $\rho_{\text{0}}=0.15$ fm$^{-3}$; 2)  $L=7, A=66$ where the density is $53\%$ of the saturation density. The particle numbers of $38$ and $66$ are chosen due to the closed shell in momentum lattice.  

\subsection{Realistic neutron matter}
\begin{figure}[H]
    \centering
    \includegraphics[width=0.78\linewidth]{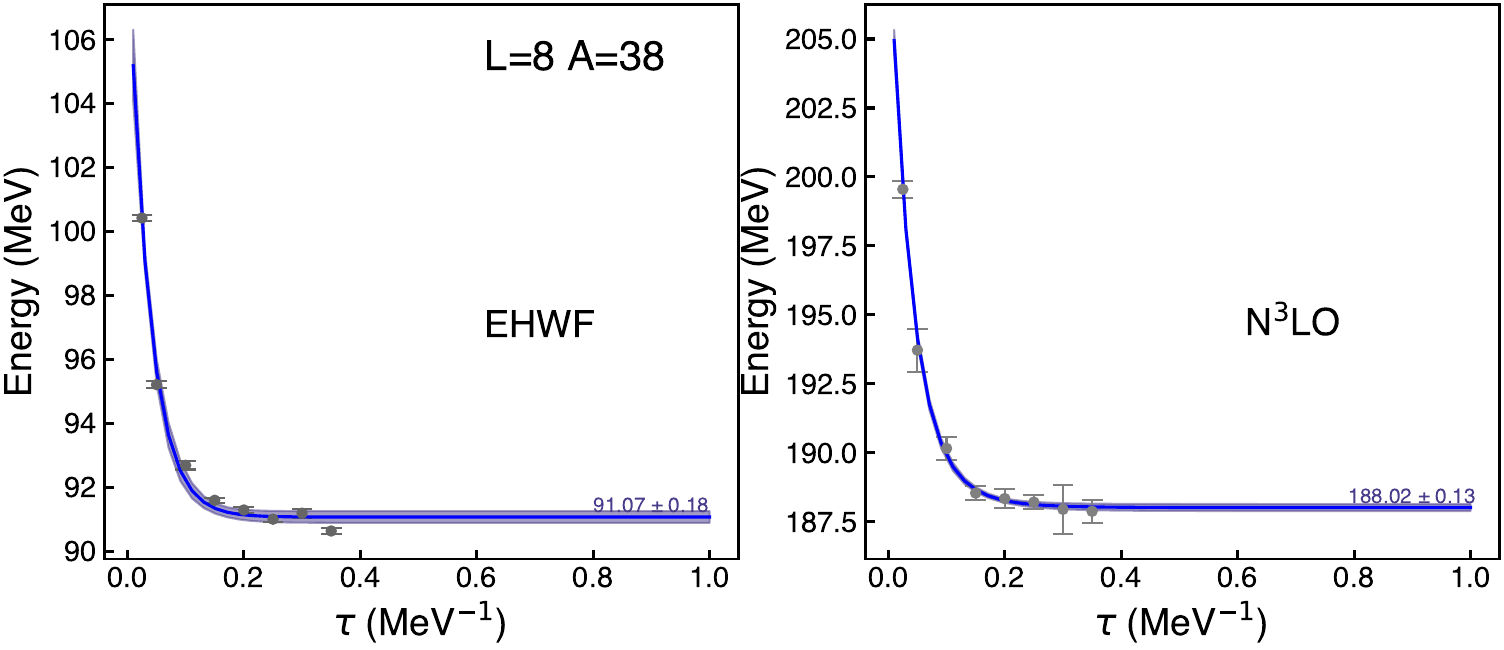}
    \caption{\textbf{Euclidean-time extrapolation of N$^3$LO energies.} Ground-state energy for the $L^3=8^3, A=38$ system as a function of Euclidean time, obtained from \textit{ab initio} lattice calculations. Left: energy obtained with the nonperturbative Hamiltonian $H^{\text{S}}$ (EHWF). Right: energy obtained with the perturbative high-fidelity chiral N$^3$LO Hamiltonian $H'$.
    \label{fig:N3LO_E_L8A38}}
\end{figure}

We first consider a system with $L^3=8^3$, $A=38$, corresponding to a density $\rho = 0.0326$ fm$^{-3}$, which is approximately $20\%$ of the nuclear saturation density $\rho_{\text{0}}=0.16$ fm$^{-3}$. After performing the Euclidean-time extrapolation, we obtain a total energy of $E_{\text{N$^3$LO}}=188.02 (0.13)$ MeV. 
As shown in Fig~\ref{fig:N3LO_E_L8A38}, the energy reaches a plateau starting at $\tau \approx 0.2,\text{MeV}^{-1}$. 
The computation cost, particularly for the perturbative operator measurements, increases significantly with larger Euclidean time $\tau$. 
For the measurement of s-wave, p-wave and quartet signals, we fix $\tau=0.2$ MeV$^{-1}$, which already requires approximately $4\times2048$ GPU node hours on the \textit{Frontier} supercomputer.    

\begin{figure}[H]
    \centering
    \includegraphics[width=0.78\linewidth]{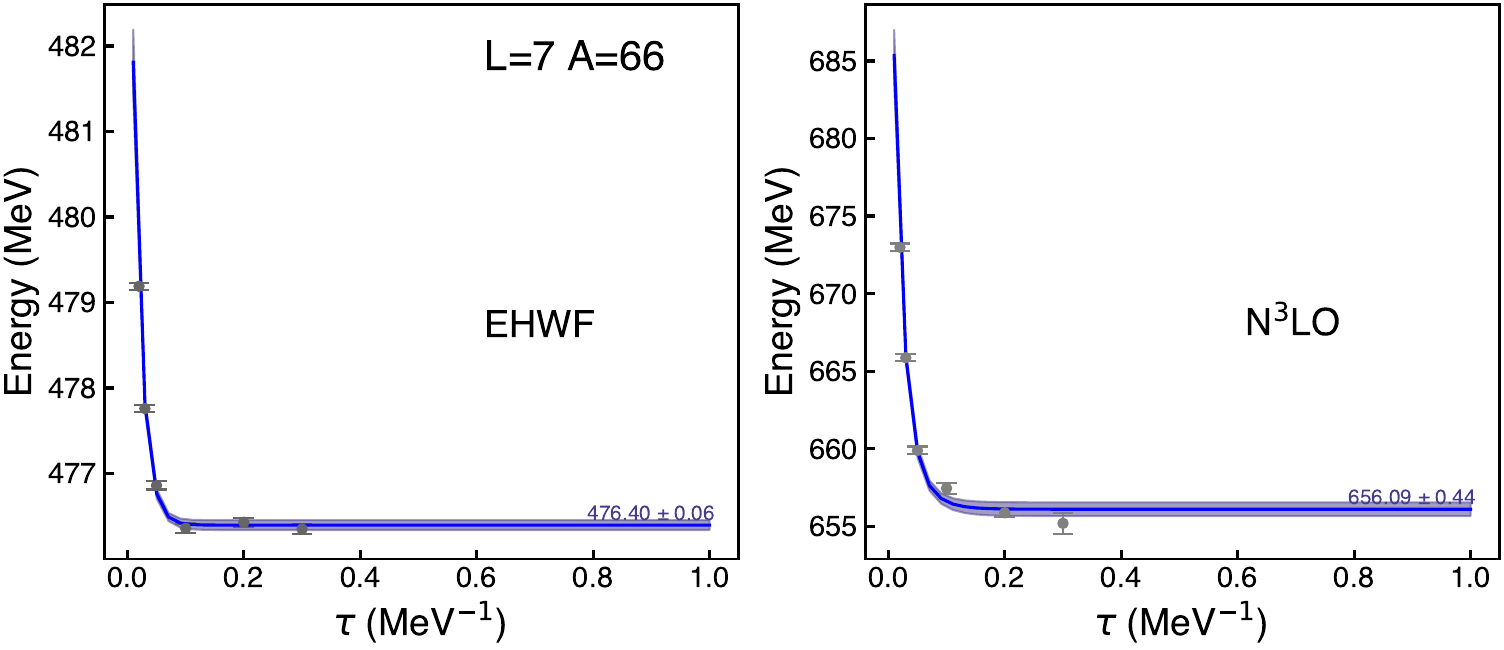}
    \caption{\textbf{Euclidean-time extrapolation of N$^3$LO energies.} Ground-state energy for the $L^3 = 7^3$, $A = 66$ system as a function of Euclidean time, obtained from \textit{ab initio} lattice calculations. Left: energy obtained with the nonperturbative Hamiltonian $H^{\text{S}}$ (EHWF). Right: energy obtained with the perturbative high-fidelity chiral N$^3$LO Hamiltonian $H'$.
    \label{fig:N3LO_E_L7A66}}
\end{figure}

We next consider a system with $L^3 = 7^3$ and $A = 66$, corresponding to a density of $\rho = 0.0845~\text{fm}^{-3}$, which is approximately $53\%$ of the nuclear saturation density $\rho_0 = 0.16~\text{fm}^{-3}$. 
After performing the Euclidean time extrapolation, we obtain a total energy of $E_{\text{N$^3$LO}}=656.09 (0.44)$ MeV. 
As shown in Fig.~\ref{fig:N3LO_E_L7A66}, the energy again exhibits a plateau beginning at $\tau = 0.2~\text{MeV}^{-1}$.

In Table~\ref{tab:En3lo_L8_kf119} and Table~\ref{tab:En3lo_L7_kf269}, we present energies with one, two and four more particles above the closed shell $A=38$ and $A=66$. All energy values are obtained from Euclidean time extrapolation. The s-wave pairing gap, the p-wave pairing gap and the quartet gap are obtained from the three-point formula. Besides the dominant s-wave pairing gaps, positive p-wave pairing gaps for $^3P_0$ and $^3P_2$ channels also be observed, which can be traced back to the attractive p-wave phase shifts for those two channels. As shown in Table~\ref{tab:SU2_O2}, both $E$ and $T_2$ cubic representations on lattice can contribute to the $^3P_2$ channel. The quartet signals are obtained by comparing energy of $A=38$, $^1S_0$ energy of $A=40$ and energy of $A=42$.
The Fermi surface for $L^3=8^3$ $A=38$ is $k_F=119.88$ MeV and  for $L^3=7^3$ $A=66$ is $k_F=269.28$ MeV. Differences in the Fermi surface can explain the difference of pairing gaps, by looking at the phase shifts in Fig~\ref{fig:N3LO_PhaseShifts}.

\begin{table}[htbp]
\centering
\caption{\textbf{Energy signatures of multimodal superfluidity at N$^3$LO.} Ground-state energies and extracted pairing gaps for an $L = 8$ system obtained from \textit{ab initio} lattice calculations with high-fidelity chiral N$^3$LO interactions. The density of closed shell $A=38$ is $\rho = 0.0326$ fm$^{-3}$. Shown are the dominant pairing gaps in the $^1S_0$ and $^3P_J$ channels at $A=40$  , together with the quartet gap at $A=42$.}
\label{tab:En3lo_L8_kf119}
\begin{tabular}{lllll}
\toprule
$A$ & $E_{\text{N$^3$LO}}$ (MeV) & channel  
& 2$\Delta$ (MeV) 
& 4$\Delta_{\text{Q}}$ (MeV) \\
\hline
$38$ & 188.02 (13)  &                 &     & \\
$39$ & 199.41 (39)  &                 &      & \\
$40$ & 205.22 (47)  & $^1S_0$         & 5.58 (92) & \\
$40$ & 209.54 (35)  & $^3P_0$         & 1.25 (86) &  \\
$40$ & 210.20 (81)  & $^3P_1$         & 0.60 (113) &  \\
$40$ & 209.44 (33)  & $^3P_2[T_2^-]$  & 1.36 (86) &  \\
$40$ & 208.64 (42)  & $^3P_2[E^-]$     & 2.16 (90) & \\
$42$ & 218.52 (138) &                 &  & 3.90 (167)\\
\bottomrule
\end{tabular}
\end{table}

\begin{table}[htbp]
\centering
\caption{\textbf{Energy signatures of multimodal superfluidity at N$^3$LO.} Ground-state energies and extracted pairing gaps for an $L = 7$ system obtained from \textit{ab initio} lattice calculations with high-fidelity chiral N$^3$LO interactions. The density of closed shell $A=66$ is $\rho = 0.0845$ fm$^{-3}$. Shown are the dominant pairing gaps in the $^1S_0$ and $^3P_J$ channels at $A=68$ , together with the quartet gap at $A=70$.}
\label{tab:En3lo_L7_kf269}
\begin{tabular}{lllll}
\toprule
$A$ & $E_{\text{N$^3$LO}}$ (MeV) & channel  
& 2$\Delta$ (MeV) 
& 4$\Delta_{\text{Q}}$ (MeV) \\
\hline
$66$ & 655.85 (39)  &                 &     & \\
$67$ & 685.80 (27)  &                 &      & \\
$68$ & 713.15 (36)  & $^1S_0$         & 2.61 (76) & \\
$68$ & 713.80 (68)  & $^3P_0$         & 1.95  (95) &  \\
$68$ & 715.99 (45)  & $^3P_1$         & -0.24 (81) &  \\
$68$ & 713.65 (44)  & $^3P_2[T_2^-]$  & 2.10 (80) &  \\
$68$ & 713.72 (56)  & $^3P_2[E^-]$     & 2.03 (87) & \\
$70$ & 768.43 (26) &                 &  & 2.01 (86)\\
\bottomrule
\end{tabular}
\end{table}

\begin{figure}[H]
    \centering
    \includegraphics[width=0.78\linewidth]{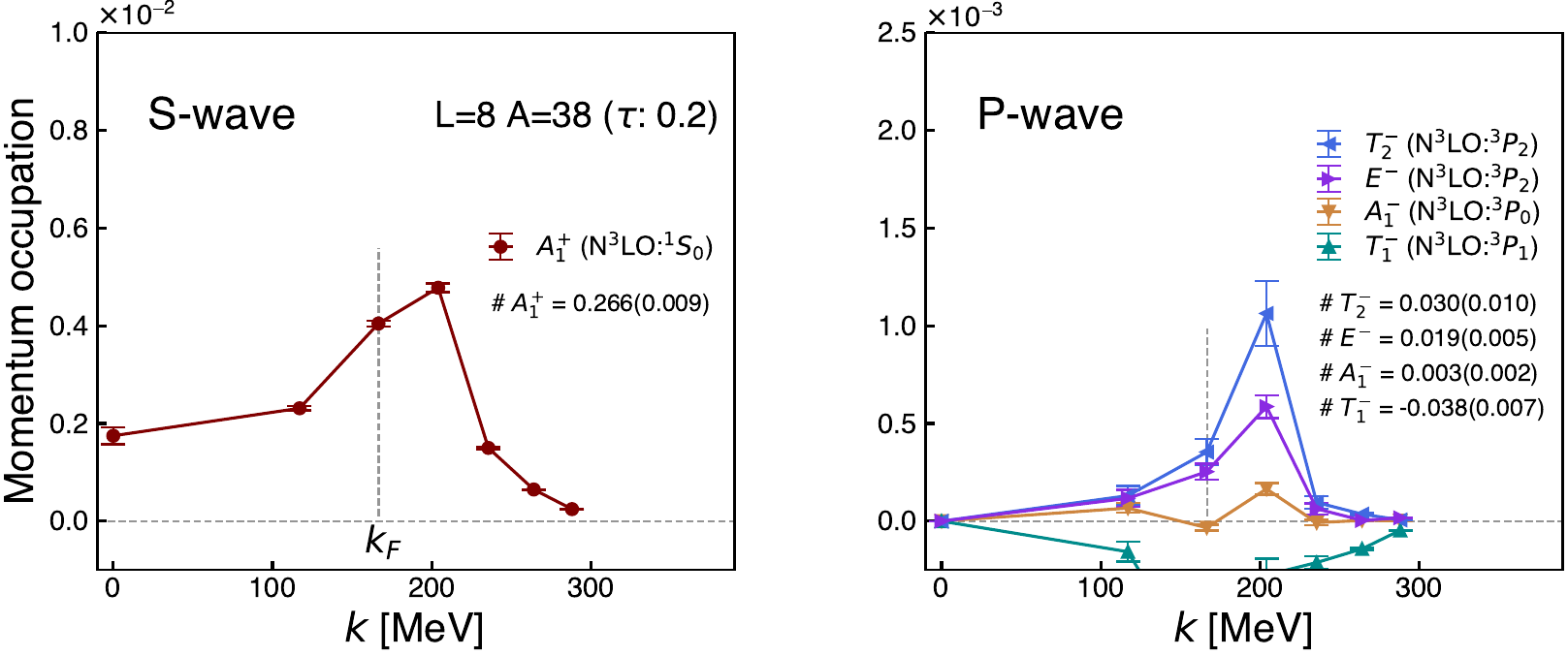}
    \caption{\textbf{Momentum-space pairing at N$^3$LO with rotation and projection for $L=8$.} Momentum-space s-wave (left) and p-wave (right) pair observables for the $L=8, A=38$ system obtained from \textit{ab initio} lattice calculations with the chiral N$^3$LO interaction and rotation and projection. For the p-wave channel, different irreducible representations are shown.
    The quantities labeled by ``\#'' indicate the pair numbers, obtained by summing the pair contributions over all momentum modes for each channel.
    \label{fig:N3LO_SP_ROT_L8A38}}
\end{figure}

In Fig~\ref{fig:N3LO_SP_ROT_L8A38}, we present the projected momentum-space pair occupations in the s-wave and p-wave channels at the N$^3$LO level. Higher-order contributions from the chiral interaction are incorporated by including first-order perturbative corrections to the wavefunction, and further details can be found in Ref~\cite{Lu:2022, Ma:2024}.
From the figure, we clearly observe the coexistence of s-wave and p-wave pairing in cold neutron matter at approximately one-quarter of nuclear saturation density.
To project onto the irreducible representations of the octahedral group, 48 rotational operations must be applied to each momentum pair.
Consequently, the computational cost is increased by a factor of 48 compared to calculations without rotation and projection.
Thus we restrict the momentum $|\vec{k}| < 300$ MeV to reduce computational cost. 
Both s-wave and p-wave pairing signals increase with momentum, reaching a peak slightly above the Fermi surface before decreasing at higher momenta. 
The last data point is already relatively small, for both channels.
The peak position does not coincide exactly with the Fermi surface due to lattice discretization effects. 
As shown in the right figure of Fig~\ref{fig:SU2_L10A66_1B}, the lattice degeneracy factor $N_{\text{latt}}$ for the fourth momentum shell is smaller than that of the third shell, resulting in a larger ratio $\rho^{II}(k)/N_{\text{latt}}$.
According to the mapping between the SU(2) group and the octahedral group (shown in Table~\ref{tab:SU2_O2}), the positive parity $A^+_1$ corresponding to the $^1S_0$ scattering channel (shown in Fig~\ref{fig:N3LO_PhaseShifts}). Therefore, the observed s-wave pairing can be traced back to the strong attractive interaction in the $^1S_0$ channel.

Compared to the s-wave pairing signal, the p-wave pairing is about five times smaller in magnitude but exhibits a more complicated structure. 
In the legend of the p-wave plots, the mapping between SU(2) channels and octahedral representations is indicated as $^3P_0: A^-_1$ , $^3P_1: T^-_1$  and $^3P_2: T^-_2, E^-$. 
In contrast to Fig~\ref{fig:SU2_SP_ROT_L8A38}, where SU(2) symmetry is preserved, the N$^3$LO interaction explicitly breaks this symmetry. 
The most pronounced difference appears in the $T_1^-$ component, which becomes negative. This behavior originates from the repulsive nature of the $^3P_1$ interaction in chiral potentials.
In principle, repulsive interaction will lead to zero irreducible two-body densities not negative values. However, the first-order perturbation only keeps linear terms which will cause the negative values. This can be seen from a simple example: considering a function $f(x) = -x^2+1$, the first-order approximation at $x=0$ is linear $\bar{f}(x)=-2x+1$ and the prediction at $x=1$ thereby overshoots to $-1$.  
Furthermore, the $^3P_2$ channel exhibits a larger contribution than the $^3P_0$ channel.
This is because particle pairs with opposite momenta near the Fermi surface have large relative momenta, where the $^3P_2$ interaction is more attractive than the $^3P_0$ interaction.

\begin{figure}[H]
    \centering
    \includegraphics[width=0.78\linewidth]{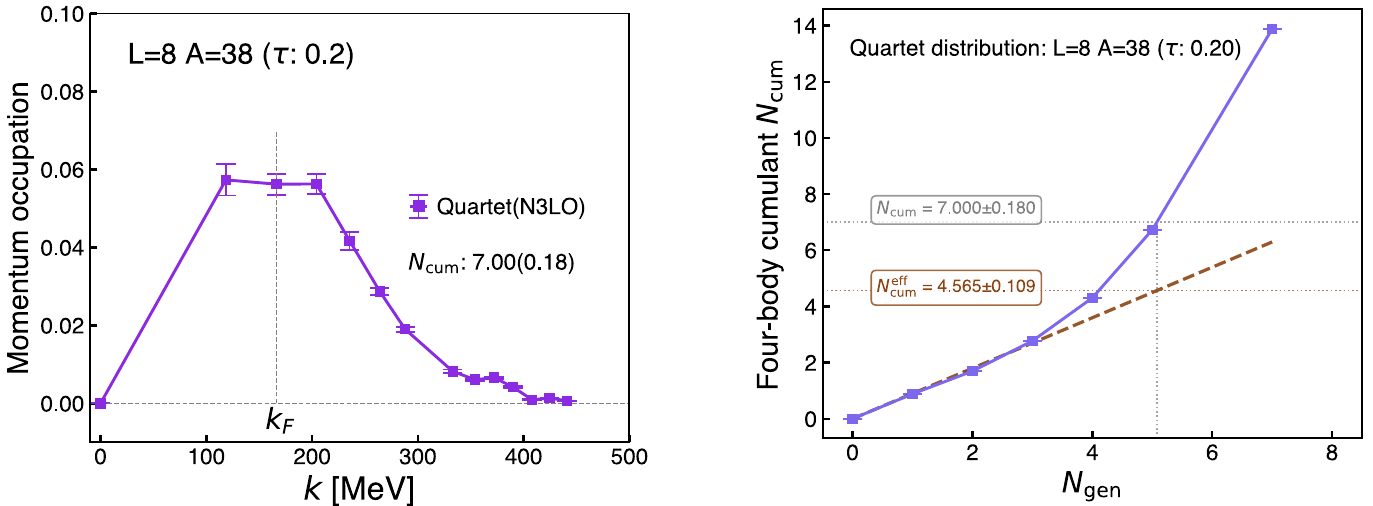}
    \caption{\textbf{Quartet correlations at N$^3$LO for $L=8$.} Left: momentum-space irreducible quartet distribution for the $L=8, A=38$ system obtained from \textit{ab initio} lattice calculations with the chiral N$^3$LO interaction, where  ${N_{\text{cum}}}$ denotes the sum of all four-body cumulants. Right: numerical correlation between ${N_{\text{cum}}}$ and the generated quartet number $N_{\text{gen}}$ obtained from random Monte Carlo sampling. The dashed line indicates the linear contribution of Eq~\ref{eq:Ncum_Ngen} and $N^{\text{eff}}_{\text{cum}}$ denotes the estimated quartet number in the system.
    \label{fig:N3LO_QT_L8A38}}
\end{figure}

For the irreducible quartet measurement, we use the momentum pinhole method + perturbation theory as discussed before. After generating enough pinhole configurations, we can perform statistics afterwards, which allows us to reach higher momentum modes.
In Fig~\ref{fig:N3LO_QT_L8A38} we can see that the irreducible quartet signal has a peak at the Fermi surface and a wide range of distribution. When summing up all the 1211808 $\{k_1, k_2, k_3, k_4\}$ configurations, we find the total quartet number is $4.57(0.11)$ which indicates a large fraction of particles $4.57/(38/4)=48\%$ contribute to the quartet correlations.

\begin{figure}[H]
    \centering
    \includegraphics[width=0.78\linewidth]{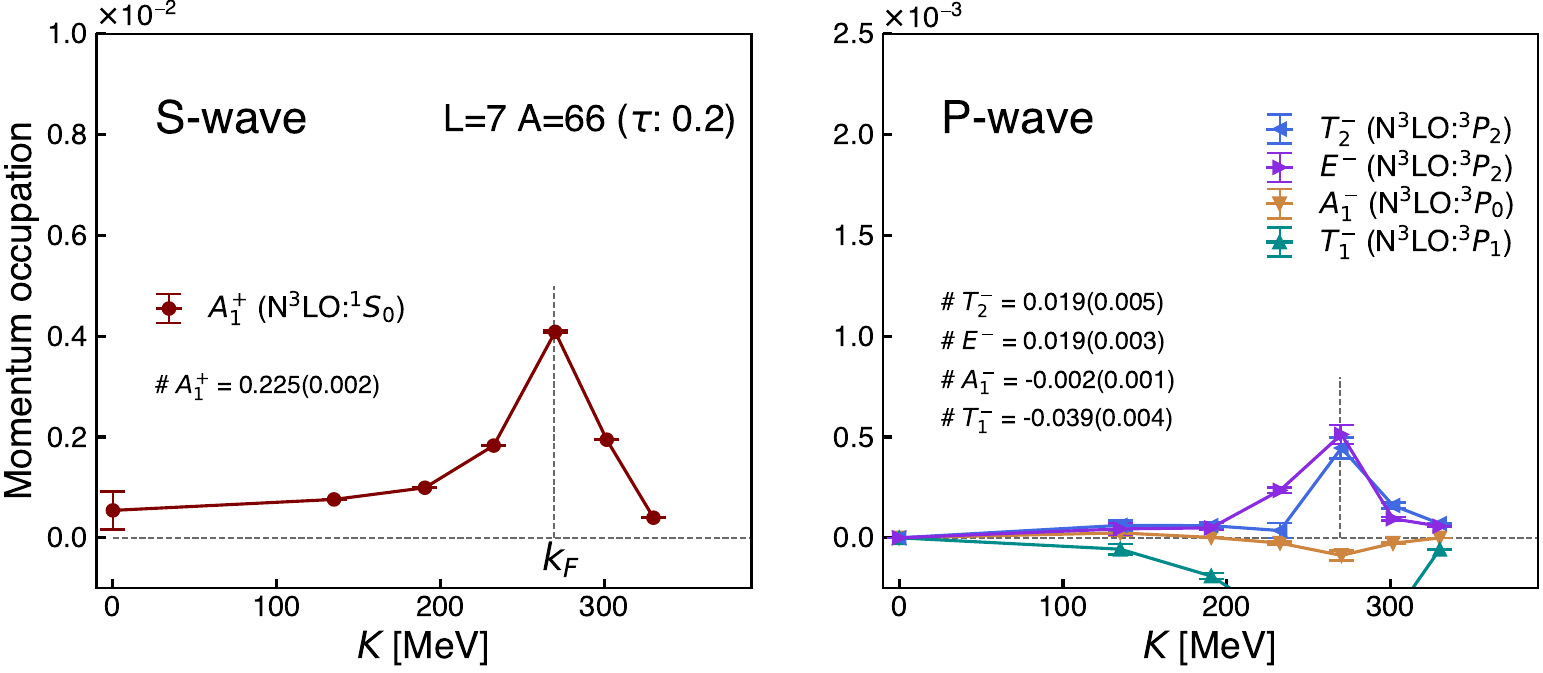}
    \caption{\textbf{Momentum-space pairing at N$^3$LO with rotation and projection for $L=7$.} Momentum-space s-wave (left) and p-wave (right) pair observables for the $L=7, A=66$ system obtained from \textit{ab initio} lattice calculations with the chiral N$^3$LO interaction and rotation and projection. For the p-wave channel, different irreducible representations are shown.
    The quantities labeled by ``\#'' indicate the pair numbers, obtained by summing the pair contributions over all momentum modes for each channel. \label{fig:N3LO_SP_ROT_L7A66}}
\end{figure}

\begin{figure}[H]
    \centering
    \includegraphics[width=0.78\linewidth]{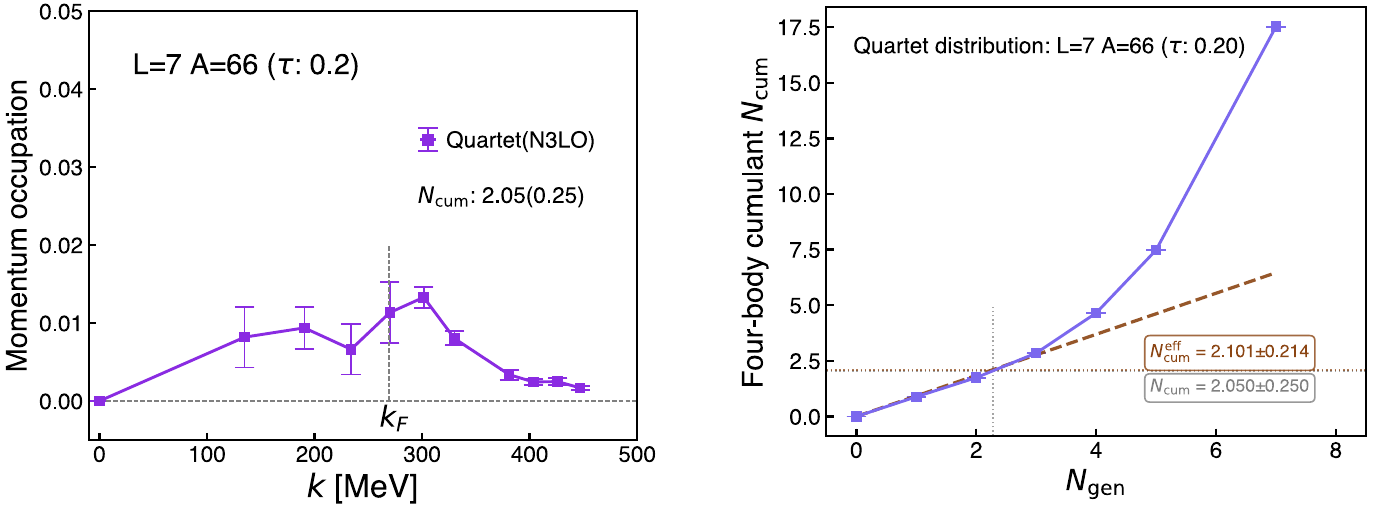}
    \caption{\textbf{Quartet correlations at N$^3$LO for $L=7$.} Left: momentum-space irreducible quartet distribution for the $L=7, A=66$ system obtained from \textit{ab initio} lattice calculations with the chiral N$^3$LO interaction, where  ${N_{\text{cum}}}$ denotes the sum of all four-body cumulants. Right: numerical correlation between ${N_{\text{cum}}}$ and the generated quartet number $N_{\text{gen}}$ obtained from random Monte Carlo sampling. The dashed line indicates the linear contribution of Eq~\ref{eq:Ncum_Ngen} and $N^{\text{eff}}_{\text{cum}}$ denotes the estimated quartet number in the system. \label{fig:N3LO_QT_L7A66}}
\end{figure}

In Fig~\ref{fig:N3LO_SP_ROT_L7A66} and Fig~\ref{fig:N3LO_QT_L7A66}, we present the superfluidity signal of s-wave, p-wave and quartets for a $L^3=7^3, A=66$ system with chiral N$^3$LO interactions. 
Due to the higher density and larger Fermi momentum, all the superfluid signals are weaker than those in the $L^3=8^3, A=38$ system, consistent with the energy-gap trends reported in Tables~\ref{tab:En3lo_L8_kf119} and \ref{tab:En3lo_L7_kf269}.
While the positive $^3P_0$ gap appears in the energy analysis, it remains difficult to identify directly in the momentum pair measurements because of current computational limitations.

\subsection{Connection to other nuclear \textit{ab initio} methods}
To place the present lattice formulation in a broader context, we briefly summarize other major nuclear \textit{ab initio} frameworks that have been developed over the past two decades. Much of this progress has been driven by the development and systematic improvement of nuclear forces derived from chiral effective field theory~\cite{Weinberg:1990, Weinberg:1991, Kolck:1994, Epelbaum:2008ga, Machleidt:2011, Entem:2017, Hammer:2020, Epelbaum:2020}. The inclusion of higher-order two-nucleon interactions~\cite{Epelbaum:2008ga, Machleidt:2011, Entem:2017, Ekstrom:2015} together with consistent three-nucleon forces~\cite{Kolck:1994, Navratil:2007, Roth:2011, Hebeler:2020ocj} has enabled increasingly accurate descriptions of nuclear systems. In practical calculations, short-range repulsion is often softened through renormalization techniques~\cite{Bogner:2006pc, Bogner:2010}.

In addition to the \textit{ab initio} lattice method, a broad spectrum of complementary \textit{ab initio} methods have been developed for few- and many-nucleon systems. Large-scale configuration interaction approaches such as the no-core shell model (NCSM)~\cite{Bruce:2013, Roth:2014, Jurgenson:2013, LENPIC:2022cyu}, symmetry-adapted NCSM~\cite{Dytrych:2020, Dreyfuss:2020}, and various quantum Monte Carlo techniques~\cite{Carlson:2015, Pastore:2017uwc, Lynn2019, Gandolfi:2020pbj, Schiavilla:2021dun, Jin:2025} have extended calculations into the lower $sd$ shell.
Controlled truncation schemes have also led to computationally efficient many-body frameworks, including self-consistent Green’s functions~\cite{Soma:2020xhv}, many-body perturbation theory~\cite{Roth:2010, Tichai:2016}, coupled-cluster theory~\cite{Hagen:2013nca}, the in-medium similarity renormalization group (IMSRG)~\cite{Hergert:2016}, and the \textit{ab initio} no-core Monte Carlo shell model~\cite{Abe:2021sky, Otsuka:2022bcf}. Effective valence-space Hamiltonians can furthermore be derived using open-shell MBPT~\cite{Morten:1995, Coraggio:2009, Tichai:2018}, valence-space IMSRG~\cite{Bogner:2014, Stroberg:2016ung}, or shell-model coupled-cluster techniques~\cite{Jansen:2014,Sun2018}.
To incorporate continuum effects in weakly bound nuclei, a number of approaches have been developed, including the no-core shell model with continuum (NCSMC)~\cite{Navratil:2016} and complex coupled-cluster theory~\cite{Hagen2012}. 
The Gamow shell model (GSM), originally introduced in Ref.~\cite{Michel:2002}, has since been extended to no-core implementations~\cite{Papadimitriou:2013, Li:2019pmg}, to IMSRG implementations~\cite{Hu:2019}, and to formulations with realistic interactions~\cite{Sun:2017,Ma:2020}. More recently, related treatments of continuum dynamics have been explored using emulator-based frameworks~\cite{ZhangXL:2025}. Beyond finite nuclei, \textit{ab initio} descriptions of infinite neutron matter and symmetric nuclear matter have been achieved with lattice formulations similar to the one presented here~\cite{Ma:2024} as well as continuum Quantum Monte Carlo (QMC) methods~\cite{Gandolfi:2015jma,Tews:2025},MBPT techniques~\cite{alp:2025} and SCGF formulations~\cite{ding:2016}.

%%%%%%%%%%%%%%%%%%%%%%%%%%%%%%%%%%%%%%%%%%%%%%

\section{Bardeen-Cooper-Schrieffer Theory Calculations}

The mean-field description of fermionic superfluidity corresponds to the Bardeen Cooper Schrieffer (BCS) theory which describes the ground state of a superfluid as a condensation of pairs. The symmetries carried by the pair wavefunctions often lend the name to the type of superfluidity, e.g., a singlet superfluid is described as a condensation of pairs in spin-singlet states. In this section we present a brief overview of the BCS theory of singlet and triplet superfluidity and focus on its application to the a spin-balanced system and a fully polarized system. Finally, we review the mean-field description of mixed spin, i.e., singlet and triplet, superfluidity.

The general Hamiltonian of a system of $N$ particles in a volume $\Omega$ interacting via a two-body spin-independent potential $V(\mf{r},\mf{r}')$, formulated in a grand canonical ensemble, is
\begin{align}
H&=\sum_{\alpha\mfk} (\epsilon_{\mfk}-\mu_\sigma) c_{\mfk\sigma}^\+c_{\mfk\sigma}+\frac{1}{2\Omega}\sum_{\substack{\mfk\mfk'\mf{q}\\ \sigma\sigma'}} V_{\sigma\sigma'}(\mfk,\mfk';\mfq) p^\+_{\mfk\sigma\sigma';\mf{q}}p_{\mfk'\sigma\sigma';\mf{q}} \label{eq:gen_ham}
\end{align}
where the operator $c_{\mfk\alpha}^\+$ ($c_{\mfk\alpha}$) creates (annihilates) a particle with momentum $\mfk$ and spin $z$-projection $\alpha$ and $p_{\mfk\alpha\beta;\mf{q}} = c_{\mf{q}/2-\mfk \alpha}c_{\mf{q}/2+\mfk \beta}$.
The single-particle dispersion relation is assumed the same for both species (i.e., spin projections) and equal to that in free-space, $\epsilon_{\mfk}=\hslash^2 \mfk^2/2m$, and $\mu_\sigma$ is the chemical potential of species $\sigma$. The interaction matrix element can be separated in orbital angular momentum channels
\begin{align}
    V_{\sigma\sigma'} (\mfk,\mfk';\mfq) = \bra{\frac{\mfq}{2}-\mfk \sigma; \frac{\mfq}{2}+\mfk \sigma'}V\ket{\frac{\mfq}{2}-\mfk' \sigma; \frac{\mfq}{2}+\mfk' \sigma'}
\end{align}
In its simplest form, the BCS theory assumes pairs with vanishing centre-of-mass momentum and so from the sum in Eq.~(\ref{eq:gen_ham}) only the $\mf{q}=0$ terms are kept. Hence we focus on the pair operators
$p_{\mfk\alpha\beta}=p_{\mfk\alpha\beta;\mf{q}=0}$ and the interaction matrix elements:
\begin{align}
    V_{\sigma\sigma'} (\mfk,\mfk') &= \bra{-\mfk \sigma; \mfk \sigma'}V\ket{-\mfk' \sigma; \mfk' \sigma'} \\
    &=\frac{(4\pi)^2}{\Omega}\sum_{l m} Y_{l m}(\hat{\mfk})V_{lm}(k,k') Y^\star_{lm}(\hat{\mfk}') \\
    V_{\up\dn} (\mfk,\mfk')& =  \frac{(4\pi)^2}{\Omega} V_{00}(k,k') Y_{00}(\hat{\mfk})Y^\star_{00}(\hat{\mfk}') + \cdots \\
    V_{\sigma\sigma} (\mfk,\mfk')& =  \frac{(4\pi)^2}{\Omega} V_{1}(k,k')\sum_{m=-1,0,1} Y_{1m}(\hat{\mfk})Y^\star_{1m}(\hat{\mfk}') + \cdots \label{eq:v1} \\
    V_l(k,k') &= \int dr r^2 j_l(kr) V(r) j_{l}(k'r)
\end{align}
and $j_l$ is a spherical Bessel function.
The pair operators have non-zero expectation value only on a paired state $\ket{\psi}$, i.e.,$
f_{\sigma\sigma'}(\mfk) = \bra{\psi}p_{\mfk\alpha\beta}\ket{\psi}\neq 0$.
The mean-field paired ground state is determined by defining the gap functions $\Delta_{\sigma\sigma'}(\mfk) = \sum_{\mfk'} V_{\sigma\sigma'}(\mfk,\mfk') f_{\mfk' \sigma\sigma'}$
and diagonalizing the mean-field Hamiltonian defined by bi-linearizing Eq.~(\ref{eq:gen_ham}) in $\Delta_{\sigma\sigma'}$,
\begin{align}
H_{\textrm{MF}}=\sum^{'}_{\mfk} \psi(\mfk)^\+ \left(\begin{array}{cc}
\xi(\mfk) I_{\sigma\sigma'} & \Delta_{\sigma\sigma'}(\mfk) \\ \Delta^*_{\sigma\sigma'}(\mfk) & -\xi(\mfk) I_{\sigma\sigma'} 
\end{array}\right)
\psi(\mfk)~,
\end{align}
where we also defined the Nambu vector
\begin{align}
\psi(\mfk) =\left(\begin{array}{c}
c_{\mfk\up} \\
c_{\mfk\dn} \\
c^\+_{-\mfk\up}\\
c^\+_{-\mfk\dn}
\end{array}\right)~,\quad \psi^\+ = \left(c^\+_{\mfk\up},c^\+_{\mfk\dn}, c_{-\mfk\up},c_{-\mfk\dn}\right)~.
\end{align}
Note that for weakly-coupled systems, the gap function is approximated by a constant that is its value close to the Fermi surface. However, in the strongly-coupled regime where cold-atomic and nuclear systems reside, the gap function's momentum-dependence must be retained.

Writing $H_{\rm MF}$ explicitly and diagonalizing it, we get the pair wavefunctions
\begin{align}
\left(\begin{array}{cccc}
\xi(\mfk) & 0 & \Delta_{\up\up}(\mfk) & \Delta_{\up\dn}(\mfk) \\
0 & \xi(\mfk) & \Delta_{\dn\up}(\mfk) & \Delta_{\dn\dn}(\mfk) \\
\Delta^*_{\up\up}(\mfk) & \Delta^*_{\dn\up}(\mfk) & -\xi(\mfk) & 0 \\
\Delta^*_{\up\dn}(\mfk) & \Delta^*_{\dn\dn}(\mfk) & 0& -\xi(\mfk)
\end{array}\right)  \label{eq:bcs_ham}
\left(\begin{array}{c}
u_{\up n}(\mfk) \\
u_{\dn n}(\mfk) \\
v_{\up n}(\mfk) \\
v_{\dn n}(\mfk) \\
\end{array}\right) = E_{\mfk n} \left(\begin{array}{c}
u_{\up n}(\mfk) \\
u_{\dn n}(\mfk) \\
v_{\up n}(\mfk) \\
v_{\dn n}(\mfk) \\
\end{array}\right)
\end{align}
In the absence of a mismatch in the species' Fermi surfaces, the eigenvalues of $H_\textrm{MF}$ correspond to the quasiparticle energies:
\begin{align}
{E^2}_{1(2)}(\mfk) &= \xi^2(\mfk) +\frac{1}{2}\sum_{\sigma\sigma'}\left|\Delta_{\sigma\sigma'}(\mfk)\right|^2 +(-) \sqrt{\left(\frac{1}{2}\sum_{\sigma\sigma'}\left|\Delta_{\sigma\sigma'}(\mfk)\right|^2\right)^2-\left|\Delta_{\dn\up}(\mfk)\Delta_{\up\dn}(\mfk)-\Delta_{\up\up}(\mfk)\Delta_{\dn\dn}(\mfk)\right|^2} 
\end{align} 
The eigenvectors provide the Bogolyubov transformation that defines the quasiparticle operators 
\begin{equation}
b_{\mfk n} = \sum_{\sigma}[u^*_{\sigma n}(\mfk) c_{\mfk \sigma} - v^*_{\sigma n}(\mfk) c^\+_{-\mfk \sigma}]    
\end{equation}
and the BCS ground state as the vacuum of the quasiparticle operators~\cite{vollhardt:book}:
\begin{align}
\ket{\psi} = \prod_{\mfk}^{'} \prod_{\sigma} \left(u_{\sigma\sigma}(\mfk) + \sum_{\sigma'} v_{\sigma\sigma'}(\mfk) p^\+_{\mfk\sigma\sigma'}\right)\ket{0}~,
\end{align}
where the product over $\mfk$ runs over half the momentum space to avoid double counting. Finally, in this parametrization, the pair expectation values are $f_{\sigma\sigma'}(\mfk) = v_{\sigma\sigma'}(\mfk)u^*_{\sigma\sigma}(\mfk)$ and the number density of particles of species $\alpha$ is $n_{\sigma}(\mfk) =  \sum_{\sigma'} v_{\sigma\sigma'}(\mfk)v^*_{\sigma\sigma'}(\mfk)$

\subsection{A low-momentum approximation to the NLEFT potentials}
\begin{figure}
\centering
\begin{subfigure}{.5\textwidth}
  \centering
  \includegraphics[width=.8\linewidth]{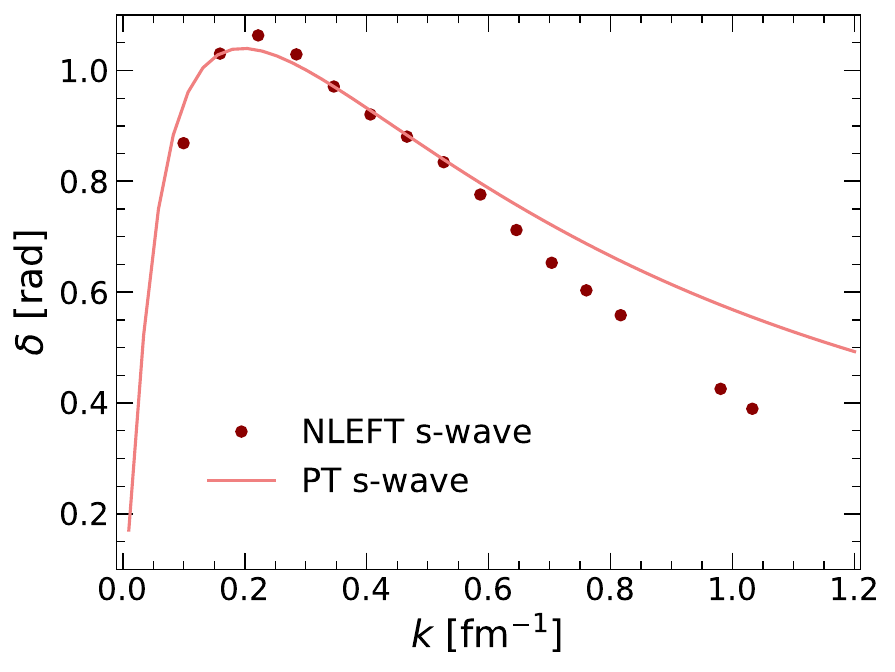}
\end{subfigure}%
\begin{subfigure}{.5\textwidth}
  \centering
  \includegraphics[width=.9\linewidth]{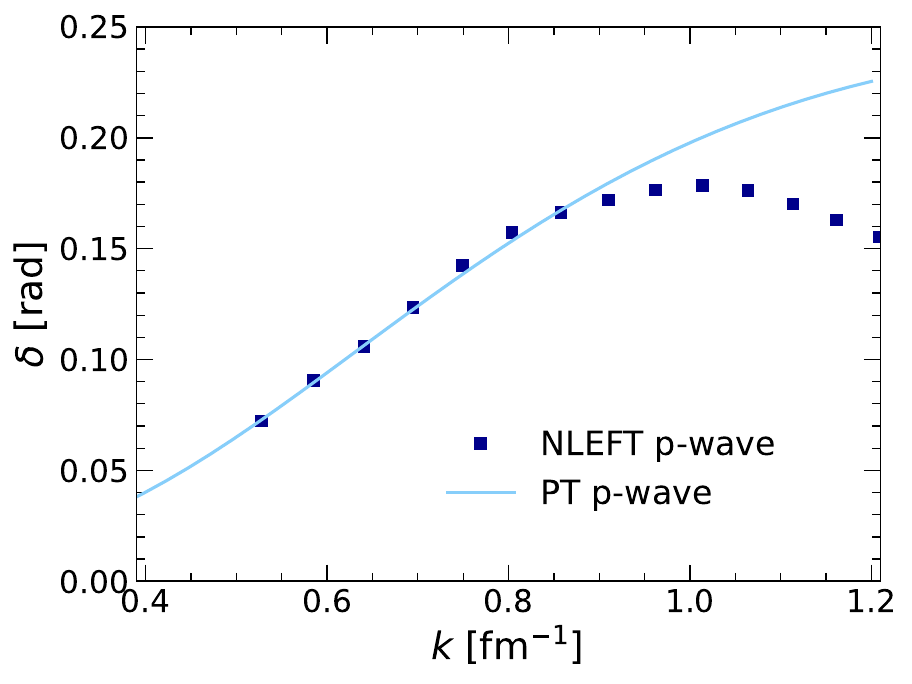}
\end{subfigure}
\caption{\textbf{The low-momentum approximation the NLEFT phase-shifts.} The s-wave and p-wave phase-shifts of the coordinate potential used to match the NLEFT potential in each channel.
\label{fig:phase_shifts_pt}}
\end{figure}

We can tune a simplified potential to reproduce the low-momentum behavior of the potentials used in NLEFT. Given a tunable form, and using an effective range expansion, we can match the low-momentum scattering properties of the potential to reproduce those of the potential used by NLEFT. We use a modified P{\"o}schl-Teller potential,
\begin{align}
    V(r) = \frac{\hslash^2}{m} \frac{\beta^2\lambda(\lambda-1)}{\cosh^2(\beta r)}~,
\end{align}
where $\lambda$ and $\beta$ are tunable parameters, tuned to match the scattering length, $a^{(l)}$ and effective range, $r_e^{(l)}$ of the effective range expansion
\begin{align}
    k^{2l+1}\cot[\delta_l(k)] = -\frac{1}{a^{(l)}} + \frac{1}{2}r_{\textrm{e}}^{(l)}k^2 +\mathcal{O}(k^4)~.
\end{align}
For the NLEFT potentials used for the 3D GAE Hubbard Model, we tune one set of parameters $(\lambda,\beta)$ for each of the S-wave and the P-wave channels, yielding the potentials with phase shifts plotted in Fig.~\ref{fig:phase_shifts_pt}.

%\begin{figure}
%    \centering
%    \includegraphics[width=0.5\linewidth]{Figures/kfre_gapIepern__0.pdf}
%    \caption{\label{fig:kf_gapIepern__0}}
%\end{figure}
%\begin{figure}
%    \centering
%    \includegraphics[width=0.5\linewidth]{Figures/kfre_gapIepern__1.pdf}
%    \caption{\label{fig:kf_gapIepern__1}}
%\end{figure}

\subsection{Spin-balanced system with s-wave interactions}
In a spin-balanced system with only s-wave interactions only s-wave pairs can appear due to Pauli's exclusion principle. The gap equations can be derived by setting $\Delta_{\up\up}(\mfk)=\Delta_{\dn\dn}(\mfk)=0$ in Eq.~(\ref{eq:bcs_ham}) yielding
\begin{align}
H^{(s)}_{MF}=\left(
\begin{array}{cccc}
    \xi(\mfk) & 0 & 0& \Delta(\mfk) \\
0 & \xi(\mfk) & \Delta(\mfk) & 0 \\
0 & \Delta^*(\mfk) & -\xi(\mfk) & 0 \\
\Delta^*(\mfk) & 0 & 0& -\xi(\mfk)
\end{array}\right) 
\end{align}
This is the textbook case of superfluid pairing leading to the gap equations~\cite{tinkham:book}
\begin{align}
    \Delta(\mfk) &= -\frac{1}{2}\sum_{\mfk'} V(\mfk,\mfk') \frac{\Delta(\mfk')}{E(\mfk')}\label{eq:s_gap1} \\
    \mv{N} &=  \sum_\mfk \left(1-\frac{\xi(\mfk)}{E(\mfk)}\right)\label{eq:s_gap2}
\end{align}
where $E(\mfk)=\sqrt{\xi^2(\mfk)+\Delta^2(\mfk)}$ is the quasiparticle excitation spectrum and we have included the equation for the average particle number often referred to as the second gap equation. Equations (\ref{eq:s_gap1}) and (\ref{eq:s_gap2}) are often projected on the spherical components of the gap function as,
\begin{align}
    \Delta(\mfk) = \sum_{lm}\sqrt{\frac{4\pi}{2l+1}}\Delta_{lm}(k) Y_{lm}(\hat{\mfk})~. \label{eq:gap_exp}
\end{align}
For pure s-wave interactions, we retrieve a radial gap equation for the s-wave amplitude of the gap function, $\Delta_0(k)$
\begin{align}
    \Delta_0(k) &= -\frac{1}{\pi}\int dk' {k'}^2 V(k,k') \frac{\Delta_0(k')}{E(k')}~.
\end{align}
The order parameter of the phase is then defined as $\Delta_S=\Delta_0(k_F)$, also known as simply the pairing gap, with $k_F=(3\pi^2 \rho)^{1/3}$, or equivalently as the minimum of the quasiparticle excitation spectrum $E(k)$.

\begin{figure}
\centering
\begin{subfigure}{.5\textwidth}
  \centering
  \includegraphics[width=.8\linewidth]{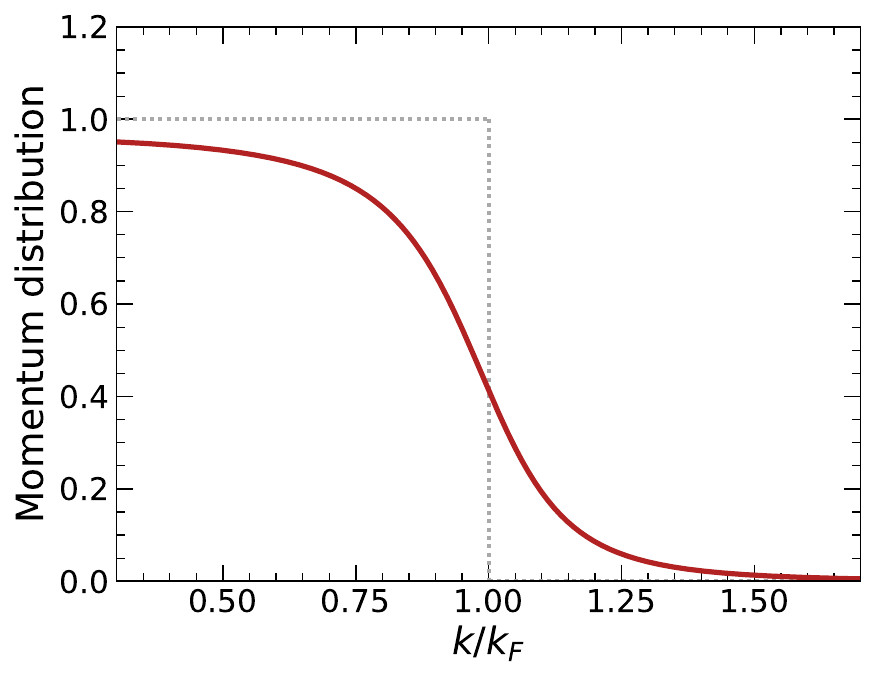}
\end{subfigure}%
\begin{subfigure}{.5\textwidth}
  \centering
  \includegraphics[width=.85\linewidth]{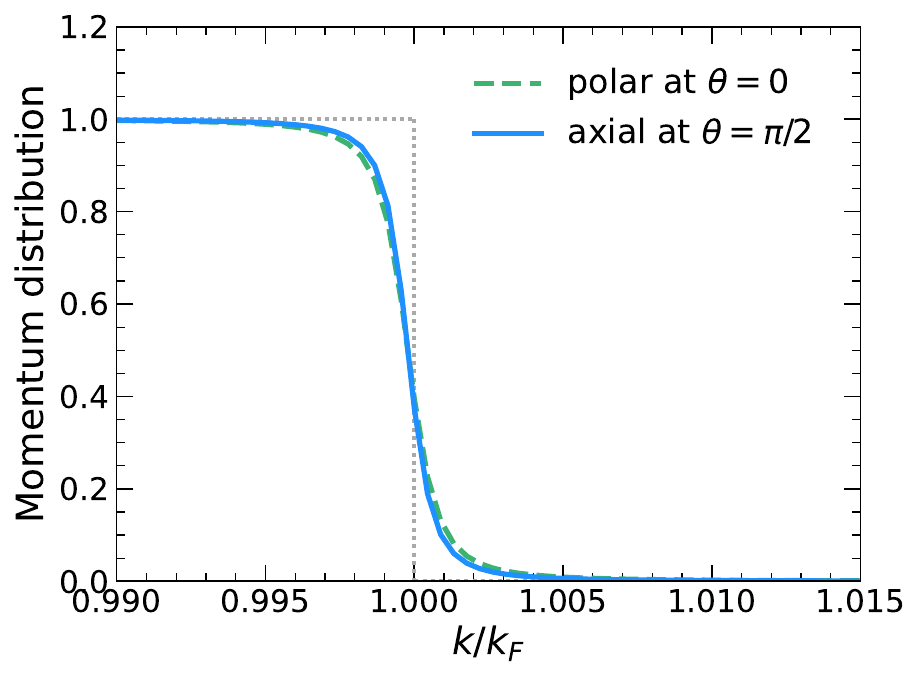}
\end{subfigure}
\caption{\textbf{The s- and p-wave momentum distributions from BCS.} The left panel shows the quasiparticle momentum distribution $v^2(k)$ of the S-wave paired spin-balanced system, at $k_\textrm{F}\approx 0.45~\textrm{fm}^{-1}$. The right panel shows the momentum distribution $v^2(k)$ for the p-wave superfluid phases of the spin-polarized system, at $k_\textrm{F}\approx 1.39~\textrm{fm}^{-1}$, and at $\theta=0$ for the polar phase and at $\theta=\pi/2$ for the axial phase.}
\label{fig:k_mom}
\end{figure}

In this framework we can analyze the mean-field properties of a balanced s-wave superfluid driven by the interactions used in the NLEFT calculations. In the left panel of Fig.~\ref{fig:k_mom} we show the momentum distribution of the quasiparticles, $v^2(k)$.  The pairing gap for the s-wave superfluid phase of the spin-balanced system is shown in left panel of Fig.~\ref{fig:kf_gap_unitless}.

\subsection{Spin-polarized system with p-wave interactions}
In a spin-polarized system only odd-$l$ interactions are active, where $l$ is the relative orbital angular momentum between two particles. For purely attractive interactions, the p-wave channel dominates due to the centrifugal barrier. Hence, we'll approximate the pairing in a polarized system by considering only p-wave pairs, ignoring higher $l$-pairs. Starting from the description of a spin-balanced system with p-wave same-spin pairs, we set $\Delta_{\up\dn}(\mfk)=\Delta_{\dn\up}(\mfk)=0$ in Eq.(\ref{eq:bcs_ham}). To make the connection apparent, we will work out the description of equal-spin p-wave pairing in the balanced system and arrive at the description of the fully polarized case by subsequently constraining the pair binding and density of one species to 0, i.e, $\Delta_{\up\up}(\mfk)=0$ and $\rho_\up=0$ [or $\Delta_{\dn\dn}(\mfk)=0$ and $\rho_\dn=0$]. The mean-field Hamiltonian is 
\begin{align}
H^{(p)}_{MF}=\left(\begin{array}{cccc}
\xi(\mfk) & 0 & \Delta_{\up\up}(\mfk) &0 \\
0&\xi(\mfk)  & 0 & \Delta_{\dn\dn}(\mfk) \\
\Delta^\star_{\up\up}(\mfk) & 0 & -\xi(\mfk) & 0 \\
0 & \Delta^\star_{\dn\dn}(\mfk) & 0 & -\xi(\mfk)\\ 
\end{array}\right)~.
\label{eq:bcs_ham_p}
\end{align}
The eigenvalues and the normalized eigenvectors are
\begin{align}
E_{0} = - E_{\dn\dn}(\mfk);\quad {\alpha}^0 =& \left(0,-\left|v_{\dn\dn}(\mfk)\right|e^{-i\phi_{\dn\dn}},0,|u_{\dn\dn}(\mfk)| \right)~,  \\
E_{1} = E_{\dn\dn}(\mfk);\quad {\alpha}^1 =& \left(0,|u_{\dn\dn}(\mfk)|e^{-i\phi_{\dn\dn}},0,|v_{\dn\dn}(\mfk)| \right)~, \\
E_{2} = -E_{\up\up}(\mfk);\quad {\alpha}^2 =& \left(-|v_{\up\up}(\mfk)|e^{-i\phi_{\up\up}},0,|u_{\up\up}(\mfk)|,0 \right)~,\\
E_{3} = E_{\up\up}(\mfk);\quad {\alpha}^3 =& \left(|u_{\up\up}(\mfk)|e^{-i\phi_{\up\up}},0,|v_{\up\up}(\mfk)|,0 \right)~,
\end{align}
where $\phi_{\sigma\sigma}=\arg(\Delta_{\sigma\sigma})$ and 
\begin{align}
E_{\sigma\sigma}(\mfk) &= \sqrt{\xi^2(\mfk)+|\Delta_{\sigma\sigma}(\mfk)|^2}~, \\ \quad v_{\sigma\sigma}(\mfk)=e^{i\beta_{\sigma\sigma}} \sqrt{\frac{1}{2}\left(1-\frac{\xi(\mfk)}{E_{\sigma\sigma}(\mfk)}\right)}~,&\quad 
u_{\sigma\sigma}(\mfk) =e^{i\beta'_{\sigma\sigma}} \sqrt{\frac{1}{2}\left(1+\frac{\xi(\mfk)}{E_{\sigma\sigma}(\mfk)}\right)}
\end{align}
from which it also follows that
\begin{align}
|u_{\sigma\sigma}(\mfk)v_{\sigma\sigma}(\mfk)| = \sqrt{\frac{1}{4}\left(1-\frac{\xi(\mfk)}{E_{\sigma\sigma}(\mfk)}\right)\left(1+\frac{\xi(\mfk)}{E_{\sigma\sigma}(\mfk)}\right)} = \frac{1}{2}\frac{|\Delta_{\sigma\sigma}(\mfk)|}{E_{\sigma\sigma}(\mfk)}
\end{align}
Due to the particle-hole symmetry of the Bogoliubov-de Gennes formalism, the Hamiltonian in Eq.~(\ref{eq:bcs_ham_p}) has two negative and two positive eigenvalues corresponding to quasiholes and quasiparticles, respectively. Focusing on quasiparticles, we assign a spin to each by examining the action of $\alpha^n\cdot \psi$ outside of the Fermi sea at the limit of no pairing. With these we define the Bogolyubov matrices as 
\begin{align}
U_{\sigma\sigma'}(\mfk) = \left(\begin{array}{cc}
|u_{\up\up}(\mfk)| e^{-i\phi_{\up\up}} & 0 \\ 
0 & |u_{\dn\dn}(\mfk)|e^{-i\phi_{\dn\dn}}
\end{array} \right)~,\quad V_{\sigma\sigma'}(\mfk) = \left(\begin{array}{cc}
|v_{\dn\dn}(\mfk)| & 0 \\ 
0 & |v_{\up\up}(\mfk)|
\end{array} \right) 
\end{align}
From these we write the pairing amplitudes $f_{\sigma\sigma'}(\mfk)$ and occupations $n_\sigma(\mfk)$ as
\begin{align}
f_{\sigma\sigma'}(\mfk) &= V_{\sigma\sigma'}(\mfk)U^\star_{\sigma'\sigma'}(\mfk) =\delta_{\sigma\sigma'}|v_{\sigma\sigma}(\mfk)u_{\sigma\sigma}(\mfk)|e^{i\phi_{\sigma\sigma}} = \delta_{\sigma\sigma'}\frac{1}{2}\frac{|\Delta_{\sigma\sigma}(\mfk)|}{E_{\sigma\sigma}(\mfk)}e^{i\phi_{\sigma\sigma}}~,\label{eq:ampl_sol} \\
n_{\sigma}(\mfk) &= \sum_{\sigma'}V_{\sigma\sigma'}(\mfk)V^\star_{\sigma\sigma'}(\mfk) =v_{\sigma\sigma}^2(\mfk)~.
\end{align}
From these the gap equations for a balanced system with only same-spin pairing follow as
\begin{align}
\Delta_{\sigma\sigma'}(\mfk) &= -\delta_{\sigma\sigma'}\sum_{\mfk'} V_{\sigma\sigma}(\mfk,\mfk')f_{\sigma\sigma} (\mfk')= -\frac{\delta_{\sigma\sigma'}}{2}\sum_{\mfk'}V_{\sigma\sigma}(\mfk,\mfk')\frac{\Delta_{\sigma\sigma}(\mfk')}{E_{\sigma\sigma}(\mfk')}\label{eq:p_gap1}\\
\rho_\sigma&=\sum_{\mfk} n_\sigma(\mfk)=\frac{1}{2}\sum_\mfk \left[1-\frac{\xi(\mfk)}{E_{\sigma\sigma}(\mfk)}\right] \label{eq:p_gap2}
\end{align}
where we have included the superfluid density of component $\sigma$, often referred to as the second gap equation. Now the description of the polarized system composed of only $\sigma=\up$ is retrieved by setting $\Delta_{\dn\dn}(\mfk)=0$ and we set $\Delta_{\up\up}(\mfk)=\Delta(\mfk)$ for the rest of this section.

\begin{figure}
\centering
\begin{subfigure}{.5\textwidth}
  \centering
  \includegraphics[width=.8\linewidth]{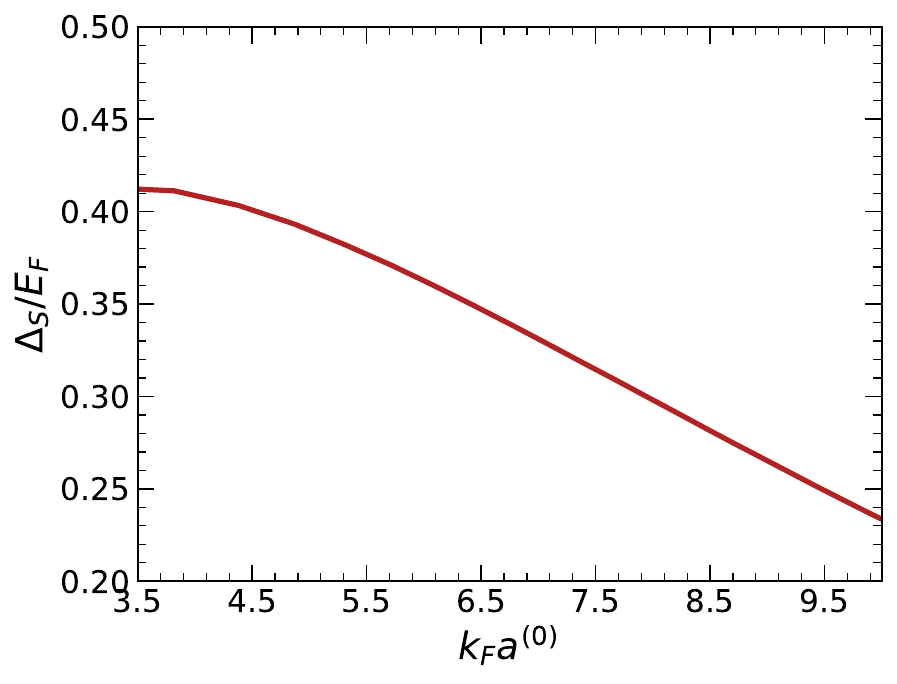}
\end{subfigure}%
\begin{subfigure}{.5\textwidth}
  \centering
  \includegraphics[width=.8\linewidth]{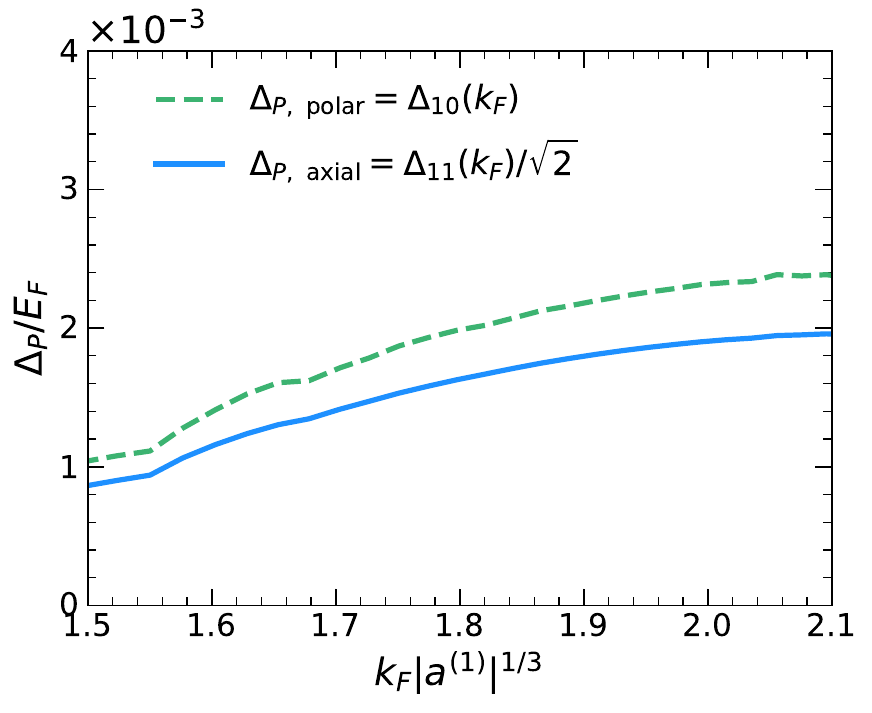}
\end{subfigure}
\caption{\textbf{The BCS s-wave and p-wave pairing gaps for the spin-balanced and the spin-polarized systems respectively in dimensionless units.} The order parameter of the s-wave superfluid in the spin-balanced system (left panel) and the p-wave superfluid phases of the spin-polarized system (right panel), in units of $E_\textrm{F}$. $a^{(0)}$ is the s-wave scattering length and $a^{(1)}$ is the p-wave scattering volume. \label{fig:kf_gap_unitless}}
\end{figure}

\begin{figure}
\centering
\begin{subfigure}{.5\textwidth}
  \centering
  \includegraphics[width=.8\linewidth]{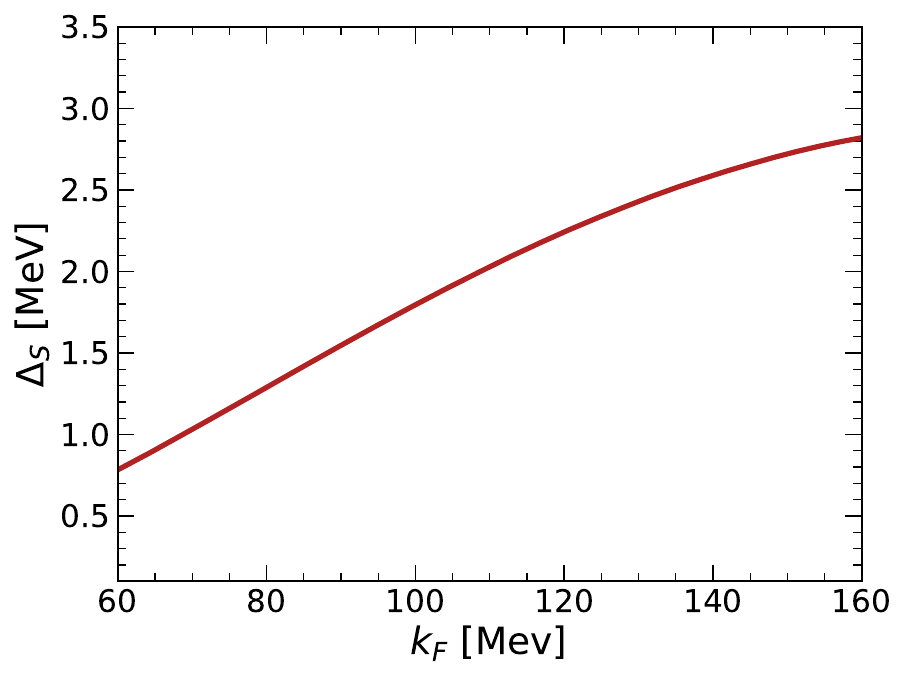}
\end{subfigure}%
\begin{subfigure}{.5\textwidth}
  \centering
  \includegraphics[width=.8\linewidth]{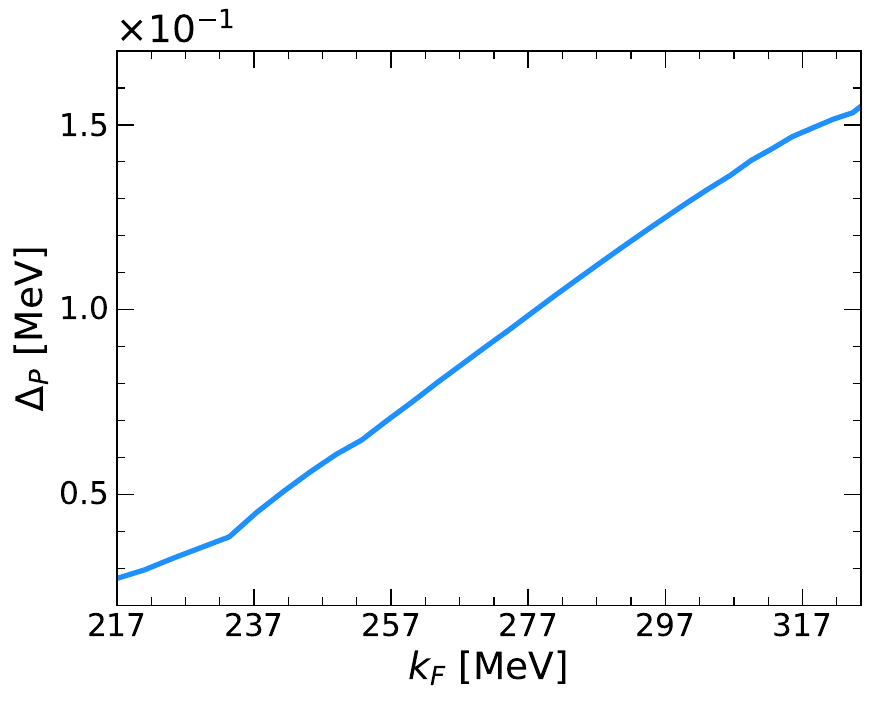}
\end{subfigure}
\caption{\textbf{The BCS s-wave and p-wave pairing gaps for the spin-balanced and the spin-polarized systems respectively in units of MeV.} The pairing gap of the s-wave superfluid in the spin-balanced system (left panel) and the p-wave superfluid phases of the spin-polarized system (right panel), in units of MeV. For the p-wave system the ground state corresponds to the axial phase (see main text).  \label{fig:kf_gap_MeV}}
\end{figure}

Eqs.~(\ref{eq:p_gap1}) and (\ref{eq:p_gap2}) are often projected on the spherical components of the gap function using Eq.~(\ref{eq:gap_exp}). We explore two possible ground states for the system corresponding to a polar phase, with $\Delta(\mfk)\sim Y_{10}(\hat{\mfk})$, and an axial phase, with $\Delta(\mfk)\sim Y_{11}(\hat{\mfk})$, akin to the Anderson-Brinkman-Morrel (ABM) phase of $^3$He~\cite{vollhardt:book,leggett:book}. For the polar phase the gap equation reduces to one for its radial component for $\Delta_{10}(k)$, 
\begin{align}
    \Delta_{10}(k) &= -\frac{1}{\pi} \int dk' k'^2 V_1(k,k')\Delta_{10}(k')\int d^2\hat{\mfk}'\frac{|Y_{10}(\hat{\mfk}')|^2}{\sqrt{\xi^2(k')+\frac{4\pi}{3}|Y_{10}(\hat{\mfk}')|^2\Delta_{10}^2(k')}} \\
    &= -\frac{1}{\pi} \int dk' k'^2 V_1(k,k')\Delta_{10}(k')\left[\frac{3}{2}\int d\theta \frac{\sin\theta\cos^2\theta}{\sqrt{\xi^2(k')+\cos^2\theta\Delta_{10}^2(k')}}\right]~.
\end{align}
For the axial phase, the radial component follows 
\begin{align}
    \Delta_{11}(k) &= -\frac{1}{\pi} \int dk' k'^2 V_1(k,k')\Delta_{11}(k')\int d^2\hat{\mfk}'\frac{|Y_{11}(\hat{\mfk}')|^2}{\sqrt{\xi^2(k')+\frac{4\pi}{3}|Y_{11}(\hat{\mfk}')|^2\Delta_{11}^2(k')}} \\
    &= -\frac{1}{\pi} \int dk' k'^2 V_1(k,k')\Delta_{11}(k')\left[\frac{3}{4}\int d\theta \frac{\sin^3\theta}{\sqrt{\xi^2(k')+\frac{1}{2}\sin^2\theta\Delta_{11}^2(k')}}\right]~.
\end{align}
The second gap equation for the two phases respectively becomes
\begin{align}
    \rho_{\textrm{polar}} &= \frac{1}{4\pi^2} \int dk k^2\left[1-\frac{1}{2}\frac{\xi(k)}{\sqrt{\xi^2(k)+\cos^2\theta \Delta^2_{10}(k)}}\right] \\
    \rho_{\textrm{axial}} &= \frac{1}{4\pi^2} \int dk k^2\left[1-\frac{1}{2}\frac{\xi(k)}{\sqrt{\xi^2(k)+\frac{1}{2}\sin^2\theta \Delta^2_{11}(k)}}\right]
\end{align}

Note that performing a standard angle-averaging in the excitation spectrum, replacing its angle-dependence with an averaged uniform value as is often done in neutron matter~\cite{Dean:2003}, washes away the structure of the spectrum, makes the two cases indistinguishable and yields a smaller effective gap~\cite{annett:book}. For these anisotropic pairing states $\Delta(\mfk_\textrm{F})$ is angle dependent,
\begin{align}
    |\Delta_{\textrm{polar}}(\mfk_\textrm{F})| = \Delta_{10}(k_\textrm{F}) \cos\theta~,\quad |\Delta_{\textrm{axial}}(\mfk_\textrm{F})| = \Delta_{11}(k_\textrm{F}) \sqrt{\frac{1}{2}}\sin\theta~,
\end{align}
where $k_F=(6\pi^2\rho)^{1/3}$, and so we define the pairing gap for the p-wave paired states, $\Delta_P$, as the peak value of $\Delta(\mfk_\textrm{F})$:
\begin{align}
    \Delta_{P,\textrm{~polar}} = \Delta_{10}(k_\textrm{F})~,\quad \Delta_{P,\textrm{~axial}} = \frac{1}{\sqrt{2}}\Delta_{11}(k_\textrm{F})
\end{align}
In Fig.~\ref{fig:k_mom} we show the momentum distributions of the quasiparticles, $v^2(k)$, at $\theta=0$ and $\pi/2$ for the polar and the axial phases, respectively.

The s-wave pairing gap seen in Fig.~\ref{fig:kf_gap_unitless} takes the typical values for a strongly paired system on the BCS side of the BCS-BEC crossover, which can be parametrized with the dimensionless coupling $\eta=1/[k_F a^{(0)}]$ with the unitary Fermi gas situated at the crossover's center, at $\eta=0$. The strong pairing of the state can be also seen in the substantial smearing of the momentum distribution in the left panel of Fig.~\ref{fig:k_mom} compared to the Fermi distribution shown with a gray dotted line.

In the spin-polarized system, although the polar state exhibits a taller peak gap amplitude to compensate for its pairing vanishing along an entire equator, the axial state is the true thermodynamic ground state because its point-node geometry leaves more of the Fermi surface gapped, thereby maximizing the total macroscopic condensation energy~\cite{Leggett1975}. The order parameter of the p-wave superfluid ground state is then $\Delta_P=\Delta_{P,~\textrm{axial}}$. This system, less strongly coupled than the spin-balanced case, exhibits  a pairing gap of a small fraction of its Fermi energy, as seen in the right panel of Fig.~\ref{fig:kf_gap_unitless}. Again, this is corroborated by the weaker smearing of the momentum distribution shown in the right panel of Fig.~\ref{fig:k_mom} compared to the corresponding Fermi distribution plotted with a gray dotted line.

In Fig.~\ref{fig:kf_gap_MeV} we show the pairing s-wave and p-wave pairing gaps, $\Delta_S$ and $\Delta_P$ in units of MeV to facilitate the comparison with the calculations within the 3D GAE Hubbard Model. We find that the self-consistent lattice Cooper model, which solves the BCS gap equations in a finite volume and on the lattice, is qualitatively consistent with the mean-field BCS calculations. In detail, for the s-wave pairing gap, both the BCS calculations and the self-consistent lattice Cooper model overestimate the pairing gap extracted from the many-body calculation by a factor of $\sim 1.5$, consistent with similar comparisons in studies with continuum QMC~\cite{Gezerlis2008}. Similarly, for the p-wave pairing gap, the self-consistent lattice Cooper model and BCS calculations are qualitatively consistent at the density range available to both approaches.
%%%%%%%%%%%%%%%%%%%%%%%%%%%%%%%%%%%%%%%%%%%%%%

\section{Composite Boson Theory, Pauli Repulsion, and Multimodal Coexistence}

This section summarizes the composite boson (``coboson'') formalism developed by
Combescot and collaborators \cite{Combescot2003,Combescot2008,Combescot2015} and explains
how Pauli exclusion between fermionic constituents governs the energetics, saturation,
and coexistence of bound clusters observed in our lattice calculations. 
The purpose of this section is not to construct an effective action, but to provide a
microscopic many-body interpretation of induced quartet correlations and their density
dependence.

\subsection{Many-Body Energy of Composite Bosons}

Consider a composite boson creation operator $B^\dagger$ that creates a normalized bound
state of fermions with binding energy $E_0<0$.
The $N$-boson state is
\begin{equation}
    |\Phi_N\rangle = (B^\dagger)^N |0\rangle .
\end{equation}
Because $B^\dagger$ is constructed from fermionic operators, it does not satisfy
canonical bosonic commutation relations. The many-body energy is defined as
\begin{equation}
    E(N) =
    \frac{\langle \Phi_N | H | \Phi_N \rangle}
         {\langle \Phi_N | \Phi_N \rangle},
\end{equation}
where $H$ is the underlying fermionic Hamiltonian.
For ideal bosons, one would have $E(N)=N E_0$.
For composite bosons, deviations from ideal behavior arise solely from Pauli exclusion
between fermions belonging to distinct copies of the composite boson. By scaling to the thermodynamic limit, the energy per composite boson can be expressed as a well-behaved density expansion:
\begin{equation}
    \frac{E(N)}{N} = E_0 + \frac{1}{2}\rho_B\,\mathcal{U}_{\mathrm{eff}} + \mathcal{O}(\rho_B^2),
\end{equation}
where $\rho_B = N/V$ is the composite boson density and $\mathcal{U}_{\mathrm{eff}}$ is the volume-independent effective interaction strength between pairs of composite bosons. 

\subsection{Pauli and Interaction Scatterings}

The effective interaction $\mathcal{U}_{\mathrm{eff}}$ is not a phenomenological parameter but a
specific combination of two microscopic contributions:

\begin{itemize}
\item \textbf{Pauli scattering ($\bar{\lambda}_2$):}
A volume-independent parameter that measures the overlap between fermionic wavefunctions
belonging to two different composite bosons. It arises from the non-orthogonality of the multi-fermion states and defined as
\begin{equation}
    \frac{\langle \Phi_2 | \Phi_2 \rangle}{2!}
    = 1 - \frac{\bar{\lambda}_2}{V}.
\end{equation}

\item \textbf{Interaction scattering ($\bar{\xi}_2$):}
A dynamical, volume-independent contribution arising from interactions between fermions belonging to
different composite bosons, distinct from their internal binding energy $E_0$.
\end{itemize}
Through exact cancellations of the macroscopic normalization factors, the leading-order effective interaction in the thermodynamic limit reduces to \cite{Combescot2015}
\begin{equation}
    \mathcal{U}_{\mathrm{eff}} = \bar{\xi}_2 - 2 \bar{\lambda}_2 E_0 .
\end{equation}
Since $E_0<0$, the Pauli term $-2\bar{\lambda}_2 E_0$ is strictly positive.
This contribution represents a Pauli-induced kinetic repulsion, which exists even
in the absence of any explicit repulsive interaction between the composite bosons.

\subsection{Scaling from Pairs to Quartets}

The magnitude of Pauli repulsion depends strongly on the number of constituent fermions
and the number of allowed exchange processes.

\subsubsection{Exchange Channel Counting}

The Pauli scattering parameter $\bar{\lambda}_2$ is obtained by summing over all possible
fermion-exchange diagrams between two composite bosons.

\begin{enumerate}
\item \textbf{S-wave pairs ($n_f=2$):}
Each pair contains one spin-up and one spin-down fermion.
Between two pairs, exchange can occur either between the two spin-up fermions or between
the two spin-down fermions.
Thus $\bar{\lambda}_2^{(S)}$ receives contributions from two exchange channels.

\item \textbf{Quartets ($n_f=4$):}
Each quartet contains two spin-up and two spin-down fermions.
Between two quartets, there are $2\times2=4$ possible exchanges in the spin-up sector and
$2\times2=4$ in the spin-down sector.
Thus $\bar{\lambda}_2^{(Q)}$ receives contributions from eight exchange channels.
\end{enumerate}

As a result,
\begin{equation}
    \bar{\lambda}_2^{(Q)} \sim 4\,\bar{\lambda}_2^{(S)},
\end{equation}
up to differences in the spatial structure of the bound-state wavefunctions.

\subsubsection{Saturation Mechanism}

Because the Pauli term in $\mathcal{U}_{\mathrm{eff}}$ scales with $\bar{\lambda}_2 |E_0|$, the effective
repulsion between quartets grows rapidly with density.
Although quartets may be energetically favored at very low density due to their large
binding energy, the steep increase of Pauli-induced repulsion causes the quartet chemical
potential
\begin{equation}
    \mu_Q(\rho_Q) = \frac{\partial (E/V)}{\partial \rho_Q}
\end{equation}
to rise quickly (here $\rho_Q$ is the quartet density).
At sufficiently high density, this repulsion renders further quartet occupation
energetically unfavorable, leading to saturation and coexistence with weaker pairing
channels.  We note that a similar analysis could be extended beyond quartets to sextets, octets, and even higher-body correlations.

\subsection{Synthesis: Multimodal Coexistence and Phase Space}

The phenomenology of the multimodal superfluid state emerges from the interplay of
energetics and Pauli exclusion across three distinct sectors: s-wave pairs, quartets,
and p-wave pairs. 

\subsubsection{Energetic Hierarchy vs.\ Pauli Saturation}

\begin{enumerate}
\item \textbf{Quartets (Strongest Binding, Highest Repulsion):}
Quartets represent the most deeply bound state ($|E_0^{(Q)}| > 2 |E_0^{(S)}|$). 
However, as derived above, they suffer from the strongest Pauli repulsion due to the
large number of internal fermion exchange channels ($\bar{\lambda}_2^{(Q)} \sim 4 \bar{\lambda}_2^{(S)}$).
This leads to saturation: quartets dominate at low density but cannot consume
the entire Fermi surface.

\item \textbf{S-wave Pairs (Intermediate Binding, Moderate Repulsion):}
S-wave pairs have intermediate binding energy. Their Pauli repulsion is weaker
than that of quartets ($\bar{\lambda}_2^{(S)} < \bar{\lambda}_2^{(Q)}$). When the quartet channel is saturated, additional fermions find it energetically favorable
to form s-wave pairs.

\item \textbf{P-wave Pairs (Weakest Binding, Minimal Overlap):}
P-wave pairs have the least amount of binding. However, they occupy a region of
Hilbert space that is largely orthogonal to the s-wave and quartet sectors. When the quartet and s-wave channels are saturated, additional fermions find it energetically favorable
to form p-wave pairs.
\end{enumerate}

\subsubsection{The Role of P-wave Orthogonality}

To see this orthogonality explicitly, consider the pair creation operators in momentum space:
\begin{align}
    P_S^\dagger &= \sum_{\mathbf{k}} \varphi_S(\mathbf{k}) \frac{1}{\sqrt{2}}
    (a^\dagger_{\mathbf{k}\uparrow} a^\dagger_{-\mathbf{k}\downarrow} - a^\dagger_{\mathbf{k}\downarrow} a^\dagger_{-\mathbf{k}\uparrow}), \\
    P_P^\dagger &= \sum_{\mathbf{k}, \alpha, \beta} \varphi_P(\mathbf{k}) Y_{1m}(\hat{\mathbf{k}}) (i\sigma_y \sigma_m)_{\alpha\beta} a^\dagger_{\mathbf{k}\alpha} a^\dagger_{-\mathbf{k}\beta}.
\end{align}
The overlap integral governing Pauli exchange between an s-wave and a p-wave pair vanishes:
\begin{equation}
    \sum_{\mathbf{k}} \varphi_S^*(\mathbf{k}) \varphi_P(\mathbf{k}) Y_{1m}(\hat{\mathbf{k}}) \times \Tr[\text{Spin}] = 0.
\end{equation}
This vanishing is enforced by both parity (even $\times$ odd orbital functions) and spin
(singlet versus triplet).

Because Pauli scattering is proportional to wavefunction overlap, the effective repulsion
between the p-wave sector and the dominant s-wave/quartet sectors is parametrically suppressed.
Thus, even though p-wave pairs have lower binding energy, they experience a ``free'' region of
phase space that is not Pauli-blocked by the denser s-wave condensate.
This unique feature enables the coexistence of p-wave correlations alongside strong
s-wave and quartet order, as observed in our lattice results.

%%%%%%%%%%%%%%%%%%%%%%%%%%%%%%%%%%%%%%%%%%%%%%

\section{Effective Action Description of Multimodal Superfluidity}

\subsection{Physical Summary}

The multimodal superfluid phase described in this work arises from the coexistence of
attractive s-wave and p-wave channels. The primary instability depends on the density. 
At low densities and assuming that the p-wave attraction is enough to bind two s-wave pairs into a quartet, the extra binding makes quartet condensation the 
primary instability. At higher densities, conventional s-wave pairing can also become 
a primary instability, leading to co-condensation. 

Two p-wave pairs can combine into a scalar configuration with total intrinsic spin $s_{\mathrm{tot}}=0$ and total orbital angular momentum $l_{\mathrm{tot}}=0$. This
minimizes Pauli blocking and maximizes spatial overlap of the underlying fermions. When both s-wave and p-wave interactions are attractive, the corresponding quadratic stiffness for this channel is reduced, making the scalar p-wave double pair the most energetically favorable of the p-wave configurations. We also have mixing between the quartet formed by the bound state of two s-wave pairs and the scalar p-wave double pairs. All three condensates are phase locked with each other.

%-------------------------------------------------
\subsection{Framework and Scope}

To characterize the ground state of the generalized extended Hubbard model and of
realistic neutron matter under competing s-wave and p-wave interactions, we employ
the quantum effective action formalism.
The effective action $\Gamma[\Phi]$ is defined as the Legendre transform of the
generating functional of connected Green’s functions $W[J]$ and provides a
nonperturbative description of spontaneous and induced order.

The stationarity condition
\begin{equation}
    \frac{\delta \Gamma}{\delta \Phi(x)} = 0
\end{equation}
determines the ground state expectation values of macroscopic fields in the
thermodynamic limit.
This framework is essential for distinguishing between primary condensates and order
parameters that acquire expectation values solely through higher-order composite
correlations enforced by symmetry. The effective action constructed here is a long-wavelength
description of the collective correlations measured directly in our {\it ab initio} lattice calculations.

%-------------------------------------------------
\subsection{Auxiliary Pair Fields and Global Symmetry}

The microscopic theory possesses global symmetry
\begin{equation}
    G = \mathrm{U}(1) \times \mathrm{SU}(2)_{\mathrm{spin}} \times \mathrm{SO}(3)_{\mathrm{orb}} .
\end{equation}
We can produce attractive pairing interactions by introducing auxiliary bosonic fields $\phi$ and $A_{\mu j}$, where $\phi$ couples to the spin-singlet, parity-even ($^1S_0$) fermion bilinear and
$A_{\mu j}$ couples to the spin-triplet, parity-odd ($^3P_J$) fermion bilinears.
Here $\mu = 1,2,3$ denotes the vector index for orbital angular momentum ($l=1$) and $j=1,2,3$ denotes the vector index for spin ($s=1$).

At the level of the functional integral, these fields are auxiliary. After integrating out the gapped fermionic quasiparticles, however, they acquire induced dynamics and encode the collective pairing correlations. They can be viewed as interpolating fields in the Lehmann–Symanzik–Zimmermann reduction framework \cite{Lehmann:1954rq,Lehmann:1957zz}.

%-------------------------------------------------
\subsection{S-wave Sector}

The scalar field $\phi(x)$ transforms as $(s,l)=(0,0)$ and carries charge $2$ with respect to the U(1) symmetry.  We also introduce a composite field ${Q}_s(x)$ that serves as an interpolating field for the quartet bound state, which is an asymptotic state in our theory that carries charge $4$ with respect to the U(1) symmetry.  We can choose ${Q}_s(x) \propto [\phi(x)]^2$.

%-------------------------------------------------
\subsection{P-wave Sector and Absence of Vector Condensation}

The p-wave field $A_{\mu j}(x)$ carries one orbital and one spin index and transforms as
$(s,l)=(1,1)$.
Vector condensation corresponds to
\begin{equation}
    \langle A_{\mu j} \rangle \neq 0 ,
\end{equation}
which necessarily breaks continuous rotational symmetries or locks spin and orbital
rotations. In the multimodal phase we instead find
\begin{equation}
    \langle A_{\mu j} \rangle = 0 ,
\end{equation}
so that no vector order parameter condenses.

%-------------------------------------------------
\subsection{Quartic Invariants and Group Theory Counting}

The number of independent quartic terms in the effective potential is determined by the number of irreducible scattering channels allowed by the global symmetry $G = \mathrm{SO}(3)_{\mathrm{orb}} \times \mathrm{SU}(2)_{\mathrm{spin}}$.
The p-wave order parameter $A_{\mu j}$ describes a boson with $l=1$ and $s=1$. The interaction between two such bosons is governed by the tensor product of two pairs:
\begin{equation}
    (1_l \otimes 1_s) \otimes (1_l \otimes 1_s) = (1_l \otimes 1_l) \otimes (1_s \otimes 1_s).
\end{equation}
The product of two vectors decomposes into scalar ($J=0$), vector ($J=1$), and tensor ($J=2$) irreducible representations:
\begin{equation}
    1 \otimes 1 = 0 \oplus 1 \oplus 2.
\end{equation}
Since the order parameter describes identical bosons, the two-body wavefunction must be totally symmetric under particle exchange. The symmetry of the spatial part must match the symmetry of the spin part. Symmetric representations ($J=0, 2$) must combine with symmetric representations.  Antisymmetric representations ($J=1$) must combine with antisymmetric representations.
This selection rule allows for 5 independent channels:
\begin{enumerate}
    \item $(l=0) \otimes (s=0)$: symmetric scalar $\times$ symmetric scalar.
    \item $(l=2) \otimes (s=0)$: symmetric tensor $\times$ symmetric scalar.
    \item $(l=0) \otimes (s=2)$: symmetric scalar $\times$ symmetric tensor.
    \item $(l=2) \otimes (s=2)$: symmetric tensor $\times$ symmetric tensor.
    \item $(l=1) \otimes (s=1)$: antisymmetric vector $\times$ antisymmetric vector.
\end{enumerate}
Consequently, there are exactly 5 independent quartic invariants in the effective action.

%-------------------------------------------------
\subsection{Composite Quartet Channels and Irreducible Decomposition}

From the quartic invariants we identify composite operators that serve as interpolating
fields for quartet correlations.
To avoid double counting and to enforce irreducible representations, the tensor operators
are defined with traceless components. We define the minimal basis of 5 independent operators as:
\begin{align}
    {D}_1 & \equiv \Tr (A A^T) = \sum_{\mu, j} A_{\mu j} A_{\mu j}, \\
    {D}_2 & \equiv \Tr (A^\dagger A) = \sum_{\mu, j} A^*_{\mu j} A_{\mu j}, \\
    {T}^{(3)}_{jk}
    & \equiv \sum_\mu A^*_{\mu j} A_{\mu k} - \frac{1}{3} \delta_{jk} {D}_2, \quad \text{(Spin Density Tensor)}\\
    {T}^{(4)}_{\mu\nu}
    & \equiv \sum_j A_{\mu j} A_{\nu j} - \frac{1}{3} \delta_{\mu\nu} {D}_1, \quad \text{(Orbital Pair Tensor)}\\
    {T}^{(5)}_{jk}
    & \equiv \sum_\mu A_{\mu j} A_{\mu k} - \frac{1}{3} \delta_{jk} {D}_1. \quad \text{(Spin Pair Tensor)}
\end{align}
\noindent
We note that a sixth candidate operator, the Orbital Density Tensor ${T}^{(6)}_{\mu\nu} \equiv \sum_j A_{\mu j} A^*_{\nu j} - \frac{1}{3} \delta_{\mu\nu} {D}_2$, is redundant. Its invariant norm is identical to that of the Spin Density Tensor due to the cyclic property of the trace, $|{T}^{(6)}|^2 = |{T}^{(3)}|^2$. We therefore exclude it from the basis.

%-------------------------------------------------
\subsection{Effective Potential and Mixing Structure}

To determine the uniform ground state of the system, we evaluate the effective action for translationally invariant fields. Consequently, all derivative terms vanish, and the symmetry-allowed effective potential is constructed by retaining only the most relevant non-derivative operators of the interpolating fields, organized by their scaling dimensions and consistent with the global symmetry group $G$:
\begin{align}
    {V}_{\mathrm{eff}}
    &=
    \mu_s |\phi|^2
    + \frac{g_H}{2} |\phi|^4
    + \alpha_s |{Q}_s|^2
    + \frac{g_Q}{2} |{Q}_s|^4
    + r_p {D}_2
    + \sum_{i=1}^2 \beta_i |{D}_i|^2 \nonumber \\
    &\quad
    + \beta_3 \Tr[({T}^{(3)})^2]
    + \beta_4 \Tr[({T}^{(4)})^\dagger {T}^{(4)}]
    + \beta_5 \Tr[({T}^{(5)})^\dagger {T}^{(5)}] \nonumber \\
    &\quad
    - \frac{\eta_s}{2}
    \big[ (\phi^\dagger)^2 {Q}_s + \mathrm{h.c.} \big]
    + u_{sp} |\phi|^2 {D}_2 \nonumber \\
    &\quad
    - \frac{\lambda_{sp}}{2}
    \big[ {Q}_s^\dagger {D}_1 + \mathrm{h.c.} \big] 
    - \frac{\eta_{sp}}{2}
    \big[ (\phi^\dagger)^2 {D}_1 + \mathrm{h.c.} \big]. \label{eq:eff_action}
\end{align}
In this effective action, we note that ${Q}_s$ is the macroscopic field for the quartet and $\phi^2$ is a product of two macroscopic fields for two s-wave pairs at the same point.  They no longer represent the same quantity in the effective action. The coefficients $\alpha_s$, $\mu_s$, and $\beta_k$ represent the effective stiffness of the respective composite modes. As quadratic coefficients, their signs and magnitudes are governed by the chemical potential of the system:
\begin{itemize}
    \item $\alpha_s$: Because the s-wave quartet is a bound state, this coefficient becomes negative at a lower density than the s-wave coefficient $\mu_s$.
    \item $\beta_1$: This parameter governs the scalar p-wave double-pair channel. Because Pauli blocking is minimized and the underlying interaction is attractive, $\beta_1$ is algebraically lower than the other $\beta_k$ coefficients. Even if the chemical potential is such that $\beta_1$ remains positive, the scalar p-wave double pair will still acquire an induced expectation value driven by the primary condensates.
\end{itemize}
The effective quartic couplings $g_H$ and $g_Q$ for the s-wave and quartet fields are positive, ensuring stability.

%-------------------------------------------------
\subsection{Minimization and Induced Quartet Order}

Let us assume that $\beta_1$ remains positive.  Minimization of ${V}_{\mathrm{eff}}$ with respect to the fields yields the ground state expectation values. Let $v \equiv \langle \phi \rangle$, $\Sigma_s \equiv \langle {Q}_s \rangle$, and $\Sigma_1 \equiv \langle {D}_1 \rangle$. Assuming the other tensor expectations vanish ($\langle {T}^{(k)} \rangle = 0$) and the ground state expectation values are real, the stationarity conditions are:
\begin{align}
    \frac{\partial {V}_{\mathrm{eff}}}{\partial v} &= 2v \left( \mu_s + g_H v^2 - \eta_s \Sigma_s - \eta_{sp} \Sigma_1 \right) = 0, \\
    \frac{\partial {V}_{\mathrm{eff}}}{\partial \Sigma_s} &= 2\alpha_s \Sigma_s + 2g_Q \Sigma_s^3 - \eta_s v^2 - \lambda_{sp} \Sigma_1 = 0, \\
    \frac{\partial {V}_{\mathrm{eff}}}{\partial \Sigma_1} &= 2\beta_1 \Sigma_1 - \lambda_{sp} \Sigma_s - \eta_{sp} v^2 = 0.
\end{align}
From the third equation, the p-wave double pair $\Sigma_1$ condensate is induced by the other condensates:
\begin{equation}
    \Sigma_1 = \frac{\lambda_{sp} \Sigma_s + \eta_{sp} v^2}{2\beta_1}.
\end{equation}
Substituting this back yields two coupled equations for the s-wave and quartet fields, defining two distinct physical regimes based on the density.

\paragraph*{Low-Density Regime (Quartet-Driven).}
At low densities, assuming that the p-wave attraction is enough to bind two s-wave pairs into a quartet, the binding of the quartet state causes its quadratic coefficient to become negative ($\alpha_s < 0$) while the s-wave coefficient remains positive ($\mu_s > 0$). In this regime, the primary instability is the quartet field, which acquires a ground state expectation value approximately given by $\Sigma_s \approx \sqrt{-\alpha_s/g_Q}$. The s-wave field $v$ remains massive but acquires an induced expectation value driven by the quartet through the $\eta_s$ coupling.

\paragraph*{Higher-Density Regime (Co-condensation).}
As density increases, the s-wave quadratic coefficient also becomes negative ($\mu_s < 0$). In this regime, both the s-wave and quartet fields possess direct instabilities and co-condense. They mutually reinforce one another through the interconversion term proportional to $\eta_s$, and the system is governed by the fully coupled non-linear equations. Neither field can be viewed simply as a rigid, static background for the other.

\paragraph*{Phase Locking Hierarchy.}
The interconversion terms, such as $(\phi^\dagger)^2 {Q}_s + \mathrm{h.c.}$, explicitly break independent phase rotation symmetries. By writing the phases of the condensates as $v = |v|e^{i\theta}$, $\Sigma_s = |\Sigma_s|e^{i\theta_{\Sigma_s}}$, and $\Sigma_1 = |\Sigma_1|e^{i\theta_{\Sigma_1}}$, the free energy is minimized when the phases are locked according to:
\begin{equation}
    \theta_{\Sigma_s} = \theta_{\Sigma_1} = 2\theta.
\end{equation}
This rigid phase-locking hierarchy holds regardless of whether the primary instability is driven by the quartet or the s-wave pair.

%-------------------------------------------------
\subsection{Relation to the A and B Phases of Superfluid $^3$He}

Superfluid $^3$He provides the canonical example of spin-triplet, p-wave superfluidity.
Its order parameter is the p-wave field $A_{\mu j}$ itself, which condenses directly. In the B phase, spin and orbital rotations are locked into a combined $\mathrm{SO}(3)_J$
symmetry, producing an isotropic, fully gapped superfluid. In the A phase, the condensate is anisotropic and chiral, breaking rotational and
time-reversal symmetries and supporting nodal quasiparticles.

Both phases are defined by vector condensation that is linear in the pairing field.  By contrast, multimodal superfluidity preserves spin and orbital symmetries and instead
realizes strong p-wave correlations through induced scalar double-pair order.
It is therefore a distinct state of matter, not a variant of the A or B phase of
superfluid $^3$He.

%-------------------------------------------------
\subsection{Neutron Matter and Spin–Orbit Coupling}

In realistic models of neutron matter based on chiral effective field theory ($\chi$EFT),
the nuclear force includes significant spin–orbit coupling terms. This explicitly reduces the continuous symmetry from
$\mathrm{SU}(2)_{\mathrm{spin}}\times \mathrm{SO}(3)_{\mathrm{orb}}$ to the diagonal subgroup $\mathrm{SO}(3)_J$
of total angular momentum. Our \emph{ab initio} lattice calculations for neutron matter reveal simultaneous
superfluid correlations in the $^1S_0$, $^3P_0$, and $^3P_2$ channels.

The reduction in symmetry allows for additional quartic invariants where spin and orbital
indices are contracted directly.
One example of such a spin-orbit terms is
\begin{equation}
    {O}_{LS} \sim \sum_{\mu \nu} A^\dagger_{\mu \mu} A^\dagger_{\nu \nu} A_{\mu \nu} A_{\nu \mu} + \textrm{h.c.}
\end{equation}
These terms represent scattering vertices between different $J$-states, such as scattering of
two $^3P_0$ pairs into two $^3P_2$ pairs.

Despite this increased complexity, the fundamental selection rule remains the same.
The $^1S_0$ condensate $v$ carries total angular momentum $J=0$. Consequently, the s-wave quartet source ${Q}_s \sim v^2$ transforms as a $J=0$ scalar and can only mix with p-wave double-pair operators carrying $J_{\mathrm{tot}}=0$.
The coupling to the scalar composite field ${D}_1$ produces condensates for scalar $^3P_0$ double pairs and scalar $^3P_2$ double pairs. The $^3P_1$ channel does not participate because the interactions are repulsive.

\subsection{Lattice Symmetry Constraints}

On a lattice, the continuous rotation group $\mathrm{SO}(3)_{\mathrm{orb}}$ is reduced to the
cubic group $O_h$.
The s-wave condensate and ${Q}_s$ transform as the identity representation
$A_{1g}$.
The p-wave vector transforms as $T_{1u}$.
The product of two p-wave vectors decomposes as
\begin{equation}
    T_{1u} \otimes T_{1u}
    =
    A_{1g} \oplus E_g \oplus T_{1g} \oplus T_{2g}.
\end{equation}
Exactly one copy of the identity representation appears in this decomposition,
corresponding to the scalar contraction ${D}_1$.
Thus, the mechanism of induced multimodal superfluidity is symmetry-protected in both
continuum and lattice realizations.

%%%%%%%%%%%%%%%%%%%%%%%%%%%%%%%%%%%%%%%%%%%%%%

\section{Thermodynamics, Vortices, and Astrophysical Implications}

\subsection{Coupling Mechanism and Phase Structure}

We investigate the phase structure of dilute neutron matter characterized by attractive interactions in both the s-wave and p-wave channels. In the density regime of the inner crust, the s-wave attraction is known to be robust. We consider a scenario where the p-wave attraction is insufficient to produce a condensate in isolation but contributes to a stable multimodal superfluid phase through many-body correlations. In this phase, the order parameter is a coherent superposition of s-wave pairs, p-wave pairs in entangled double-pair combinations, and quartets. The stability of this phase relies on a scalar coupling between s-wave double pairs, p-wave double pairs, and quartets.

At nonzero temperature, the system exhibits two distinct superfluid regimes governed by the stability of these quartet correlations:

\begin{enumerate}
\item \textbf{Multimodal superfluid phase ($T < T^*$):} The ground state where quartets effectively couple the s-wave and p-wave condensates into a single coherent macroscopic wavefunction.

\item \textbf{S-wave superfluid phase ($T^* < T < T_c$):} The regime where thermal fluctuations dissociate the quartets. The p-wave component vanishes, and the system reverts to an s-wave superfluid.
\end{enumerate}

\subsection{Energetics and the Transition Temperature}

The thermodynamics of the transition at $T^*$ are determined by the competition between the internal energy gain from quartet formation and the entropic penalty of maintaining coherent four-body correlations. Let $B_Q$ denote the excess binding energy gained when two s-wave pairs organize into a quartet structure. The free energy difference relative to the pure s-wave phase can be approximated as

\begin{equation}
\Delta F \approx -n_Q B_Q - T \Delta \mathcal{S},
\end{equation}
where $n_Q$ is the quartet density and $\Delta \mathcal{S} < 0$ represents the entropy loss due to the reduction of phase space in the bound quartet state.

The system undergoes a phase transition when the entropic gain from dissociation overcomes the binding energy. Treating this as a chemical equilibrium problem, the transition temperature scales with the quartet binding energy:

\begin{equation}
k_B T^* \approx \frac{B_Q}{\eta},
\end{equation}
where $\eta$ is a dimensionless factor of order unity characterizing the entropy of dissociation.

This establishes a hierarchy of symmetry breaking scales. The primary s-wave transition at $T_c$ is governed by the pair-breaking energy $2\Delta_S$, with standard BCS theory predicting $k_B T_c \approx 0.57 \Delta_S$. In contrast, the secondary transition scales with the smaller quartet binding energy $B_Q$. Consequently, a cooling neutron star is expected to first enter a standard s-wave superfluid state before condensing into the multimodal phase at lower temperatures relevant to the crust.

\subsection{Thermal Consequences: Heat Capacity and Cooling}

The onset of multimodal superfluidity introduces physical mechanisms that can alter the subsequent thermal evolution of the crust. These mechanisms provide a microscopic framework for interpreting the thermal relaxation observed in transiently accreting neutron stars.

In the multimodal scenario, the quartet binding energy implies that the energy required to liberate a nucleon is governed by an effective gap
\begin{equation}
\Delta_{\text{eff}} \simeq \Delta_S + \Delta_Q \qquad (4\Delta_Q = B_Q),
\end{equation}
which is larger than the fundamental s-wave gap $\Delta_S$. At temperatures relevant to crust cooling ($k_B T \sim 0.01$--$0.1$ MeV), this leads to a strong suppression of the neutron heat capacity. Following the standard suppression formalism \cite{Levenfish1994, Yakovlev2004}, the specific heat scales schematically as
\begin{equation}
C_V^{(Q)} \propto \exp\left(-\frac{\Delta_{\text{eff}}}{T}\right)
= \exp\left(-\frac{\Delta_S}{T}\right)\exp\left(-\frac{\Delta_Q}{T}\right).
\end{equation}
This expression should be interpreted as a quasiparticle activation behavior rather than a uniform-matter calculation, since the inner crust contains band structure and lattice scattering effects.

Such additional suppression of the specific heat provides a natural microscopic mechanism for the low thermal inertia inferred from the rapid thermal relaxation of transient sources such as KS 1731-260 \cite{Shternin:2007md,Brown2009,Cackett2010}. While this does not replace the need for an additional energy source to explain the total heat deposited during accretion, it is possible that the transition into the multimodal phase itself contributes to the required ad hoc shallow heating. As accretion drives crustal material to higher densities, the continuous reorganization of neutrons into bound quartets and entangled p-wave double-pairs would release latent heat. If this phase transition occurs at the lower densities found near the neutron drip line, this steady release of energy during accretion episodes could potentially serve as a microscopic candidate for the observed shallow heating. Furthermore, the multimodal phase governs the fast early-time cooling timescale by suppressing the thermal inertia. Standard models can reproduce this rapid cooling only if the bare s-wave gap is pushed near the upper limits of microscopic predictions \cite{Hebeler:2009iv, Gezerlis:2009iw, Drischler:2016cpy,Vidana:2021wna,Gandolfi:2022dlx}. Multimodal superfluidity relaxes this requirement by generating a larger effective excitation gap.

However, late-time observations of these systems during quiescence instead suggest an enhancement of the neutron specific heat in the deep crust \cite{Deibel2017}. This behavior can be explained by a gapless superfluid state produced by sustained superflow within the crust \cite{Allard2024}. Because the larger effective gap $\Delta_{eff}$ increases the Landau critical velocity, triggering this gapless state requires a higher macroscopic superfluid velocity. Furthermore, the larger gap reduces the coherence length, making proximity effects around crustal nuclei less efficient at washing out pairing inhomogeneities. Maintaining such a high superflow therefore requires pinning forces significantly stronger than typically expected for a conventional neutron superfluid. The complex topological defects present in the multimodal phase—such as bound vortex dimers and percolated domain-wall networks—provide a natural source of enhanced pinning to support the required superflow.  In this way, multimodal superfluidity offers a unified mechanism capable of accommodating both the early-time suppression of thermal inertia and the late-time enhancement of heat capacity.\footnote{We are grateful to Nicolas Chamel for a discussion of these points.}

While a larger effective gap suppresses the heat capacity, it also suppresses the standard pair breaking and formation (PBF) neutrino emission process \cite{Page2006}.  Without an alternative cooling channel the crust would become radiatively inert at late times.  The multimodal condensate provides such a channel. A scalar $^1S_0$ superfluid couples primarily to the conserved vector current, strongly suppressing neutrino emission at low momenta \cite{Kundu:2004mz}. However, the multimodal condensate contains spin-triplet ($^3P_2$) correlations. While some energy is required to decouple these entangled $^3P_2$ pairs, this is much lower than the energy required for pair breaking. Collective excitations involving internal spin structure couple to the axial-vector current, which is not conserved. As shown by Leinson \cite{Leinson2010}, such modes can efficiently radiate neutrino pairs at low momenta, keeping the crust thermally active even when the PBF process is frozen out.

\subsection{Vortex Structure and Topology}

The phase-locking condition derived in our effective action,

\begin{equation}
\theta_{\Sigma_s} = \theta_{\Sigma_1} = 2\theta,
\end{equation}
imposes non-trivial topological constraints on rotation. While the quartet condensate supports vortices with $2\pi$ winding (circulation $h/4m_n$), the coupled s-wave order parameter winds by only $\pi$. Such fractional winding requires a domain wall across which the s-wave phase jumps.

Two structural regimes follow:

\begin{itemize}
\item \textbf{Dimer phase:} half-quantum vortices bind into dimers connected by a short soliton, mimicking conventional vortices.

\item \textbf{Percolated phase:} domain walls form a connected network behaving as a rigid shear structure capable of collective failure events.
\end{itemize}
These structures introduce new elastic moduli and pinning mechanisms absent from standard glitch models and may help explain the triggering and sustainability of pulsar glitches.

\subsection{Implications for Superfluid Stiffness and Entrainment}

Standard glitch models require an angular momentum reservoir in the crustal neutron superfluid \cite{Baym1969}. A challenge is entrainment, where Bragg scattering from the lattice produces an effective neutron mass $m^* \gg m_n$ and reduces the available superfluid fraction \cite{Chamel2012}. Although calculations including pairing and lattice motion continue to suggest strong entrainment \cite{Chamel2025a, Chamel2025b}, some studies suggest that the problem could be mitigated by interband processes \cite{Almirante2025} or uncertainties in the equation of state \cite{Piekarewicz:2014lba}.

Independently of the final strength of entrainment, the multimodal phase provides an additional mechanism. The scalar coupling among s-wave pairs, p-wave double pairs, and quartets enforces rigid phase locking. A macroscopic phase twist therefore generates currents in all coupled sectors simultaneously, increasing the free-energy cost of phase gradients and enhancing the effective superfluid stiffness.

This behavior is analogous to coupled condensates in multigap superconductors \cite{Leggett1966,Leggett1975}. By increasing stiffness, multimodal correlations can mitigate the effective suppression of the superfluid density and potentially restore a sufficient angular momentum reservoir for large pulsar glitches.

%%%%%%%%%%%%%%%%%%%%%%%%%%%%%%%%%%%%%%%%%%%%%%

\section{Signatures of Multimodal Superfluidity in Nuclei}

The phenomenology of nuclear superfluidity is dominated by the s-wave pairing channel. This robust mode is ubiquitous throughout the nuclear chart, manifesting clearly in systematic odd-even mass staggering. In contrast, the experimental signatures of p-wave pairing and the predicted quartet condensate are much more subtle, appearing only under specific conditions of nuclear structure.

In uniform neutron matter, translational and rotational invariance ensure that momentum and orbital angular momentum remain good quantum numbers, allowing p-wave correlations to develop unimpeded. Finite nuclei, however, present a complex environment that typically includes configuration mixing among orbital sub-shells, and this mixing of orbitals acts as an effective source of disorder. 

The survival of s-wave pairing in this environment is explained by Anderson's theorem \cite{Anderson1959}. Conventional s-wave pairing is uniquely robust against disorder because it is protected by time reversal symmetry without any requirement of spatial symmetries. Even within the complex, mixed-parity environment of a heavy nucleus, every single-particle eigenstate $\ket{\psi_n}$ retains an exact energetic degeneracy with its time-reversed partner $\ket{\mathbb{T}\psi_n}$. The isotropic s-wave interaction couples these partners efficiently, maintaining a stable gap regardless of the complexity of the wavefunction and energy landscape.

Seeing p-wave gaps and quartet gaps associated with multimodal superfluidity requires that the relevant p-wave interactions are sufficiently strong and attractive to produce a signal and that it is not obscured by other effects such as subshell closures or s-wave pairing.  Complicating these conditions further is the fact that the $^3P_1$ channel is repulsive, and only the $^3P_0$ and $^3P_2$ are attractive. Two-body rescattering mediated by these attractive p-wave interactions is often restricted to a few energetically accessible orbitals. Consequently, the signatures of multimodal superfluidity beyond simple s-wave pairing depend critically on the magnitude of the two-body matrix elements for these orbitals. Distinct p-wave pair and quartet gaps are therefore observable only in nuclei where configuration mixing is weak and the relevant two-body matrix elements are sufficiently strong and attractive to generate a clear signal.

\subsection{Difference Formulas for Pairing and Quartetting}

Let us consider a finite sequence of nuclear isotopes ${}^{A}_{Z}\textrm{X}_{N},{}^{A+d}_{\quad Z}\textrm{X}_{N+d}, \cdots, {}^{A+(k-1)d}_{\qquad \quad Z}\textrm{X}_{N+(k-1)d}$ where $d$ is the stride length in neutron number.  We will take $d=1$ for pairing and $d=2$ for quartetting.  By default when we indicate an isotope such as ${}^{A}_{Z}\textrm{X}_{N}$, we are referring to the ground state.  To indicate a general state, we give the corresponding spin and parity as ${}^{A}_{Z}\textrm{X}_{N}(J^\pi)$.  By default, we are referring to the lowest $J^\pi$ state.

Let $f$ be some general function of the nuclear states in our finite sequence of isotopes.  We define the $k$-point difference formula as
\begin{equation}
f\{{}^{A}_{Z}\textrm{X}_{N},{}^{A+d}_{\quad Z}\textrm{X}_{N+d}, \cdots, {}^{A+(k-1)d}_{\qquad \quad Z}\textrm{X}_{N+(k-1)d}\}_\pm \equiv  \pm \sum_{j=0}^{k-1}  (-1)^{k-1-j}\binom{k-1}{j} f({}^{A+jd}_{\quad \; \; Z}\textrm{X}_{N+jd}). \label{eqn:k-point}
\end{equation}
The $\pm$ indicates whether the term with the largest number of neutrons has a positive or negative sign.  We can apply this staggered difference to nuclear binding energies $B$,
\begin{equation}
B\{{}^{A}_{Z}\textrm{X}_{N},{}^{A+d}_{\quad Z}\textrm{X}_{N+d}, \cdots, {}^{A+(k-1)d}_{\qquad \quad Z}\textrm{X}_{N+(k-1)d}\}_\pm \equiv  \pm \sum_{j=0}^{k-1} (-1)^{k-1-j}\binom{k-1}{j} B({}^{A+jd}_{\quad \; \; Z}\textrm{X}_{N+jd}).
\end{equation}
We can apply the difference formula to $d$-neutron separation energies, $S_{dn}$.  We note that the $k$-point formula for the binding energies equals the $(k-1)$-point formula for the separation energies
\begin{align}
B\{&{}^{A}_{Z}\textrm{X}_{N},{}^{A+d}_{\quad Z}\textrm{X}_{N+d}, \cdots, {}^{A+(k-1)d}_{\qquad \quad Z}\textrm{X}_{N+(k-1)d}\}_\pm \nonumber \\ 
= &S_{dn}\{{}^{A+d}_{\quad Z}\textrm{X}_{N+d}, {}^{A+2d}_{\quad \; \; Z}\textrm{X}_{N+2d},\cdots, {}^{A+(k-1)d}_{\qquad \quad Z}\textrm{X}_{N+(k-1)d}\}_\pm
\end{align}

\begin{figure}
    \centering
    \includegraphics[height=6cm]{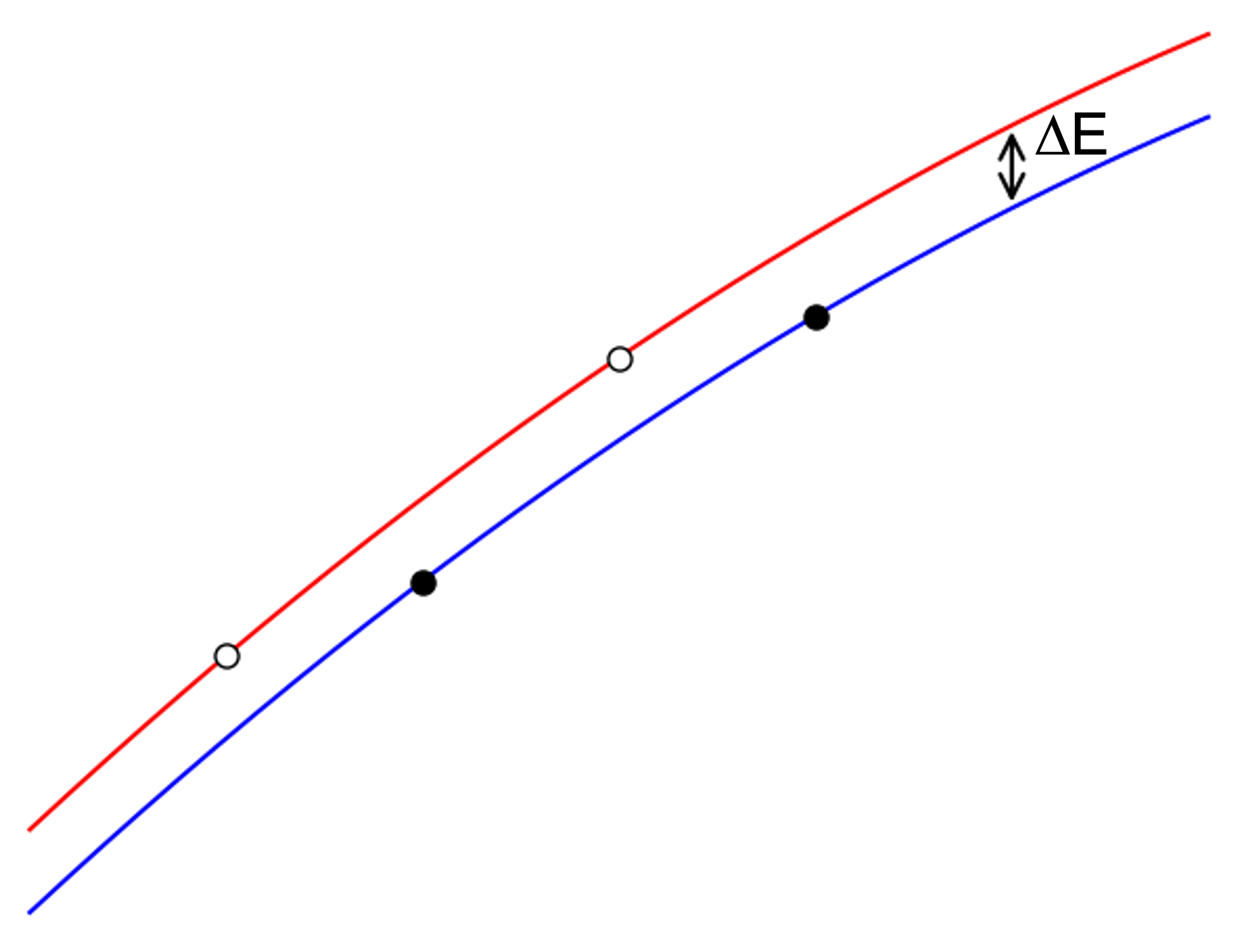}  
    \caption{\textbf{Sketch of the $k$-point difference formula for the binding energies.}  We are plotting binding energy versus neutron number.  Starting from $N$ neutrons, the two parallel curves go through the even and odd multiples of $d$ neutrons and are separated by a constant shift value of $\Delta E$.  The curves are polynomials of degree $k-2$.  \label{fig:energy_diff}}
\end{figure}

We can make a geometric interpretation of this difference formula.  Let us define $\Delta E$ as
\begin{equation}
    \Delta E= 2^{2-k} B\{{}^{A}_{Z}\textrm{X}_{N},{}^{A+d}_{\quad Z}\textrm{X}_{N+d}, \cdots, {}^{A+(k-1)d}_{\qquad \quad Z}\textrm{X}_{N+(k-1)d}\}_\pm.
\end{equation}
The desired choice for $\pm$ depends on whether the last point $N+kd$ is on the upper curve ($+$) or the lower curve ($-$). With this definition, $\Delta E$ corresponds to the energy difference between two parallel curves that are polynomials of degree $k-2$ that go through the binding energies for the even and odd multiples of $d$ neutrons, starting from $N$ neutrons.  This is shown in Fig.~\ref{fig:energy_diff}.  We note that the pairing binding energies and quartet binding energies that we define the next sections are twice this energy $\Delta E$.  

\subsection{Derivation of the Geometric Interpretation}

To derive the relationship between the difference formula and the energy spacing $\Delta E$, we model the binding energies as lying on two parallel curves. Let the lower curve be a polynomial $P(x)$ of degree $k-2$. The upper curve is then $P(x) + \Delta E$. Since the data points alternate between these curves, we can write the binding energy at step $j$ as:
\begin{equation}
    B_j = P(x_j) + \frac{\Delta E}{2} \left[ 1 - (-1)^j \right] (-1)^\sigma,
\end{equation}
where $\sigma$ determines the phase of the staggering, specifying whether the even-indexed points lie on the upper or lower energy curve. 

We identify that the $k$-point difference formula defined in Eq.~\eqref{eqn:k-point} is mathematically equivalent to $\pm 1$ times the finite difference operator of order $k-1$, denoted here as $D^{k-1}$. We apply this linear operator to $B_j$:
\begin{equation}
    D^{k-1}[B] = D^{k-1}[P] + D^{k-1}\left[ \text{const} \right] - (-1)^\sigma \frac{\Delta E}{2} D^{k-1} [(-1)^j].
\end{equation}
Two key properties of finite differences simplify this expression:
\begin{enumerate}
    \item The $(k-1)$-th difference of a polynomial of degree $k-2$ is exactly zero. Thus, $D^{k-1}[P] = 0$.
    \item The difference of a constant is zero.
\end{enumerate}
We are left evaluating the operator on the alternating term $(-1)^j$. The explicit summation over the $k$ points gives
\begin{align}
    D^{k-1} [(-1)^j] &= \sum_{j=0}^{k-1} \binom{k-1}{j} (-1)^{k-1-j} (-1)^j \nonumber \\
    &= (-1)^{k-1} \sum_{j=0}^{k-1} \binom{k-1}{j} = (-1)^{k-1} 2^{k-1}.
\end{align}
Substituting this back into the expression for the binding energy difference, we get
\begin{equation}
    B\{{}^{A}_{Z}\textrm{X}_{N},{}^{A+d}_{\quad Z}\textrm{X}_{N+d}, \cdots, {}^{A+(k-1)d}_{\qquad \quad Z}\textrm{X}_{N+(k-1)d}\}_\pm = \pm \frac{\Delta E}{2} 2^{k-1} = \pm \Delta E \cdot 2^{k-2}.
\end{equation}
Solving for the energy spacing yields the geometric formula claimed above,
\begin{equation}
    \Delta E = 2^{2-k} B\{{}^{A}_{Z}\textrm{X}_{N},{}^{A+d}_{\quad Z}\textrm{X}_{N+d}, \cdots, {}^{A+(k-1)d}_{\qquad \quad Z}\textrm{X}_{N+(k-1)d}\}_\pm .
\end{equation}

\subsection{S-Wave Pair Binding}
For the $0^+$ ground states of a nucleus with even $Z$ and even $N$, we can define the s-wave pair binding energy, $2\Delta_S$ using the $k$-point formula for the binding energies or the $(k-1)$-point for the one-neutron separation energies,
\begin{align}
    2\Delta_S= 2\Delta E = & 2^{3-k} B\{{}^{A}_{Z}\textrm{X}_{N},{}^{A+1}_{\quad Z}\textrm{X}_{N+1}, \cdots, {}^{A+(k-1)}_{\qquad \; \; Z}\textrm{X}_{N+(k-1)}\}_\pm \nonumber \\
    = & 2^{3-k} S_{n}\{{}^{A+1}_{\quad Z}\textrm{X}_{N+1}, {}^{A+2}_{\quad Z}\textrm{X}_{N+2},\cdots, {}^{A+(k-1)}_{\qquad \; \; Z}\textrm{X}_{N+(k-1)}\}_\pm.
\end{align}
In the following, we use the $(k-1)$-point separation energy formula since it will more easily generalize to the p-wave pairing case.  In Fig.~\ref{tab:s-wave_binding}, we compare results using two-point, three-point, and four-point difference formulas for the separation energies $S_n$.  The values are calculated using Atomic Mass Evaluation 2020 (AME2020) masses \cite{Huang2021,Wang2021}.  The two-point difference formula implicitly assumes that $S_n$ for the even and odd neutron number curves remains fixed as more neutrons are added.  The three-point difference formula corrects for this by taking into account a linear dependence of the separation energy on neutron number.  The four-point difference formula includes a quadratic dependence for the separation energy.  In general, higher-order formulas do a better job of subtracting out the background dependence for the separation energies.  However, one should not cross orbital shell or subshell closures or other effects that produce significant changes in the underlying nuclear structure.  

\begin{table}[h]
    \centering
    \caption{\textbf{S-wave pairing energies in units of MeV.} We show results using two-point, three-point, and four-point difference formulas for the separation energies $S_n$. The data without one-sigma error bars have uncertainties smaller than the significant digits shown. \label{tab:s-wave_binding}}
    \vspace{0.2cm}
    \begin{tabular}{lll}
        \toprule
        Nuclei & Predominant Orbitals & $2\Delta_S$ \\
        \midrule
        $S_n\{{}^{17}_{\;\, 8}\textrm{O}_{9},{}^{18}_{\;\, 8}\textrm{O}_{10} \}_+$  & $(1d_{5/2})^2$   & $3.902$ \\
        $\frac{1}{2}S_n\{{}^{17}_{\;\, 8}\textrm{O}_{9},{}^{18}_{\;\, 8}\textrm{O}_{10}, {}^{19}_{\;\, 8}\textrm{O}_{11} \}_-$  & $(1d_{5/2})^2$   & $3.996(1)$ \\ 
        $\frac{1}{4}S_n\{{}^{17}_{\;\, 8}\textrm{O}_{9},{}^{18}_{\;\, 8}\textrm{O}_{10}, {}^{19}_{\;\, 8}\textrm{O}_{11}, 
        {}^{20}_{\;\, 8}\textrm{O}_{12}\}_+$  & $(1d_{5/2})^2$   & $3.934(2)$ \\        
        
        $S_n\{{}^{41}_{20}\textrm{Ca}_{21},{}^{42}_{20}\textrm{Ca}_{22} \}_+$  & $(1f_{7/2})^2$   & $3.118$ \\
        $\frac{1}{2}S_n\{{}^{41}_{20}\textrm{Ca}_{21},{}^{42}_{20}\textrm{Ca}_{22}, {}^{43}_{20}\textrm{Ca}_{23} \}_-$  & $(1f_{7/2})^2$   & $3.333$ \\    
        $\frac{1}{2}S_n\{{}^{41}_{20}\textrm{Ca}_{21},{}^{42}_{20}\textrm{Ca}_{22}, {}^{43}_{20}\textrm{Ca}_{23}, 
        {}^{44}_{20}\textrm{Ca}_{24}\}_-$  & $(1f_{7/2})^2$   & $3.353$ \\

        $S_n\{{}^{209}_{\;\;82}\textrm{Pb}_{127},{}^{210}_{\;\;82}\textrm{Pb}_{128} \}_+$  & $(2g_{9/2})^2$   & $1.248(2)$ \\
        $\frac{1}{2}S_n\{{}^{209}_{\;\;82}\textrm{Pb}_{127},{}^{210}_{\;\;82}\textrm{Pb}_{128}, {}^{211}_{\;\;82}\textrm{Pb}_{129} \}_-$  & $(2g_{9/2})^2$   & $1.299(2)$ \\        
        $\frac{1}{4}S_n\{{}^{209}_{\;\;82}\textrm{Pb}_{127},{}^{210}_{\;\;82}\textrm{Pb}_{128}, {}^{211}_{\;\;82}\textrm{Pb}_{129},
        {}^{212}_{\;\;82}\textrm{Pb}_{130}
        \}_+$  & $(2g_{9/2})^2$   & $1.309(2)$ \\            
        
        \bottomrule
    \end{tabular}
\end{table}

\subsection{P-Wave Pair Binding}
In order to find a clean signal for p-wave pairing, we need to consider nuclear states where s-wave pairing cannot also contribute to the binding.  For this purpose, we look at states with unnatural parity.  In the following, we consider $1^+$ states for nuclei with even $Z$ and even $N$.  We look for pairs where one neutron is mostly in the ground state orbital and one neutron is mostly in an excited state orbital.  We use say mostly because the actual many-body wavefunction is a superposition of many combinations. 

In order to compute the p-wave binding energy, we use difference formulas for the separation energies associated with the removal of the excited state orbital.  Since the amount of binding energy is quite weak, we use the highest-order difference formula possible.  In Fig.~\ref{tab:p-wave_binding} we show p-wave binding energies for several examples.  We observe that the p-wave binding energies in nuclei are about an order of magnitude smaller than the s-wave binding energies.  The values are calculated using AME2020 masses \cite{Wang2021,Huang2021} and the Evaluated Nuclear Structure Data File (ENSDF) from the National Nuclear Data Center (NNDC) \cite{NNDC, Kondev2021}. The spin and parity assignments for ${}^{18}_{\;\, 8}\textrm{O}_{10}(1^+)$ and ${}^{56}_{26}\textrm{Fe}_{30}(1^+)$ require further confirmation.  We note that ${}^{56}_{26}\textrm{Fe}_{30}(1^+)$ and ${}^{58}_{28}\textrm{Ni}_{30}(1^+)$ involve paired neutrons from nearly degenerate orbitals due to pseudospin symmetry \cite{Hecht:1969zz,Arima:1969zz,Bahri:1992zz,Ginocchio:1997mb}. 

\begin{table}[h]
    \centering
    \caption{\textbf{P-wave pairing energies in units of MeV.} We show results using two-point and three-point difference formulas for the separation energies $S_n$ for the removal of an excited state orbital. The data without one-sigma error bars have uncertainties smaller than the significant digits shown. \label{tab:p-wave_binding}}
    \vspace{0.2cm}
    \begin{tabular}{lll}
        \toprule
        Nuclei & Predominant Orbitals & $2\Delta_P$ \\
        \midrule
        $S_n\{{}^{17}_{\;\, 8}\textrm{O}_{9}(\frac{3}{2}^+),{}^{18}_{\;\, 8}\textrm{O}_{10}(1^+) \}_+$  & $1d_{5/2} \otimes 1d_{3/2}$   & $0.172(12)$ \\
        
        $\frac{1}{2}S_n\{{}^{55}_{26}\textrm{Fe}_{29}(\frac{5}{2}^-),{}^{56}_{26}\textrm{Fe}_{30}(1^+), {}^{57}_{26}\textrm{Fe}_{31}(\frac{5}{2}^-) \}_-$  & 
        $2p_{3/2} \otimes 1f_{5/2}$    & $0.139(3)$ \\    

        $\frac{1}{2}S_n\{{}^{57}_{28}\textrm{Ni}_{29}(\frac{5}{2}^-),{}^{58}_{28}\textrm{Ni}_{30}(1^+), {}^{59}_{28}\textrm{Ni}_{31}(\frac{5}{2}^-) \}_-$  & 
        $2p_{3/2} \otimes 1f_{5/2}$    & $0.245(1)$ \\ 
        
        \bottomrule
    \end{tabular}
\end{table}

\subsection{Quartet Binding}
We can compute the quartet binding energy using the difference energy formulas with stride $d=2$ for the binding energies and the two-neutron separation energies,
\begin{align}
    4\Delta_Q= 2\Delta E = & 2^{3-k} B\{{}^{A}_{Z}\textrm{X}_{N},{}^{A+2}_{\quad Z}\textrm{X}_{N+2}, \cdots, {}^{A+2(k-1)}_{\qquad \quad Z}\textrm{X}_{N+2(k-1)}\}_\pm \nonumber \\
    = & 2^{3-k} S_{2n}\{{}^{A+2}_{\quad Z}\textrm{X}_{N+2}, {}^{A+4}_{\quad Z}\textrm{X}_{N+4},\cdots, {}^{A+2(k-1)}_{\qquad \quad Z}\textrm{X}_{N+2(k-1)}\}_\pm.
\end{align}
In Table~\ref{tab:quartet_binding}, we show results using two-point and three-point difference formulas for the two-neutron separation energies $S_{2n}$.  The values are calculated using AME2020 masses \cite{Wang2021,Huang2021}.  The quartet binding strongly depends on the overlap between the valence neutron orbitals, and the signature is weaker in heavier systems.

For the tin isotopes above $^{100}$Sn, the odd-neutron isotopes do not have a simple pattern for their ground state spin values.  This suggests that a combination of the $1g_{7/2}$, $2d_{7/2}$, $3s_{1/2}$ orbitals are being filled in some correlated combinations.  This would increase the overlap of the neutron orbitals and may contribute to the extended pattern of quartet binding in the two-neutron separation energies.  The odd-neutron isotopes for lead below $^{208}$Pb also do not have a simple pattern for their ground state spin values.  In this case, however, the quartet pattern may be driven by the filling of the $1i_{13/2}$ orbitals for the even-neutron isotopes.  It may be that near the neutron driplines of some isotopes, the conditions for multimodal superfluidity are enhanced by the reduction in the splitting between orbitals with different parity. This seems indicated by data for $^{28}$O and $^{30}$F \cite{Kondo:2023lty,SAMURAI21-NeuLAND:2024kah}.  This could lead to more examples of multimodal superfluidity as more rare isotopes are discovered.

\begin{table}[h]
    \centering
    \caption{\textbf{Quartet binding energies in units of MeV.} We show results using two-point and three-point difference formulas for the two-neutron separation energies $S_{2n}$. \label{tab:quartet_binding}}
    \vspace{0.2cm}
    \begin{tabular}{lll}
        \toprule
        Nuclei & Predominant Orbitals & $4\Delta_Q$ \\
        \midrule

        $S_{2n}\{{}^{6}_{2}\textrm{He}_{4},{}^{8}_{ 2}\textrm{He}_{4} \}_+$  & $(1p_{3/2})^4$   & $1.150    $ \\
        
        $\frac{1}{2}S_{2n}\{{}^{18}_{\;\, 8}\textrm{O}_{10},{}^{20}_{\;\, 8}\textrm{O}_{12},
        {}^{22}_{\;\, 8}\textrm{O}_{14} \}_-$  & $(1d_{5/2})^4$   & $0.142(28)$ \\      
        
        $\frac{1}{2}S_{2n}\{{}^{42}_{20}\textrm{Ca}_{22},{}^{44}_{20}\textrm{Ca}_{24}, {}^{46}_{20}\textrm{Ca}_{26} \}_-$  & $(1f_{7/2})^4$   & $0.236(1)$ \\  

        $\frac{1}{2}S_{2n}\{{}^{44}_{20}\textrm{Ca}_{24}, {}^{46}_{20}\textrm{Ca}_{26}, 
        {}^{48}_{20}\textrm{Ca}_{28} \}_+$ & $(1f_{7/2})^4$   & $0.332(3)$ \\           

        $\frac{1}{2}S_{2n}\{{}^{106}_{\;\;50}\textrm{Sn}_{56},{}^{108}_{\;\;50}\textrm{Sn}_{58}, {}^{110}_{\;\;50}\textrm{Sn}_{60} \}_+$  & combination  & $0.033(11)$ \\  

        $\frac{1}{2}S_{2n}\{{}^{108}_{\;\;50}\textrm{Sn}_{58},{}^{110}_{\;\;50}\textrm{Sn}_{60}, {}^{112}_{\;\;50}\textrm{Sn}_{62} \}_-$  & combination  & $0.007(17)$ \\  

        $\frac{1}{2}S_{2n}\{{}^{110}_{\;\;50}\textrm{Sn}_{60},{}^{112}_{\;\;50}\textrm{Sn}_{62}, {}^{114}_{\;\;50}\textrm{Sn}_{64} \}_+$  & combination  & $0.025(16)$ \\  

        $\frac{1}{2}S_{2n}\{{}^{112}_{\;\;50}\textrm{Sn}_{62},{}^{114}_{\;\;50}\textrm{Sn}_{64}, {}^{116}_{\;\;50}\textrm{Sn}_{66} \}_-$  & combination  & $0.015(7)$ \\    
        
        $\frac{1}{2}S_{2n}\{{}^{114}_{\;\;50}\textrm{Sn}_{64},{}^{116}_{\;\;50}\textrm{Sn}_{66}, {}^{118}_{\;\;50}\textrm{Sn}_{68} \}_+$  & combination  & $0.050(30)$ \\           

        $\frac{1}{2}S_{2n}\{{}^{194}_{\;\;82}\textrm{Pb}_{112},{}^{196}_{\;\;82}\textrm{Pb}_{114}, {}^{198}_{\;\;82}\textrm{Pb}_{116} \}_+$  & combination   & $0.047(22)$ \\  

        $\frac{1}{2}S_{2n}\{{}^{196}_{\;\;82}\textrm{Pb}_{114},{}^{198}_{\;\;82}\textrm{Pb}_{116}, {}^{200}_{\;\;82}\textrm{Pb}_{118} \}_-$  & combination   & $0.057(16)$ \\  
        
        $\frac{1}{2}S_{2n}\{{}^{198}_{\;\;82}\textrm{Pb}_{116},{}^{200}_{\;\;82}\textrm{Pb}_{118}, {}^{202}_{\;\;82}\textrm{Pb}_{120} \}_+$  & combination  & $0.021(15)$ \\   

        $\frac{1}{2}S_{2n}\{{}^{200}_{\;\;82}\textrm{Pb}_{118},{}^{202}_{\;\;82}\textrm{Pb}_{120}, {}^{204}_{\;\;82}\textrm{Pb}_{122} \}_-$  & combination  & $0.013(13)$ \\  

        $\frac{1}{2}S_{2n}\{{}^{202}_{\;\;82}\textrm{Pb}_{120},{}^{204}_{\;\;82}\textrm{Pb}_{122}, {}^{206}_{\;\;82}\textrm{Pb}_{124} \}_+$  & combination   & $0.014(7)$ \\  

        $\frac{1}{2}S_{2n}\{{}^{204}_{\;\;82}\textrm{Pb}_{122},{}^{206}_{\;\;82}\textrm{Pb}_{124}, {}^{208}_{\;\;82}\textrm{Pb}_{126} \}_-$  & combination   & $0.110(2)$ \\         
        
        $\frac{1}{2}S_{2n}\{{}^{210}_{\;\;82}\textrm{Pb}_{128},{}^{212}_{\;\;82}\textrm{Pb}_{130}, {}^{214}_{\;\;82}\textrm{Pb}_{132} \}_-$  & $(2g_{9/2})^4$   & $0.014(2)$ \\  

        $\frac{1}{2}S_{2n}\{{}^{212}_{\;\;84}\textrm{Po}_{128},{}^{214}_{\;\;84}\textrm{Po}_{130}, {}^{216}_{\;\;84}\textrm{Po}_{132} \}_-$  & $(2g_{9/2})^4$   & $0.019(1)$ \\  

        $\frac{1}{2}S_{2n}\{{}^{214}_{\;\;84}\textrm{Po}_{130},{}^{216}_{\;\;84}\textrm{Po}_{132}, {}^{218}_{\;\;84}\textrm{Po}_{134} \}_+$  & $(2g_{9/2})^4$   & $0.015(2)$ \\            
        
        $\frac{1}{2}S_{2n}\{{}^{214}_{\;\;86}\textrm{Rn}_{128},{}^{216}_{\;\;86}\textrm{Rn}_{130}, {}^{218}_{\;\;86}\textrm{Rn}_{132} \}_-$  & $(2g_{9/2})^4$   & $0.079(12)$ \\            
        
        \bottomrule
    \end{tabular}
\end{table}

\subsection{Tetraneutron resonances versus multimodal quartets}

The possibility of a low-energy tetraneutron resonance composed of four neutrons has been the subject of intense theoretical and experimental investigation for over two decades. The tetraneutron is postulated as a low-energy resonance existing in vacuum, independent of any many-body environment. A recent experiment reported the observation of a resonance-like structure in the missing-mass spectrum of the reaction $^8\text{He}(p, p\alpha)4n$~\cite{Duer:2022ehf}. This result builds upon earlier experimental searches~\cite{Shimoura:2002ns,Kisamori:2016jie} and has stimulated a significant body of new theoretical work aimed at understanding the nature of the four-neutron continuum.  Although some studies suggest the existence of a low-energy resonance~\cite{Shirokov:2016ywq,Li:2019pmg,Fossez:2016dch}, other calculations find no evidence for such a state, instead predicting a non-resonant continuum governed by the large neutron-neutron scattering length~\cite{Pieper:2003dc,Hiyama:2016nwn,Deltuva:2018xoa,Higgins:2020avy,Wu:2026tis}.  An important element in the interpretation of these experimentally-observed energy peaks is the role of reaction mechanisms and initial state correlations. 

A recent analysis in Ref.~\cite{Lazauskas:2022mvq} provides an explanation for the energy peak observed in Ref.~\cite{Duer:2022ehf}. They demonstrate that the observed enhancement in the cross section can be reproduced as a kinematic consequence of the large neutron-neutron scattering length and the shallow binding energy of the valence neutrons in the $^8\text{He}$ projectile.  This peak in the energy spectrum persists even when the nuclear interaction is modified to omit p-wave interactions~\cite{Lazauskas:2022mvq}.  Their proposed reaction mechanism for the energy peak observed in these knockout reactions is therefore not directly related to multimodal superfluidity, which requires p-wave attraction.

What then is the structural signature of multimodal quartetting in an atomic nucleus, aside from an oscillation in the two-neutron separation energy as a function of neutron number?  Quartet correlations arising from multimodal superfluidity are not a localized clustering analogous to alpha particle clustering.  This was verified in a recent lattice effective field theory calculation of the valence neutrons in $^8$He \cite{Zhang:2025uin}.  This finding is also consistent with previous theoretical calculations \cite{Caurier:2005rb,Pieper:2007ax,Kanada-Enyo:2007iri,Papadimitriou:2011jx}, and the more balanced configuration of valence neutrons in $^8$He explains the decrease in the charge radius from $^6$He to $^8$He \cite{Wang:2004ze,Mueller:2007dhq}.

The multimodal quartet is a more subtle correlation among the four neutrons, delocalized in space and stabilized by the interplay of s-wave and p-wave interactions in the many-body environment.  The signature is particularly strong in uniform neutron matter, as we have shown from the four-body density cumulants in momentum space.  An analogous correlation, albeit weaker in magnitude, must also appear in four-body cumulants for single-particle orbitals in atomic nuclei.  Investigations to calculate these correlations are currently under way.

\bibliography{References}

\end{document}